\documentclass[english, 12pt]{article}
\usepackage[T1]{fontenc}
\usepackage[utf8]{inputenc}
\usepackage[english]{babel}
\usepackage{lmodern}
\usepackage[top=1in, bottom=1in, left=1in, right=1in]{geometry}
\usepackage{verbatim}
\usepackage[width=0.8\textwidth, justification=centering]{caption}
\usepackage[capposition=top]{floatrow}
\usepackage{amsmath,amssymb,amsopn,dsfont,mathrsfs, bm,mathtools,bbm}
\usepackage{placeins,flafter,longtable,array,booktabs,threeparttable}
\usepackage{color}
\usepackage{natbib}
\usepackage{xr}
\usepackage{subcaption}
\usepackage[pdftex]{graphicx}
\usepackage[hidelinks]{hyperref}
\usepackage[symbol]{footmisc}
\usepackage{tikz}
\usepackage{mathdots}
\usepackage{yhmath}
\usepackage{cancel}
\usepackage{color}
\usepackage{siunitx}
\usepackage{array}
\usepackage{multirow}
\usepackage{tabularx}
\newcolumntype{Y}{>{\centering\arraybackslash}X}
\usepackage{extarrows}
\usepackage{booktabs}
\usepackage{ragged2e}
\usetikzlibrary{fadings}
\usetikzlibrary{patterns}
\usetikzlibrary{shadows.blur}
\usetikzlibrary{shapes}

\renewcommand{\thefootnote}{\fnsymbol{footnote}}
\makeatletter
\newtheorem{prop}{Proposition}
\newtheorem{lem}{Lemma}
\newtheorem{hyp}{Assumption}

\newtheorem{defin}{Definition}
\newtheorem{conj}{Conjecture}

\renewcommand{\Pr}{\mathbb{P}}
\newcommand{\E}{\mathbb{E}}
\newcommand{\V}{\mathbb{V}}

\newcommand{\N}{\mathbb N}

\newcommand{\eps}{\varepsilon}

\newcommand{\indep}{\perp \!\!\! \perp}

\newcommand{\convP}{\stackrel{P}{\longrightarrow}}
\newcommand{\convD}{\stackrel{d}{\longrightarrow}}
\newcommand{\convNor}[1]{\stackrel{d}{\longrightarrow} \mathcal{N}\left(0,#1\right)}

\renewcommand{\section}{\@startsection{section}{2}{0mm}{-1.5\baselineskip}{1\baselineskip}{\normalfont\Large\bfseries}}
\renewcommand{\subsection}{\@startsection{subsection}{2}{0mm}{-1.2\baselineskip}{1\baselineskip}{\normalfont\large\bfseries}}
\renewcommand{\subsubsection}{\@startsection{subsubsection}{3}{0mm}{-0.8\baselineskip}{0.4\baselineskip}{\normalfont\normalsize\bfseries\itshape}}
\allowdisplaybreaks

\date{}
\begin{document}

\title{\vspace{-2cm}Revisiting Randomization with the Cube Method}
\author{Laurent Davezies\thanks{CREST - ENSAE - IP Paris, laurent.davezies@ensae.fr} \and Guillaume Hollard\thanks{CREST - Ecole Polytechnique - IP Paris, guillaume.hollard@polytechnique.edu} \and Pedro Vergara Merino\thanks{CREST - ENSAE - IP Paris, pedro.vergaramerino@ensae.fr}}
\maketitle
\centering
\begin{abstract}

We introduce a new randomization procedure for experiments based on the cube method, which achieves near-exact covariate balance. This ensures compliance with standard balance tests and allows for balancing on many covariates, enabling more precise estimation of treatment effects using pre-experimental information. We derive theoretical bounds on imbalance as functions of sample size and covariate dimension, and establish consistency and asymptotic normality of the resulting estimators. Simulations show substantial improvements in precision and covariate balance over existing methods, particularly when the number of covariates is large.

\medbreak
\end{abstract}

\smallbreak

{\let\thefootnote\relax\footnote{{

We would like to thank Riccardo d'Adamo,  Isaiah Andrews, Yuehao Bai, Stéphane Bonhomme, Guillaume Chauvet, Russell Davidson, Max Farrell, Xavier d'Haultfoeuille, Marc Henry, Keisuke Hirano, Xinran Li, John List, Kirill Ponomarev, Pauline Rossi, Jean Rubin, Felix Schleef, Azeem Shaikh, Sami Stouli, Max Tabord-Meehan, Yves Tillé, Panos Toulis and many other colleagues in workshops and conferences whose comments and suggestions helped improve this paper. In particular, we would like to thank everybody at the 2023 Bristol Econometric Study Group, the 2023 Advances with Field Experiments Conference, the 18th IZA \& 5th IZA/CREST Conference, the 2023 European Winter Meeting of the Econometric Society, the AMSE Big Data and Econometrics Seminar, the 2024 RCEA International Conference in Economics, Econometrics, and Finance, the 2024 BSE Summer Forum Workshop on Microeconometrics and Policy Evaluation, the 2024 AFSE Annual Conference, the 2024 Annual Conference of the International Association for Applied Econometrics, the Econometrics Workshop at UChicago, the PhD Workshop at Penn State.}}}

 \renewcommand{\thefootnote}{\number\value{footnote}}
\setcounter{footnote}{0}
\newpage
\justifying
\section{Introduction} 
\label{sec:introduction}
 
Prior to running RCTs, researchers often have access to rich information on experimental units through baseline surveys or administrative data. Researchers would, ideally, like to create treatment and control groups that balance available covariates. How to randomize using available covariates is still a matter of debate, especially when the number of covariates is large.

We here introduce a randomization method based on the cube algorithm proposed by \citet{deville_efficient_2004} for survey sampling. 
Units are iteratively assigned to treatment or control groups. At each step, one unit is allocated, under the constraint that means of covariates in the two groups are equal. Assignment probabilities of unallocated units are updated at each step to ensure that empirical means of available covariates are the same in control and treatment.

\begin{figure}[htb]
    \centering
       \caption{Effect of Number of Covariates by Randomization Method}
    \makebox[\linewidth][c]{%
    \begin{subfigure}[h]{0.4\paperwidth}
        \centering   \includegraphics[width=\linewidth]{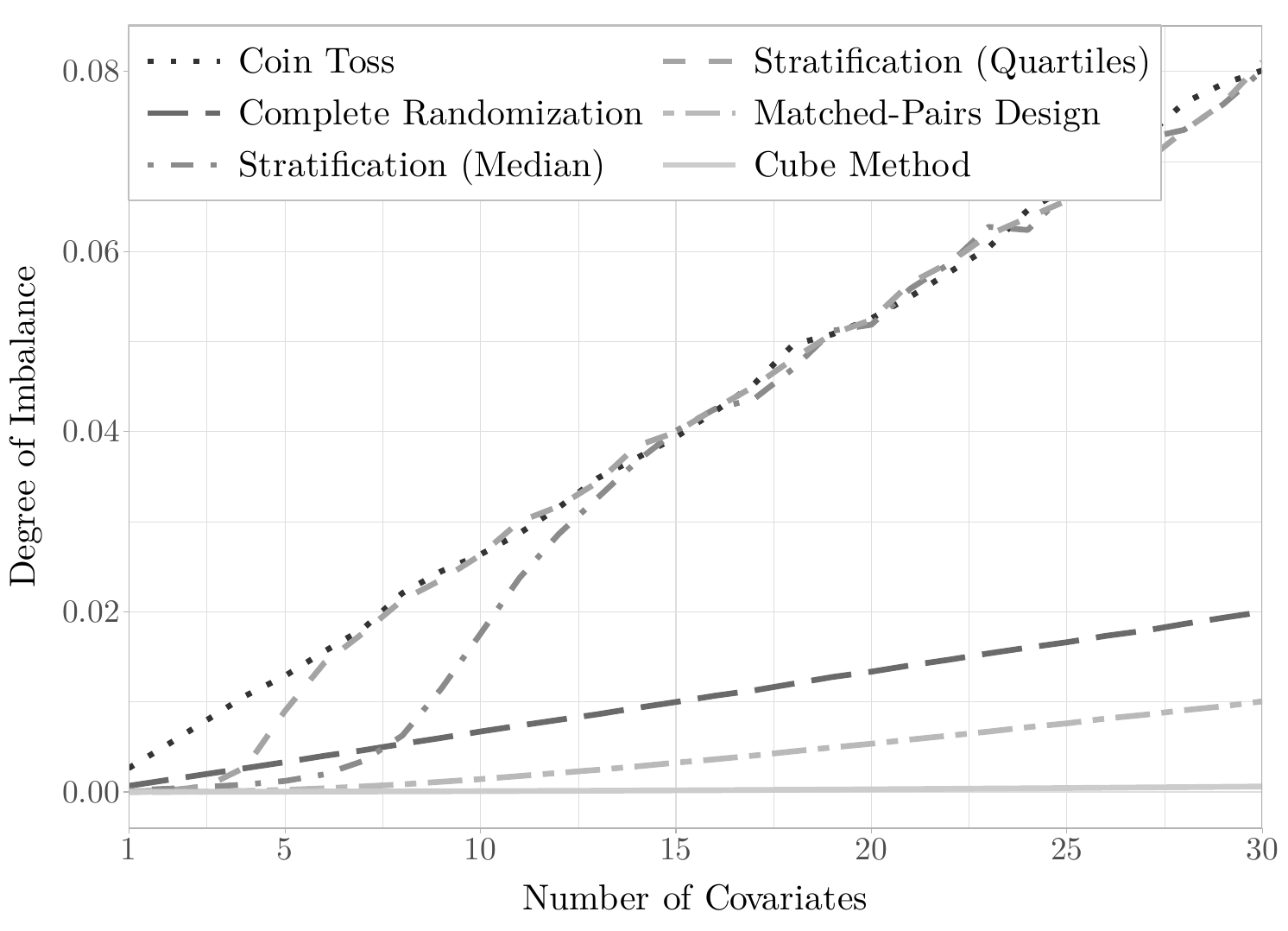} 
                \caption{Balance Quality} \label{fig:motex}
    \end{subfigure}
    \begin{subfigure}[h]{0.4\paperwidth}
        \centering
        \includegraphics[width=\linewidth]{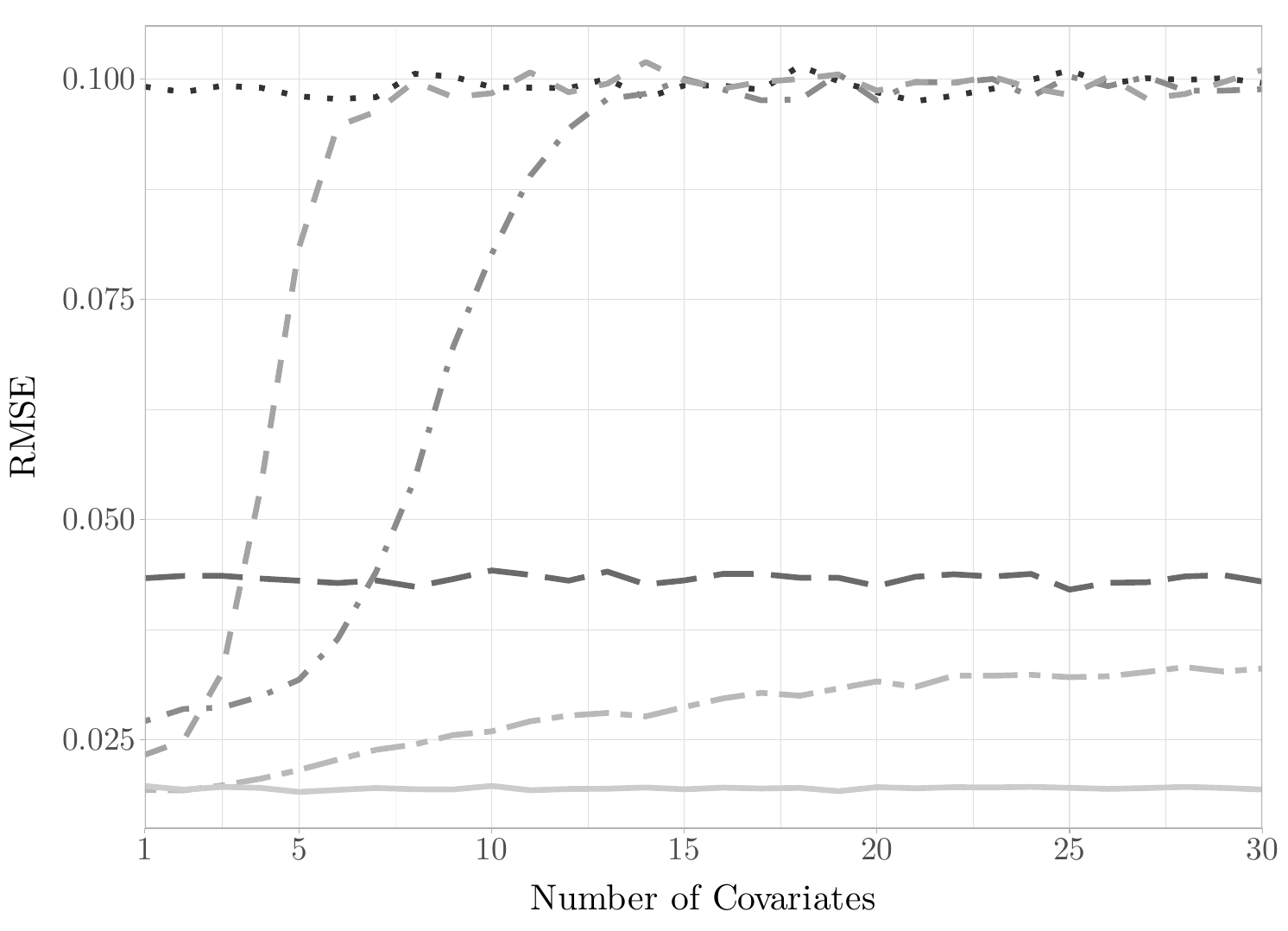} 
                \caption{Precision of ATE estimator} \label{fig:MCsim}
    \end{subfigure}}
    \begin{minipage}{\textwidth}\scriptsize
    This Figure shows the effect of the number of covariates on balancing quality and precision of ATE estimates. Panel (a) shows the degree of imbalance between control group $C$ and treatment group $T$, measured by $\operatorname{Imb}_{n,p}(X)=\E\left(\left|\left|\frac{2}{n}\sum_{i \in T} X_{i}-\frac{2}{n}\sum_{i \in C}X_{i}\right|\right|^2\right)=\E\left(\sum_{j=1}^p\left(\frac{2}{n}\sum_{i \in T} X_{j,i}-\frac{2}{n}\sum_{i \in C}X_{j,i}\right)^2\right)$. Panel (b) shows the estimated root-mean-squared error of the Horvitz-Thompson estimator for the ATE. Here, we consider the case where only one variable explains potential outcomes. For both panels, we show Monte Carlo estimates over $5,000$ simulations for $p=1,
...,30$ and $n=500$ across different randomization methods. ``Coin Toss'' refers to allocating every unit to treatment independently with probabilities equal to 1/2. ``Complete Randomization'' refers to randomly selecting one of the $\binom{500}{250}$ samples of size 250 and allocating them to treatment. For ``Stratification'', we use complete randomization within the strata defined by the intersection of quintiles of covariates. ``Matched-Pairs Design'' refers to the method in \citet{bai_inference_2022}, which creates strata of size two by minimizing the sum of the intra-strata Euclidean distance of covariates and randomly assigning one unit by pair to treatment. The ``Cube Method'' refers to the method described in Section \ref{sec:cube} with only the first moment of each covariate. See Section \ref{sec:sim} for further details about the DGP and randomization methods.
\end{minipage}
\end{figure}

The cube method aims at balancing \textit{chosen} moments, such as mean, variance, or cross-moments. Still, it does not impose balancing constraints on moments considered of little relevance (for instance, high-order moments). In contrast,  stratification and matched-pairs designs, two prominent methods proposed by \citet{bai_inference_2022,bai_optimality_2022} and \citet{cytrynbaum_optimal_2023}, aim at reducing the distance between groups' joint probability distributions of covariates to achieve balance of \textit{all} moments. Crucially, we show that reducing the distance between joint probability distributions becomes infeasible in finite samples for more than a few covariates. Stratification may even backfire, i.e., increasing imbalances, when stratifying naively on ``too many'' covariates. The cube method greatly mitigates this curse of dimensionality. Figure \ref{fig:motex} illustrates how imbalances when randomizing with the cube method are much reduced compared to other methods as the number of covariates used for balancing $p$ grows, for a fixed sample size $n$.

The set of balanced covariates has a direct impact on the precision of estimated treatment effects. Maximal precision gains are achieved by selecting the set of covariates that are the most correlated with potential outcomes. But relevant covariates are only known after the RCT was run and must be chosen under some uncertainty. Furthermore, RCTs often estimate several treatment effects and corresponding outcomes could be correlated with different covariates, notably pretreatment outcomes. Balancing on a larger set of covariates increases the likelihood of selecting the most relevant ones. The cube method allows increasing the number of covariates, while being robust to the addition of irrelevant ones. Figure \ref{fig:MCsim} displays the precision of the average treatment effect estimator in a situation in which only one covariate, out of 30, correlates with potential outcomes. The cube method allows balancing on all 30 covariates, increasing precision, while not being penalized by adding 29 irrelevant covariates. The cube method stands out as other methods suffer from precision loss when balancing on irrelevant covariates.

This paper contributes to three streams of research. First, we extend the cube method, developed first for survey sampling by \cite{deville_efficient_2004}. The cube method is routinely used for sampling by national statistical institutes \citep[see][for a review of applications of the cube method]{tille_ten_2011}. Our technical contribution is to extend the scope of the cube method beyond sampling for estimating treatment effects in RCTs. Moving from sampling to RCTs implies redefining balancing constraints and deriving asymptotic properties for estimators of the population average treatment effect (PATE) and the sample average treatment effect (SATE). For the PATE, we show that the semiparametric efficiency bound in \cite{hahn_role_1998} is attained for large $n$ and fixed $p$ under a linearity assumption of the conditional expectation of potential outcomes. We also provide valid inference strategies for the estimators. As is the case for the other methods achieving the bound, precision improves when the share of the variance of potential outcomes explained by the covariates is larger. We thus formally motivate the interest of using a method that balances almost exactly a large number of covariates $p$ for a given sample size $n$.

Our second contribution is to compare the balancing performance of existing randomization methods when $p$, the number of covariates used for balancing, increases. Asymptotic properties of randomization methods as the number of units $n$ gets large are well studied. In contrast, their asymptotic behavior when $p$ grows has been seldom studied yet, to the best of our knowledge. We show that observed patterns in Figure \ref{fig:motex} are de facto generic for bounded covariates. A key finding is the identification of three distinct regimes where stratification exhibits different behaviors. For a small number of covariates (i.e., $p\ll \ln(n)$), stratification improves balancing compared to complete randomization. When $p\approx \ln(n)$, there is a critical regime where the balancing quality deteriorates quickly. Last, for $p \gg \ln(n)$, because of the small strata issue, stratification is similar to a coin toss and worse than complete randomization. In sum, stratification exhibits different balancing properties when $p$ varies. We also derive upper and lower bounds for imbalances in matched-pair designs, showing that contrary to stratification, it always performs better than complete randomization but that imbalances increase rather quickly for $p\gg \ln(n)$. In sharp contrast, the cube method grants the balance of (selected moments of) covariates with no critical change when $p\ll n$, allowing balancing on a larger set of covariates. 

Last, we contribute to the the literature that establishes pros and cons of randomization methods to guide experimenters when choosing a randomization method \citep{bruhn_pursuit_2009, athey_chapter_2017,bai_primer_2024}. We briefly review the recent use of covariate-balancing designs in RCTs, and discuss practical implications of using the cube method for the publication process. Indeed, as explained above, the cube method removes most of the bad luck that may arise from sampling errors, as the most unfavorable samples have a null probability of being selected. Avoiding large imbalances has implications for the publication process. \cite{snyder_examining_2024} show that balance checks generates p-hacking and/or publication bias.

The remainder of this paper is structured as follows. Section \ref{sec:setup} introduces the potential outcome framework and covariate balancing. Section \ref{sec:cube} presents the cube algorithm and its application to RCTs. Section \ref{sec:results} gives the balancing properties of the cube method, compares imbalances to other methods, and provides novel asymptotic expressions for the variance of average treatment effect estimators. We then specify two ways of performing inference based on asymptotic normality and the randomization mechanism. Section \ref{sec:sim} uses simulated and experimental data to show our precision gains and how the cube method might be less constrained by the curse of dimensionality. Finally, Section \ref{sec:practical} reviews current practices in RCTs and discusses practical considerations of the cube method.

\section{Setup}
\label{sec:setup}
This section presents the potential outcomes framework, provides assumptions on the data-generating process, and formally defines covariate balancing.

\subsection{Data Generating Process and Assignment Design}

We consider the standard Neyman-Rubin framework of potential outcomes where $Y_i(0)$ is the outcome of unit $i$ when untreated and $Y_i(1)$ is the outcome when treated. We consider $X_i$ a vector of $p$ covariates. According to the literature, we assume i.i.d.ness and the existence of second-order moments for all these variables.
\begin{hyp}~\label{hyp:iid+mom} \\
	$(Y_i(0),Y_i(1),X_i)$ are i.i.d. across $i$ and $\E\left(Y(0)^2+Y(1)^2+||X||^2\right)<\infty$
\end{hyp}
The researcher observes $(X_1,\ldots,X_n)$ for a finite sample of size $n$. She wants to randomly allocate these $n$ units to treatment according to a design $\Pi$, i.e., a distribution on the set of the possible treatment allocations $\{0,1\}^n$. If the design $\Pi$ does not depend on the potential outcomes, it balances potential outcomes in the treatment and control groups \textit{in average}, avoiding selection bias. The design $\Pi$ could depend on $(X_1,...,X_n)$. For instance, the treatment probability of a unit $i$ could depend on $X_i$ for various reasons, such as efficiency, cost of the treatment depending on $X_i$, or subpopulations of particular interest. In the following, $D_i$ is the dummy variable indicating if $i$ is treated or untreated. Researchers have to choose not only each individual selection probability $\Pr_{\Pi}(D_i=1|X_1,...,X_n)$ but the full design $\Pi$ that determines $\Pr_{\Pi}\left(\cap_{i=1,...,n}D_i=d_i|X_1,...,X_n\right)$ for any potential allocations $(d_i)_{i=1,...,n}\in \{0,1\}^n$. A major issue is exploiting the knowledge of $(X_1,...,X_n)$ to define a ``good'' design $\Pi$ to go beyond the balancing of potential outcomes in average. To study this question, let us formulate the assumption on the class of design we consider in the following.
\begin{hyp}~\label{hyp:sample}The researcher observes a sample $(X_i)_{i=1,...,n}$ of size $n$ and generates a random assignment $(D_i)_{i=1,...,n}$ according to a randomization design $\Pi$ such that:  
	\begin{equation}(D_1,...,D_n)\indep (Y_1(0),Y_1(1),...,Y_n(0),Y_n(1))|X_1,...,X_n \label{eq:sample}
	\end{equation}
	and for any $i=1,...,n$
	\begin{equation}\Pr_{\Pi}(D_i=1|X_1,...,X_n)=p(X_i)\in[c,1-c],\label{eq:pi}
	\end{equation}
	where $p$ is a function chosen by the researcher and $c$ is a positive constant.
\end{hyp}
Assumption \ref{hyp:sample} is usual and necessary in the literature on treatment effects estimation.
Equation \eqref{eq:sample} means that assignment is independent of the unknown potential outcomes, conditional on the auxiliary information $X$. Equation \eqref{eq:pi} specifies that the assignment probability of unit $i$ could depend on $X_i$ but not on $X_j$ for $j\neq i$. It also states that the propensity score $p(X_i)$ fulfills a common support condition. \\
In what follows, we denote the propensity score $p(X_i)$ as $\pi_i$. 
Randomization methods are often presented under the assumption that $\pi_i=1/2$ for all $i$. However, the cube method handles heterogeneous assignment probabilities, which are of particular interest in RCTs, for at least three reasons. First, in view to minimize the variance of
the estimator of the ATE, the optimal assignment probabilities corresponding to the so-called
Neyman allocation are $\pi_i= V(Y (1)|X)^{1/2} \left(V(Y (1)|X)^{1/2} + V(Y (0)|X)^{1/2}\right)^{-1}$.  Second, even if $V(Y (1)|X) = V(Y (0)|X)$ and if the researcher
wants to minimize the variance of estimates of $E(Y (1) - Y (0))$, she could also adapt some
assignment probabilities to heterogeneous costs $c_1(X)$, $c_0(X)$ of treatment and control to
fulfill a budget constraint. Third, the researcher could be interested in treatment effects
on some subpopulations $X = x$ that would not be precisely estimated if using a constant
assignment probability. More generally, in an armed bandit perspective, researchers may
adapt assignment probabilities with respect to what they learned to maximize some objective,
explore treatment effects on some subpopulations, and minimize regret.  Our proposition of design accommodates any propensity score type, offering complete flexibility to researchers concerning its definition.\\
Assumption \ref{hyp:sample} ensures that for any variable $W$ such that $(D_1,...,D_n)\indep (W_1,...,W_n)|X_1,...,X_n$ we have:
\begin{align}\E\left(\frac{1}{n}\sum_{i=1}^n\frac{W_{i}D_i}{\pi_i}\bigg|(X_{i'}, W_{i'})_{i'=1,...,n}\right)&=\E\left(\frac{1}{n}\sum_{i=1}^n\frac{W_{i}(1-D_i)}{1-\pi_i}\bigg|(X_{i'}, W_{i'})_{i'=1,...,n}\right)\notag\\
&=\frac{1}{n}\sum_{i=1}^nW_{i}.\label{eq:bal_average}\end{align}
Equation \eqref{eq:bal_average} is true for $W_i=X_i$ and $W_i=(Y_i(0),Y_i(1))$.
\subsection{Parameters of interest and estimators}
After the experiment, the researcher observes $Y_i=Y_i(1)\times D_i+Y_i(0)\times(1-D_i)$.  She will thus never observe both potential outcomes for the same unit.
researchers are generally interested in estimating the sample and population average treatment effects given by
\begin{equation}
	\operatorname{SATE}:\quad\theta_0=\frac{1}{n}\sum_{i=1}^nY_i(1)-Y_i(0)
	\label{eq:SATE}
\end{equation}
and
\begin{equation}
	\operatorname{PATE}:\quad\theta_0^\ast=\E\left[Y_i(1)-Y_i(0)\right],
	\label{eq:PATE}
\end{equation}
respectively.\footnote{In some cases, they are interested in similar parameters for some subpopulations: $\frac{1}{\sum_{i=1}^n\mathds{1}\{X_i\in \mathcal{X}\}}\sum_{i=1}^n(Y_i(1)-Y_i(0))\mathds{1}\{X_i\in\mathcal{X}\}$ or $\E\left[Y_i(1)-Y_i(0)|X_i\in \mathcal{X}\right]$. Estimators of these quantities are defined by restricting the sample to units such that $X_i\in\mathcal{X}$ and the asymptotic properties of these estimators follow from a straightforward adaptation of what is presented below.}

In this paper, we will focus on the Horvitz-Thompson estimator (HT) and the Hájek estimator (H), which are of central interest in RCTs. The Horvitz-Thompson estimator is
\begin{equation}
	\widehat{\theta}_{HT}=\frac{1}{n}\sum_{i=1}^n\left(\frac{Y_i D_i}{\pi_i}-\frac{Y_i (1-D_i)}{1-\pi_i}\right)
	\label{eq:HT}
\end{equation}
which is unbiased under Assumption \ref{hyp:iid+mom} and \ref{hyp:sample} for both the SATE and the PATE and is the difference between the inverse probability weighting estimators on the treated and the control group.

The Hájek estimator is
\begin{equation}
\widehat{\theta}_{H}=\frac{1}{\sum_{i=1}^n\frac{D_i}{\pi_i}}\sum_{i=1}^n\frac{Y_i D_i}{\pi_i}-\frac{1}{\sum_{i=1}^n\frac{1-D_i}{1-\pi_i}}\sum_{i=1}^n\frac{Y_i (1-D_i)}{1-\pi_i}
\label{eq:Hajec}
\end{equation}
and corresponds as well to the inverse probability weighting OLS estimator $$\widehat{\theta}_H=\arg\min_{\theta}\min_a\sum_{i=1}^nw_i\left(Y_i-a-\theta D_i\right)^2$$
for $w_i=\frac{1}{\pi_i}$ if $D_i=1$ and $w_i=\frac{1}{1-\pi_i}$ if $D_i=0$. Let $n_T$ denote the number of treated units and $n_C$ the number of control units. When $\pi_i$ is constant, $\hat{\theta}_{H}=\frac{1}{n_T}\sum_{i: D_i=1}Y_i-\frac{1}{n_C}\sum_{i:D_i=0} Y_i$ is the difference between the average on the treated group and the control group whereas $\hat{\theta}_{HT}=\frac{1}{\E(n_T)}\sum_{i: D_i=1} Y_i-\frac{1}{\E(n_C)}\sum_{i: D_i=0}Y_i$ is a slight modification of this difference of averages. If, additionally, $n_T$ and $n_C$ are fixed, both estimators are identical to the difference-in-means estimator.

\subsection{Balancing Constraints}

Randomization methods generate control and treament groups that are balanced \textit{on average}. A more stringent requirement is to generate groups that are \textit{exactly} balanced: 

\begin{defin}[Exactly-balancing Design]~\label{def:perfbal}\\
	A design $\Pi$ is exactly-balancing over $X=(X_1,...,X_p)'$ if for $(D_{i})_{i=1,...,n} $ sampled in $\Pi$ we always have for any $j=1,...,p$:
	\begin{equation}
		\label{eq:bal}
		\frac{1}{n}\sum_{i=1}^n \frac{X_{ji} D_i}{\pi_i} =\frac{1}{n}\sum_{i=1}^n\frac{X_{ji}(1-D_i)}{1-\pi_i}
	\end{equation}
\end{defin}	
Equation \eqref{eq:bal} describes equality between the estimated weighted averages in the treatment and control groups. A perfectly balanced assignment eliminates any allocation to the treatment that does not balance perfectly the covariates between treatment and control groups.\\
A common practice in experiments is to form treatment and control groups of fixed sizes, $n_T$, and $n_C=n-n_T$, respectively. This is equivalent to satisfying the following constraint for any possibe allocation $(d_1,...,d_n)$:
\begin{equation}
n_T=\sum_{i=1}^nd_i=\sum_{i=1}^n\pi_i=\E(n_T)
\label{eq:balsize}
\end{equation}
In that case we also have: $n_C=\sum_{i=1}^n(1-d_i)=\sum_{i=1}^n(1-\pi_i)=\E(n_C)$.
As recommended by \citet{deville_efficient_2004}, we also balance a constant ($X_{ji}=1$), for the treatment and control groups:
\begin{equation}
    \frac{1}{n}\sum_{i=1}^n \frac{ D_i}{\pi_i} =\frac{1}{n}\sum_{i=1}^n \frac{ (1-D_i)}{1-\pi_i} =\frac{1}{n}\sum_{i=1}^n1=1
    \label{eq:balconsTC}
\end{equation}
Under such assignment $\widehat{\theta}_{HT}$ defined in (\ref{eq:HT}) is equal to $\widehat{\theta}_H$ defined in (\ref{eq:Hajec}).
Notice that we can  rewrite \eqref{eq:bal}, \eqref{eq:balsize}, \eqref{eq:balconsTC} as 
\begin{equation}
	\label{eq:balZ}
	\frac{1}{n}\sum_{i=1}^n \frac{Z_{1i} D_i}{\pi_i} =\frac{1}{n}\sum_{i=1}^n Z_{1i} \Leftrightarrow \frac{1}{n}\sum_{i=1}^n \frac{Z_{0i} (1-D_i)}{1-\pi_i} =\frac{1}{n}\sum_{i=1}^n Z_{0i}
\end{equation}
with $Z_{1i}= (1, \frac{\pi_i}{1-\pi_i},\pi_i,\frac{X_i'}{1-\pi_i})'$ and $Z_{0i}= ( \frac{1-\pi_i}{\pi_i},1,1-\pi_i,\frac{X_i'}{\pi_i})'=\frac{1-\pi_i}{\pi_i}Z_{1i}$. If assignment probabilities are homogeneous (i.e., $\pi_i=\pi$), the balancing covariates are reduced to $Z_{1i}=Z_{0i}=(1,X_i')$ due to perfect multicollinearity, but this is no more the case if the $\pi_i$ are heterogeneous.\\
It is worth noticing that exact balance is not always attainable: for instance, if $n=101$ and $\pi_i=1/2$. Imposing \eqref{eq:balsize} implies $n_T=50.5$, which is simply impossible. But statistical analysis ensures that balancing up to a $o_p\left(\frac{1}{\sqrt{n}}\right)$ is sufficient to take full advantage of the auxiliary information $X_i, \pi_i$. We here propose an "almost" exactly balancing design $\Pi$ such that for $(D_i)_{i=1,...,n}\sim \Pi$,
\begin{equation}
	\label{eq:balZop}
	\frac{1}{n}\sum_{i=1}^n \frac{Z_{1i} D_i}{\pi_i} =\frac{1}{n}\sum_{i=1}^n Z_{1i}+o_p\left(\frac{1}{\sqrt{n}}\right) \text{   for   }Z_{1i}=\left(1, \frac{\pi_i}{1-\pi_i},\pi_i,\frac{X'_i}{1-\pi_i}\right)'
\end{equation}
As we will show below, the cube method always achieves almost exact balancing. As a consequence precision gains are asymptotically equivalent from those achieved under exact balancing. 

\section{The Cube Method}
\label{sec:cube}
\citet{deville_efficient_2004} first introduced the cube method to produce samples balanced to the population. The cube method consists of an algorithm in two steps: the \textit{flight} and \textit{landing} phases. The technique gets its name from the graphical representation of a sampling problem. Equation \eqref{eq:balZ}  ensures that balancing treatment and control groups in an experimental setting for some covariates is equivalent to balancing the treatment group to the entire sample. Let us consider the $n$-cube $C=[0,1]^n$. Each vertex of $C$ (from $2^n$ possibilities) represents a possible allocation: for instance, $(1,1,...,1)$ corresponds to the situation where all units are allocated to treatment, $(1,0,1,0,...,1,0)$ corresponds to the case where the treatment group is $\{i: i \text{ odd}\}$. A sampling design $\Pi$ corresponds to how a vertex is selected. Recall that we consider a framework where researchers impose that Equation \eqref{eq:pi} holds for $\Pi$ and a vector $(\pi_i)_{i=1,...,n}$.

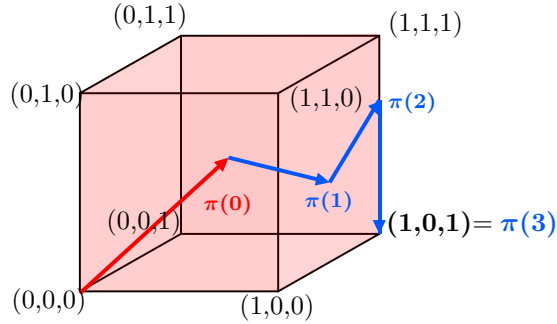
\begin{figure}[ht]
\caption{Cube Method without Balancing Constraints}
\label{fig:CubewoC}
\centering

\tikzset{every picture/.style={line width=0.75pt}} 
\begin{tikzpicture}[x=0.75pt,y=0.75pt,yscale=-1,xscale=1]
\draw  [fill={rgb, 255:red, 255; green, 0; blue, 0 }  ,fill opacity=0.03 ] (238,67) -- (338,67) -- (338,167) -- (238,167) -- cycle ;
\draw  [fill={rgb, 255:red, 255; green, 0; blue, 0 }  ,fill opacity=0.03 ] (289,38) -- (389,38) -- (389,138) -- (289,138) -- cycle ;
\draw [fill={rgb, 255:red, 255; green, 0; blue, 0 }  ,fill opacity=0.03 ]   (289,38) -- (238,67) ;
\draw [fill={rgb, 255:red, 255; green, 0; blue, 0 }  ,fill opacity=0.03 ]   (389,38) -- (338,67) ;
\draw [fill={rgb, 255:red, 255; green, 0; blue, 0 }  ,fill opacity=0.03 ]   (289,138) -- (238,167) ;
\draw [fill={rgb, 255:red, 255; green, 0; blue, 0 }  ,fill opacity=0.03 ]   (389,138) -- (338,167) ;
\draw [color={rgb, 255:red, 255; green, 0; blue, 0 }  ,draw opacity=1 ][line width=1.5]    (238,168) -- (310.05,102.2) ;
\draw [shift={(313,99.5)}, rotate = 137.59] [fill={rgb, 255:red, 255; green, 0; blue, 0 }  ,fill opacity=1 ][line width=0.08]  [draw opacity=0] (6.97,-3.35) -- (0,0) -- (6.97,3.35) -- cycle    ;
\draw  [draw opacity=0][fill={rgb, 255:red, 255; green, 0; blue, 0 }  ,fill opacity=0.17 ] (289,38) -- (389,37) -- (389,138) -- (338,167) -- (238,167) -- (238,67) -- cycle ;
\draw [color={rgb, 255:red, 0; green, 87; blue, 255 }  ,draw opacity=1 ][line width=1.5]    (313,99.5) -- (360.11,111.05) ;
\draw [shift={(364,112)}, rotate = 193.77] [fill={rgb, 255:red, 0; green, 87; blue, 255 }  ,fill opacity=1 ][line width=0.08]  [draw opacity=0] (6.97,-3.35) -- (0,0) -- (6.97,3.35) -- cycle    ;
\draw [color={rgb, 255:red, 0; green, 87; blue, 255 }  ,draw opacity=1 ][line width=1.5]    (364,112) -- (389,70.5) ;
\draw [shift={(390.5,70.5)}, rotate = 132] [fill={rgb, 255:red, 0; green, 87; blue, 255 }  ,fill opacity=1 ][line width=0.08]  [draw opacity=0] (6.97,-3.35) -- (0,0) -- (6.97,3.35) -- cycle    ;
\draw [color={rgb, 255:red, 0; green, 87; blue, 255 }  ,draw opacity=1 ][line width=1.5]    (389.38,70.5) -- (389.38,138) ;
\draw [shift={(389,138)}, rotate = 270] [fill={rgb, 255:red, 0; green, 87; blue, 255 }  ,fill opacity=1 ][line width=0.08]  [draw opacity=0] (6.97,-3.35) -- (0,0) -- (6.97,3.35) -- cycle    ;
\draw (299,116.4) node [anchor=north west][inner sep=0.75pt]  [font=\scriptsize,color={rgb, 255:red, 255; green, 0; blue, 0 }  ,opacity=1 ]  {$\boldsymbol{\pi (0)}$};
\draw (350,115) node [anchor=north west][inner sep=0.75pt]  [font=\scriptsize,color={rgb, 255:red, 0; green, 87; blue, 255 }  ,opacity=1 ]  {$\boldsymbol{\pi (1)}$};
\draw (392,65) node [anchor=north west][inner sep=0.75pt]  [font=\scriptsize,color={rgb, 255:red, 0; green, 87; blue, 255 }  ,opacity=1 ]  {$\boldsymbol{\pi (2)}$};
\draw (202,164) node [anchor=north west][inner sep=0.75pt]   [align=left] {{\footnotesize (0,0,0)}};
\draw (317,167) node [anchor=north west][inner sep=0.75pt]   [align=left] {{\footnotesize (1,0,0)}};
\draw (250,124) node [anchor=north west][inner sep=0.75pt]   [align=left] {{\footnotesize (0,0,1)}};
\draw (201,60) node [anchor=north west][inner sep=0.75pt]   [align=left] {{\footnotesize (0,1,0)}};
\draw (253,20) node [anchor=north west][inner sep=0.75pt]   [align=left] {{\footnotesize (0,1,1)}};
\draw (392,24) node [anchor=north west][inner sep=0.75pt]   [align=left] {{\footnotesize (1,1,1)}};
\draw (391,125) node [anchor=north west][inner sep=0.75pt]   [align=left] {{\footnotesize \textbf{(1,0,1)}$=\boldsymbol{\textcolor{rgb, 255:red, 0; green, 87; blue, 255}{\pi(3)}}$}};
\draw (342,62) node [anchor=north west][inner sep=0.75pt]   [align=left] {{\footnotesize (1,1,0)}};

\end{tikzpicture}
\floatfoot{\justifying This figure depicts an example of the cube method with $n=3$ when no balancing constraint is imposed. The red arrow represents the initial treatment probabilities $(\pi_i)_{i=1,...,n}$. Then, every blue arrow is a step of the \textit{flight phase}. In this example, the first unit is initially assigned to the treatment group. Then, the third unit is assigned to the treatment group. Last, the second unit is assigned the control group. Therefore, the final allocation -- in bold -- is $(1,0,1)$.}
\end{figure}

We will first describe the cube algorithm without balancing constraints before moving to the more interesting case where the balancing constraints in Equation \eqref{eq:balZ} are considered.
Whatever the set of balancing constraints, the cube method is a discrete martingale that moves in (at most) $n$ steps from the interior point $\bm{\pi}(0)=(\pi_i)_{i=0}^n$ to $\bm{\pi}(n)=(D_i)_{i=0}^n$ a vertex of $C$. Let us consider the case without constraints. At the first step, one chooses a random direction for $\bm{\pi}(1)-\bm{\pi}(0)$ and a step size such that $\bm{\pi}(1)$ belongs to a facet of $C$ and that  $\E[\boldsymbol{\pi}(1)|\boldsymbol{\pi}(0)]=\boldsymbol{\pi}(0)$. After this step, because $\bm{\pi}(1)$ belongs to a facet of $C$, one component $i_0$ of $\bm{\pi}(1)$ is equal to 0 or 1, selecting $D_{i_0}=\pi_{i_0}(1)$ one has thus assigned a first unit to either treatment or control. Because a facet of a $n$-cube is a $(n-1)$-cube, one  can then repeat the process in a $(n-1)$-cube, and so on, until landing in a vertex of $C$. At the final step $n$, one will have $(D_i)_{i=1,...,n}=\bm{\pi}(n)\in\{0,1\}^n$ and $\E[D_i]=\pi_i$ (i.e., every unit is allocated to the treatment group with the probability specified by the researcher). These successive steps are the \textit{flight phase} and for the cube method without balancing constraint, allocation $(D_i)_{i=1,...,n}$ is always determined at the end of this phase. Figure \ref{fig:CubewoC} illustrates graphically the method. \\

\begin{figure}[ht]
	\centering
	\caption{Cube Method with Fixed Sample Size}
	\label{fig:CubewSize}

\tikzset{every picture/.style={line width=0.75pt} }  

\begin{tikzpicture}[x=0.75pt,y=0.75pt,yscale=-1,xscale=1]
\draw   (237,58) -- (337,58) -- (337,158) -- (237,158) -- cycle ;
\draw   (288,29) -- (388,29) -- (388,129) -- (288,129) -- cycle ;
\draw    (288,29) -- (237,58) ;
\draw    (388,29) -- (337,58) ;
\draw    (288,129) -- (237,158) ;
\draw    (388,129) -- (337,158) ;
\draw [color={rgb, 255:red, 255; green, 0; blue, 0 }  ,draw opacity=1 ][line width=1.5]    (237,157) -- (329.12,68.27) ;
\draw [shift={(332,65.5)}, rotate = 136.08] [fill={rgb, 255:red, 255; green, 0; blue, 0 }  ,fill opacity=1 ][line width=0.08]  [draw opacity=0] (6.97,-3.35) -- (0,0) -- (6.97,3.35) -- cycle    ;
\draw  [fill={rgb, 255:red, 255; green, 0; blue, 0 }  ,fill opacity=0.17 ] (337,58) -- (388,128) -- (288,29) -- cycle ;
\draw [color={rgb, 255:red, 0; green, 87; blue, 255 }  ,draw opacity=1 ][line width=1.5]    (332,65.5) -- (353.85,82.54) ;
\draw [shift={(357,85)}, rotate = 217.95] [fill={rgb, 255:red, 0; green, 87; blue, 255 }  ,fill opacity=1 ][line width=0.08]  [draw opacity=0] (6.97,-3.35) -- (0,0) -- (6.97,3.35) -- cycle    ;
\draw [color={rgb, 255:red, 0; green, 87; blue, 255 }  ,draw opacity=1 ][line width=1.5]    (357,85) -- (385.7,125.73) ;
\draw [shift={(388,129)}, rotate = 234.83] [fill={rgb, 255:red, 0; green, 87; blue, 255 }  ,fill opacity=1 ][line width=0.08]  [draw opacity=0] (6.97,-3.35) -- (0,0) -- (6.97,3.35) -- cycle    ;
\draw (290,65) node [anchor=north west][inner sep=0.75pt]  [font=\scriptsize,color={rgb, 255:red, 255; green, 0; blue, 0 }  ,opacity=1 ]  {$\boldsymbol{\textcolor[rgb]{1,0,0}{\pi(0)}}$};
\draw (290,65) node [anchor=north west][inner sep=0.75pt]  [font=\scriptsize,color={rgb, 255:red, 255; green, 0; blue, 0}  ,opacity=1]  {$\boldsymbol{\pi}(0)$};

\draw (360,75) node [anchor=north west][inner sep=0.75pt]  [font=\scriptsize,color={rgb, 255:red, 0; green, 87; blue, 255}  ,opacity=1]  {$\boldsymbol{\pi}(1)$};
\draw (343,48) node [anchor=north west][inner sep=0.75pt]   [align=left] {{\footnotesize (1,1,0)}};
\draw (207,156) node [anchor=north west][inner sep=0.75pt]   [align=left] {{\footnotesize (0,0,0)}};
\draw (316,158) node [anchor=north west][inner sep=0.75pt]   [align=left] {{\footnotesize (1,0,0)}};
\draw (249,115) node [anchor=north west][inner sep=0.75pt]   [align=left] {{\footnotesize (0,0,1)}};
\draw (199,50) node [anchor=north west][inner sep=0.75pt]   [align=left] {{\footnotesize (0,1,0)}};
\draw (252,11) node [anchor=north west][inner sep=0.75pt]   [align=left] {{\footnotesize (0,1,1)}};
\draw (389,12) node [anchor=north west][inner sep=0.75pt]   [align=left] {{\footnotesize (1,1,1)}};
\draw (390,116) node [anchor=north west][inner sep=0.75pt]   [align=left] {{\footnotesize \textbf{(1,0,1)}$=\boldsymbol{\textcolor{rgb, 255:red, 0; green, 87; blue, 255}{\pi(2)}}$}};

\end{tikzpicture}
\floatfoot{\justifying This figure depicts an example of the cube method with $n=3$ when imposing the constraint $n_T=2$ and $\sum_{i=1}^3\pi_i=2$. The red area depicts the points $(s_1,s_2,s_3)$ in the cube satisfying the equation $\sum_{i=1}^3 s_i =2$. This condition is equivalent to imposing the balancing constraint $\sum_i \frac{Z_i s_i}{\pi_i}=\sum_i Z_i$ with $Z_i=\pi_i$. The red arrow represents the initial treatment probabilities $\left(\frac{2}{3},\frac{2}{3},\frac{2}{3}\right)$. Then, every blue arrow is a step of the \textit{flight phase}. In this example, the first unit is initially assigned to the treatment group. Then, since $n_T=2$, only one unit among the second and third units can be assigned to the treatment group. In this case, the second blue arrow shows that, in the same step, the second unit is assigned to the control group and the third one to the treatment group. Therefore, the last allocation -- in bold -- is $(1,0,1)$. }
\end{figure}
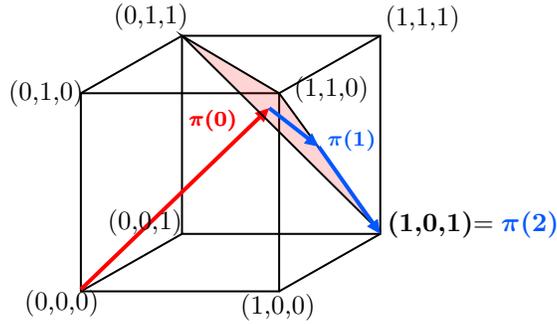

\begin{figure}[ht]
	\centering
	\caption{Cube Method with One Balancing Constraint}
	\label{fig:Cubew1C}
	\begin{subfigure}[b]{0.4\textwidth}
		\centering
		\subcaption{\textit{landing phase} not required}
		\label{fig:CubeNoLand}

\tikzset{every picture/.style={line width=0.75pt}}      

\begin{tikzpicture}[x=0.75pt,y=0.75pt,yscale=-1,xscale=1]
\draw   (243,70) -- (343,70) -- (343,170) -- (243,170) -- cycle ;
\draw   (294,41) -- (394,41) -- (394,141) -- (294,141) -- cycle ;
\draw    (294,41) -- (243,70) ;
\draw    (394,41) -- (343,70) ;
\draw    (294,141) -- (243,170) ;
\draw    (394,141) -- (343,170) ;
\draw [color={rgb, 255:red, 255; green, 0; blue, 0 }  ,draw opacity=1 ][line width=1.5]    (244,170) -- (317.98,105.63) ;
\draw [shift={(321,103)}, rotate = 138.97] [fill={rgb, 255:red, 255; green, 0; blue, 0 }  ,fill opacity=1 ][line width=0.08]  [draw opacity=0] (6.97,-3.35) -- (0,0) -- (6.97,3.35) -- cycle    ;
\draw  [fill={rgb, 255:red, 255; green, 0; blue, 0 }  ,fill opacity=0.17 ] (340,41) -- (393,95) -- (371.46,127.31) -- (343,170) -- (243,70) -- cycle ;
\draw [color={rgb, 255:red, 0; green, 87; blue, 255 }  ,draw opacity=1 ][line width=1.5]    (321,103) -- (357.17,139.17) ;
\draw [shift={(360,142)}, rotate = 225] [fill={rgb, 255:red, 0; green, 87; blue, 255 }  ,fill opacity=1 ][line width=0.08]  [draw opacity=0] (6.97,-3.35) -- (0,0) -- (6.97,3.35) -- cycle    ;
\draw [color={rgb, 255:red, 0; green, 87; blue, 255 }  ,draw opacity=1 ][line width=1.5]    (360,142) -- (345.08,166.58) ;
\draw [shift={(343,170)}, rotate = 301.26] [fill={rgb, 255:red, 0; green, 87; blue, 255 }  ,fill opacity=1 ][line width=0.08]  [draw opacity=0] (6.97,-3.35) -- (0,0) -- (6.97,3.35) -- cycle    ;
\draw (306,118.4) node [anchor=north west][inner sep=0.75pt]  [font=\scriptsize,color={rgb, 255:red, 255; green, 0; blue, 0 }  ,opacity=1 ]  {$\boldsymbol{\textcolor[rgb]{1,0,0}{\pi }\textcolor[rgb]{1,0,0}{(}\textcolor[rgb]{1,0,0}{0}\textcolor[rgb]{1,0,0}{)}\textcolor[rgb]{1,0,0}{}}$};
\draw (356,125) node [anchor=north west][inner sep=0.75pt]  [font=\scriptsize,color={rgb, 255:red, 0; green, 87; blue, 255 }  ,opacity=1 ]  {$\boldsymbol{\pi (1)}$};
\draw (346.5,65) node [anchor=north west][inner sep=0.75pt]   [align=left] {{\footnotesize (1,1,0)}};
\draw (207,167) node [anchor=north west][inner sep=0.75pt]   [align=left] {{\footnotesize (0,0,0)}};
\draw (322,170) node [anchor=north west][inner sep=0.75pt]   [align=left] {{\footnotesize \textbf{(1,0,0)}$=\boldsymbol{\textcolor{rgb, 255:red, 0; green, 87; blue, 255}{\pi(2)}}$}};
\draw (255,127) node [anchor=north west][inner sep=0.75pt]   [align=left] {{\footnotesize (0,0,1)}};
\draw (206,58) node [anchor=north west][inner sep=0.75pt]   [align=left] {{\footnotesize (0,1,0)}};
\draw (258,23) node [anchor=north west][inner sep=0.75pt]   [align=left] {{\footnotesize (0,1,1)}};
\draw (395,24) node [anchor=north west][inner sep=0.75pt]   [align=left] {{\footnotesize (1,1,1)}};
\draw (396,127) node [anchor=north west][inner sep=0.75pt]   [align=left] {{\footnotesize (1,0,1)}};

\end{tikzpicture}

	\end{subfigure}                   
	\begin{subfigure}[b]{0.4\textwidth}
		\centering
		\subcaption{\textit{landing phase} required}
		\label{fig:CubeWithLand}

\tikzset{every picture/.style={line width=0.75pt}}   

\begin{tikzpicture}[x=0.75pt,y=0.75pt,yscale=-1,xscale=1]
\draw   (244,64) -- (344,64) -- (344,164) -- (244,164) -- cycle ;
\draw   (295,35) -- (395,35) -- (395,135) -- (295,135) -- cycle ;
\draw    (295,35) -- (244,64) ;
\draw    (395,35) -- (344,64) ;
\draw    (295,135) -- (244,164) ;
\draw    (395,135) -- (344,164) ;
\draw [color={rgb, 255:red, 255; green, 0; blue, 0 }  ,draw opacity=1 ][line width=1.5]    (245,164) -- (318.98,99.63) ;
\draw [shift={(322,97)}, rotate = 138.97] [fill={rgb, 255:red, 255; green, 0; blue, 0 }  ,fill opacity=1 ][line width=0.08]  [draw opacity=0] (6.97,-3.35) -- (0,0) -- (6.97,3.35) -- cycle    ;
\draw  [fill={rgb, 255:red, 255; green, 0; blue, 0 }  ,fill opacity=0.17 ] (341,35) -- (394,89) -- (372.46,121.31) -- (344,164) -- (244,64) -- cycle ;
\draw [color={rgb, 255:red, 0; green, 87; blue, 255 }  ,draw opacity=1 ][line width=1.5]    (322,97) -- (360.1,133.24) ;
\draw [shift={(363,136)}, rotate = 223.57] [fill={rgb, 255:red, 0; green, 87; blue, 255 }  ,fill opacity=1 ][line width=0.08]  [draw opacity=0] (6.97,-3.35) -- (0,0) -- (6.97,3.35) -- cycle    ;
\draw [color={rgb, 255:red, 0; green, 87; blue, 255 }  ,draw opacity=1 ][line width=1.5]    (363,136) -- (391.8,92.34) ;
\draw [shift={(394,89)}, rotate = 123.41] [fill={rgb, 255:red, 0; green, 87; blue, 255 }  ,fill opacity=1 ][line width=0.08]  [draw opacity=0] (6.97,-3.35) -- (0,0) -- (6.97,3.35) -- cycle    ;
\draw [color={rgb, 255:red, 84; green, 153; blue, 4 }  ,draw opacity=1 ][line width=1.5]    (394,89) -- (394.91,131) ;
\draw [shift={(395,135)}, rotate = 268.75] [fill={rgb, 255:red, 84; green, 153; blue, 4 }  ,fill opacity=1 ][line width=0.08]  [draw opacity=0] (6.97,-3.35) -- (0,0) -- (6.97,3.35) -- cycle    ;
\draw (308,113.4) node [anchor=north west][inner sep=0.75pt]  [font=\scriptsize,color={rgb, 255:red, 255; green, 0; blue, 0 }  ,opacity=1 ]  {$\boldsymbol{\textcolor[rgb]{1,0,0}{\pi }\textcolor[rgb]{1,0,0}{(}\textcolor[rgb]{1,0,0}{0}\textcolor[rgb]{1,0,0}{)}\textcolor[rgb]{1,0,0}{}}$};
\draw (346,110) node [anchor=north west][inner sep=0.75pt]  [font=\scriptsize,color={rgb, 255:red, 0; green, 87; blue, 255 }  ,opacity=1 ]  {$\boldsymbol{\pi (1)}$};
\draw (397,89) node [anchor=north west][inner sep=0.75pt]  [font=\scriptsize,color={rgb, 255:red, 0; green, 87; blue, 255 }  ,opacity=1 ]  {$\boldsymbol{\pi (2)}$};
\draw (347.5,59) node [anchor=north west][inner sep=0.75pt]   [align=left] {{\footnotesize (1,1,0)}};
\draw (208,161) node [anchor=north west][inner sep=0.75pt]   [align=left] {{\footnotesize (0,0,0)}};
\draw (323,164) node [anchor=north west][inner sep=0.75pt]   [align=left] {{\footnotesize (1,0,0)}};
\draw (256,121) node [anchor=north west][inner sep=0.75pt]   [align=left] {{\footnotesize (0,0,1)}};
\draw (207,52) node [anchor=north west][inner sep=0.75pt]   [align=left] {{\footnotesize (0,1,0)}};
\draw (259,17) node [anchor=north west][inner sep=0.75pt]   [align=left] {{\footnotesize (0,1,1)}};
\draw (396,18) node [anchor=north west][inner sep=0.75pt]   [align=left] {{\footnotesize (1,1,1)}};
\draw (397,121) node [anchor=north west][inner sep=0.75pt]   [align=left] {{\footnotesize \textbf{(1,0,1)}$=\boldsymbol{\textcolor{rgb, 255:red, 84; green, 153; blue, 4 }{\pi(3)}}$}};

\end{tikzpicture}

	\end{subfigure}
\floatfoot{\justifying This figure depicts an example of the cube method with $n=3$ where we do not always get exactly-balanced allocations. We consider the initial treatment probabilities in \eqref{eq:pi} to be $\pi_1=\pi_2=\pi_3=\frac{2}{3}$. The red area depicts the points $(s_1,s_2,s_3)$ in the cube satisfying the equation $\sum_{i=1}^3 s_1+s_2-\frac{1}{2}s_3 =1$. This is equivalent to imposing the constraint in \eqref{eq:balZs} with $Z_1=Z_2=\frac{2}{3}$ and $Z_3=-\frac{1}{3}$. The red arrow represents the initial treatment probabilities. Since not every vertex of the plane is a cube vertex, we cannot always satisfy the constraint. In both panels, the algorithm assigns the first unit to the treatment group (first blue arrow). The second blue arrow corresponds to the assignment of the third unit. If the algorithm assigns the third unit to the control group (panel a), it automatically assigns the second one to treatment. However, if the algorithm assigns the third unit to the treatment group (panel b), the second unit is in neither group, even if we attain a plane vertex. In the \textit{landing phase}, the cube method will proceed by randomly allocating the second unit. In this example, the green arrow shows that the \textit{landing phase} allocates the second unit to the control group.}
\end{figure}
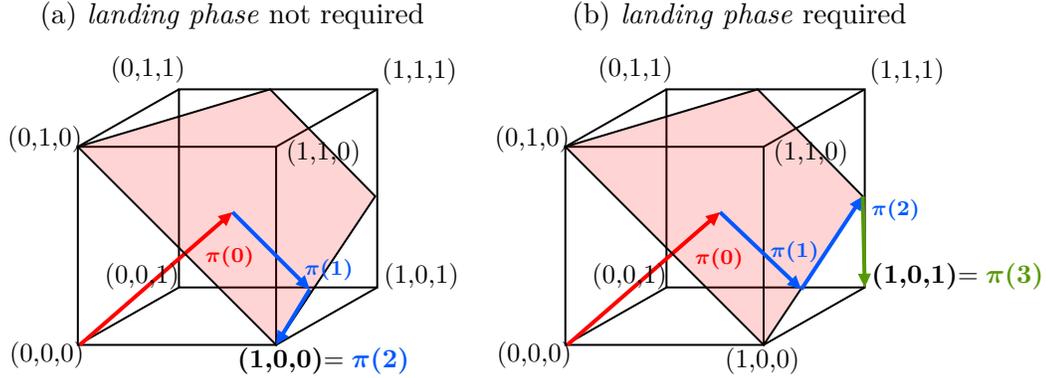

In Figure \ref{fig:CubewoC}, all vertices of the $n$-cube can be selected, meaning that all individuals could be allocated to the control group. We now consider that the researcher wants to allocate a fixed number $n_T$ of units to the treatment and $n_c$ units to the control. This can be achieved with the cube method as soon as $\sum_{i=1}^n\pi_i=n_T$. The condition that exactly $n_T$ units are assigned to the treatment can be expressed as a balancing constraint. Indeed, because $n_T=\sum_iD_i$ and $\sum_i \pi_i=n_T$, the fixed size condition is equivalent to $\sum_{i}\frac{Z_iD_i}{\pi_i}=\sum_{i}Z_i$ for $Z_i=\pi_i$. Let $K$ the set of vectors $s$ in the $n$-cube $C$ such that $\sum_i s_i=n_T$. $K$ is a closed convex set and its extreme points are vertices of $C$, that is the set of allocations respecting the fixed-size constraints. $K$ is contained in an affine subspace of dimension $n-1$ of direction $V:=\{v: \sum_{i=1}^n v_i=0\}$, we have $K=C\cap \left\{\bm{\pi}(0)+v: \sum_i v_i=0\right\}$. The cube method selects randomly an element of $V$ for the direction of $\bm{\pi}(1)-\bm{\pi}(0)$ and fixes the step size such that $\bm{\pi}(1)$ is a border point of $K$ and that $E(\bm{\pi}(1)|\bm{\pi}(0))=\bm{\pi}(0)$. After this first step, $\bm{\pi}(1)$ belongs to a facet of $C$ and a unit $i_1$ is assigned either to the treatment either to the control group. Units $i\neq i_0$ remain unassigned and we have $\sum_{i:i\neq i_0}\pi_i(1)=n_T-D_{i_0}$. We can then replicate the first step after replacing $n_T$ by $n_T-D_{i_1}$ the sample $\{1,...,n\}$ by $\{1,...,n\}\backslash\{i_0\}$ and to allocate a second unit and to update assignment probability as $\bm{\pi}(2)$. At step $n-1$, $\bm{\pi}(n-1)$ belongs to the extreme points of $K$, this ends the flight phase. If, for instance, $\pi_i=1/2$ and $n$ is even, the extreme points of $K$ are some vertices of $C$, so the assignment is achieved. Now imagine that one has 101 units to assign with equal probability to the treatment and control groups. exact balancing on the two group sizes is not possible: 101 is an odd integer and it is not feasible to assign 50.5 units to the treatment. A popular solution is to consider $\pi_i=50/101$ or $\pi_i=51/101$ and to sample randomly 50 (or 51) elements among the 101 units. However, this strategy does not accommodate easily with heterogeneous probabilities of assignment and does not generalize to take into account many balancing constraints. With the cube method described above, for each step $t$ of the flight phase we have $\sum_i \pi_i(t)=50.5$ and the extreme points of $K$ are not anymore vertices of $C$. In that case, at the end of the flight phase, $n-1=100$ units are assigned at the end of the flight phase with $(n-1)/2=50$ units to the treatment and $(n-1)/2=50$ units to the control. The cube method can be completed with a last phase that randomly assigns to the treatment of the control the remaining unit ensuring that $n_T=50$ or $51$ and $E(n_T)=50.5$. In that case, the sizes of treatment and control groups are not exactly fixed but almost fixed (in fact as fixed as possible as soon as we respect the initial assignment probabilities $\pi_i=1/2$). This second phase is called \textit{landing} phase. These two phases, the flight phase and the landing phase, can be generalized to the case where the researcher wants to impose several balancing constraints and heterogeneous probabilities of treatment.  

Let us describe the cube method with an arbitrary number of balancing constraints defined by $q$ variables $Z_i$. A point $\mathbf{s}\in C$ will satisfy an equation analog to \eqref{eq:balZ} if \begin{align}\sum_{i=1}^n\frac{Z_i s_i}{\pi_i}&= \sum_{i=1}^n Z_i.\label{eq:balZs}\end{align}
Let $A_i=\frac{Z_i}{\pi_i}$ and $A=(A_1,...,A_n)$ the matrix of size $q\times n$. Then \eqref{eq:balZs} is equivalent to
\begin{align*}
	&\quad\sum_{i=1}^nA_i s_i= \sum_{i=1}^n A_i \pi_i\\
	\Leftrightarrow&\quad A \mathbf{s}=A \boldsymbol{\pi}(0)\\
	\Leftrightarrow&\quad  \mathbf{s}\in Q \coloneqq \boldsymbol{\pi}(0) + \ker(A).
\end{align*}
$K=C\cap Q$ is, therefore, the $(n-q)$-polytope that contains all the points in $C$ such that \eqref{eq:balZs} holds. At the first step, one chooses a random direction in $v\in \ker(A)$ and we select the unique $\lambda>0$ such that $\bm{\pi}(1):=\bm{\pi}(0)+\lambda v$ is on a facet of $K$ and that $\E[\boldsymbol{\pi}(1)|\boldsymbol{\pi}(0)]=\boldsymbol{\pi}(0)$. Because any facet of $K$ is the intersection of a facet of $C$ with $Q$, a component $i_0$ of  $\boldsymbol{\pi}(1)$ is 0 or 1 and defining $D_{i_0}=\pi_{i_0}(1)$ one has assigned a first unit. Next, one applies a similar step for the facet of $K$ instead of $K$ and $\bm{\pi}(1)$ as a starting point instead of $\bm{\pi}(0)$. After $n-q$ steps, one has reached a vertex of $K$. This process corresponds to the flight phase in \citet{deville_efficient_2004}. If this vertex of $K$ is also a vertex of $C$, the flight phase allocates every unit, and the two groups are exactly balanced (see Figures \ref{fig:CubewSize} and \ref{fig:CubeNoLand}). But in many cases, the vertex of $K$ is not a vertex of $C$, and there remain at most $q$ units to assign during the landing phase \citep[according to the wording of][]{deville_efficient_2004} (see Figure \ref{fig:CubeWithLand}).

Say that at the end of the flight phase, one has not assigned $r\leq q$ units and let $\boldsymbol{\pi^\ast}=\boldsymbol{\pi}(n-q)$ be the updated treatment probabilities at this stage. The landing phase of the cube method assigns the $r$ missing units such that $\E[D_i | \boldsymbol{\pi^\ast}]=\boldsymbol{\pi^\ast}$. \citet{grafstrom_doubly_2013} describe two methods for the landing phase (these are also the options used in sampling packages): (i) Linear programming: one considers all the $2^r$ allocations for these units and assigns probabilities to each allocation to minimize a cost function and satisfy $\E[D_i | \boldsymbol{\pi^\ast}]=\boldsymbol{\pi^\ast}$. Sampling probabilities are chosen to minimize
	$$\E\left(\sum_{i\notin S}Z'_i(D_i-\pi_i^{\ast})M\sum_{i\notin S}Z_i(D_i-\pi_i^{\ast})\big|W\right),$$
where $S$ is the set of units allocated at the flight phase, $W=(S, (D_i)_{i\in S}, (\pi_i^{\ast})_{i\notin S}, (Z_{i})_{i=1,...,n})$, and $M$ is a symmetric positive-definite matrix $q\times q.$ Common choices for $M$ are the identity matrix or the inverse-covariance matrix. After solving this minimization problem, the researcher randomly draws an allocation using these probabilities. (ii) Suppression of variables: if $r>20$, solving a linear problem becomes computationally difficult. In that case, at the end of the flight phase, one can drop a covariate (i.e., a constraint) and continue with the flight phase. One can thus successively drop variables until attaining a vertex of $C$. This method, however, implies that the researcher has to define an order to drop the covariates, ideally from the least to the most important.

\section{Statistical Properties of the Cube Method}
\label{sec:results}
This Section shows how the cube method allows obtaining an ``almost-exact balance'' between the treatment and control groups and relates this balance to gains in precision for treatment effect estimators.

\subsection{Balancing Approximations}

\subsubsection{Balance Checks}

As explained above, designing an allocation mechanism that always produces exactly-balanced groups is generally impossible. However, we here prove that the cube method is successful, under certain conditions, in creating almost-exactly-balanced samples in the sense of Equation \eqref{eq:balZop}.

To check balance properties after allocating individuals according to the design $\Pi$, researchers are interested in computing the difference

\begin{equation*}
	\Delta^{\Pi}_{j,n}=\frac{1}{n}\sum_{i=1}^n \frac{X_{ji} D_i}{\pi_i} -\frac{X_{ji} (1-D_i)}{1-\pi_i}.
\end{equation*}

Because $\Pr_{\Pi}(D_i=1|(X_{i'})_{i'=1,...,n})=\pi_i$, we have $\E\left(\Delta^{\Pi}_{j,n}\right)=0$ and under weak conditions on $\Pi$, we have
\begin{align}\sqrt{n}\Delta^{\Pi}_{j,n}\convNor{\V(\Delta^{\Pi}_{j,n})},\label{eq:norm_bal}
\end{align}
where $\V(\Delta^{\Pi}_{j,n})$ is an asymptotic variance depending on $\Pi$ and the distribution of $X$.

For the so-called baseline balance tests, researchers often consider the $t$-statistic
\begin{equation*}
	t^{\Pi}_{j,n}=\sqrt{n}\frac{\Delta^{\Pi}_{j,n}}{\sqrt{\widehat{\V}(\Delta^{\Pi}_{j,n})}}
\end{equation*}
where $\widehat{\V}(\Delta^{\Pi}_{j,n})$ is a consistent estimator of the asymptotic variance of $\Delta^{\Pi}_{j,n}$ to test the null hypothesis of exact balance. $t^{\Pi}_{j,n}$ is then associated to a $p$-value $p^{\Pi}_{j,n}$ which take values between 0 and 1. As explained by \citet{snyder_examining_2024}, when creating balance tests for RCTs, small $p$-values (below $0.15$) are considered problematic and are usually underreported.

Let us first consider a naive mechanism that does not use baseline information to assign units. Such situations correspond to the case where the design $\Pi$ is a coin toss, i.e., a design where each unit $i$ is allocated to the treatment independently of the allocation of other units:
$$\Pr_{\Pi}\left(\bigcap_{i=1}^n\{D_i=d_i\}\big|(X_{i})_{i=1}^n\right)=\prod_{i=1}^n\pi_i^{d_i}(1-\pi_i)^{1-d_i}.$$
A coin toss does not balance any variable nor group sizes.

When $\pi_i=\frac{n_T}{n}$ for any $i$, another popular design is sampling without replacement of $n_T$ treated units, also known as complete randomization:
$$\Pr_{\Pi}\left(\bigcap_{i=1}^n\{D_i=d_i\}\big|(X_{i})_{i=1}^n\right)=\binom{n}{n_T}^{-1}\mathds{1}\left\{\sum_{i=1}^nd_i=n_T\right\}.$$
Complete randomization only balances constant variables. In that case, the sample of treated and control groups are fixed, and the design is also balanced on the constant $\big(\sum_{i=1}^n D_i=n_T$, $\sum_{i=1}^n (1-D_i)=n-n_T$ and $\sum_{i=1}^n\frac{D_i}{\pi_i}=n \big)$.

Under such assignments and Assumptions \ref{hyp:iid+mom} and \ref{hyp:sample}, and more generally for any design $\Pi$ such that \eqref{eq:norm_bal} holds with $\V(\Delta_{j,n}^{\Pi})>0$, we have
$\Delta_{j,n}^{\Pi}=O_p\left(\frac{1}{\sqrt{n}}\right)$,
$t_{j,n}^{\Pi}\convNor{1}$
and $p_{j,n}^{\Pi}\convD\mathcal{U}(0,1)$. This result means that if one randomizes naively, control and treatment groups will present imbalances with a strictly-positive probability. Moreover, for a confidence level of $100(1-\alpha)\%$, there exists always $100\alpha\%$ chance of obtaining significant differences. If an researcher evaluates the balance of 10 independent covariates at the $85\%$ confidence level \citep[a level over which rejection is considered problematic for publication as shown by][]{snyder_examining_2024}, there is more than $80\%$ chance of having at least one significant difference. This magnitude questions the mere implementation of
such widely used tests. Even if a multiple F-test with a confidence level of $85\%$ mitigates this rejection rate, the null hypothesis of simultaneously balanced covariates is rejected by construction with a 15\% chance.

The cube method ensures that these tests are unnecessary since we can balance control and treatment groups in any covariate $(X_j)_{j=1,...,p}$. This is achieved because $\V(\Delta^{\Pi}_{j,n})=0$ for any $j=1,...,p$ in \eqref{eq:norm_bal}. Performing these tests would not make sense since we never reject the null hypothesis by construction. However, one might report them if the editor worries about researchers randomizing badly. Usual balancing strategies are stratified or matched-pairs designs. These methods ensure $\V(\Delta^{\Pi}_{j,n})=0$ if the covariates $(X_j)_{j=1,...,p}$ are all discrete but will always generate imbalances for continuous ones since the researcher needs to discretize or aggregate them before randomizing.  

The following proposition explains how the balancing approximations are satisfied with the cube method. Because the number $q$ of balancing constraints in Equation \eqref{eq:balZop} could be large with the cube method, we are also explicit on how $q$ affects balancing approximations to allow us to consider a framework where $q$ tends to $\infty$.

\begin{prop}[Balancing Approximations with the Cube Method]~\label{prop:balance}\\
	If Assumptions \ref{hyp:iid+mom} and \ref{hyp:sample} hold, then $$\Delta^{Cube}_{j,n}=o_p\left(\frac{q}{\sqrt{n}}\right).$$	
Moreover if $\E\left[|X_{j1}|^r\right]<\infty$ for $r \geq 2$, then $\Delta^{Cube}_{j,n}=o_p\left(\frac{q}{n^{1-1/r}}\right)$,
if $X_{j1}$ is sub-Gaussian, then $\Delta^{Cube}_{j,n}=O_p\left(\frac{q\sqrt{\ln(n)}}{n}\right)$, and	if 	$X_{j1}$ has a bounded support, then
		$\big|\Delta^{Cube}_{j,n}\big| <\frac{Kq}{cn}$ for $K$ such that $|X_{j1}| <K$. As soon as $\sqrt{n}\Delta^{Cube}_{j,n}=o_p(1)$, we have
		$t^{Cube}_{j,n}\convP0$, and $p^{Cube}_{j,n}\convP1$.
\end{prop}

Proposition \ref{prop:balance} shows that, as $n$ grows, the cube method ensures the balancing Equation \eqref{eq:balZop} as soon as the second-order moments of $X$ exist. Furthermore, if moments of order $r>2$ exist for $X$, \eqref{eq:balZop} holds as soon as $q=O\left(n^{\frac{1}{2}-\frac{1}{r}}\right)$. $q$ can even be $o\left(\sqrt{\frac{n}{\ln(n)}}\right)$ if the covariates $X$ are all sub-Gaussian or $o(\sqrt{n})$ if they are bounded. This means that with probability tending to one, the $p$-values of balance tests tend to 1. Balance is thus never rejected for large $n$ contrary to randomization under a design $\Pi$ such that \eqref{eq:balZop} does not hold. 

\subsubsection{Comparison with Other Methods}
\label{sec:comparison}

We here compare the balancing properties of the cube method with other randomization methods. For the sake of simplicity, we fix $\pi_i=1/2$. We check imbalances by using the Horvitz-Thompson estimators for the average difference between the control and treatment groups $B_{n,p}(X)=\frac{2}{n}\sum_{i=1}^nX_iD_i-X_i(1-D_i)$ and looking at their squared Euclidean norm $||B_{n,p}(X)||^2=\frac{4}{n^2}\sum_{j=1}^p\left(\sum_{i=1}^nX_{ji}D_i-X_{ji}(1-D_i)\right)^2.$ 

\begin{hyp}\label{hyp:design_bal} 
	$n$ is an even positive number and $X_i=(X_{1i},...,X_{pi})'$ are some independent and identically distributed random vectors of dimension $p$ for $i=1,...,n$. $X_1$ admits a density $f_X$ with respect to the Lebesgue measure on $[0;1]^p$ such that there exists some positive constants $\underline{C}$ and $\overline{C}$ (independent of $p$) such that for any $x\in [0;1]^p$, $\underline{C}\leq f_X(x)\leq \overline{C}$. 
	\end{hyp}
Assumption \ref{hyp:design_bal} imposes mild conditions over the baseline covariates. In particular, the components of the vector $X_i$ are not assumed to be independent. Figure \ref{fig:motex} illustrates our main results for a simple case where this assumption holds: $(X_{ji})_{j=1,\ldots,p,i=1\ldots,n}$ are independent and follow a uniform distribution on $[0,1].$ In this case, we have $V(X_{ji})=1/12$, $\E(X_{ji}^2)=1/3$, and $\underline{C}=\overline{C}=1$.

A coin toss ensures that treatment and control groups are balanced on average (i.e., $\E[B_{n,p}]$=0). However, we can still have imbalances between groups for a given allocation. Additionally, a coin toss will often generate different sizes between the treatment and control groups, meaning that it fails to balance on a constant. Proposition \ref{prop:imbalance} in Appendix \ref{sec:imbalances} shows that for a coin toss, $\frac{4\underline{C}}{3}\frac{p}{n}\leq\E[||B_{n,p}(X)||^2]\leq\frac{4\overline{C}}{3}\frac{p}{n}$. For the case illustrated in Figure \ref{fig:motex}, we  thus have $\E[||B_{n,p}(X)||^2]=\frac{4p}{3n}$.

Complete randomization improves upon the coin toss procedure. In this method, the researcher fixes the group sizes. Since we here assume $\pi_i=1/2$, the researcher randomly chooses an allocation among those having an equal number of treated and untreated units. The researcher still does not use any information on the baseline covariates to refine the randomization process but manages to reduce the imbalances due to different group sizes. Indeed, under Assumption \ref{hyp:design_bal} and complete randomization $\frac{\underline{C}}{3}\frac{p}{n}\leq\E[||B_{n,p}(X)||^2]\leq\frac{\overline{C}}{3}\frac{p}{n}$, so bounds reduce by four, relative to the coin toss. For the example in Figure \ref{fig:motex}, we have $\E[||B_{n,p}(X)||^2]=\frac{p}{3n}$. This result clearly shows the advantages of using designs with fixed sample sizes such as complete randomization, but also the cube method with the constraints in Section \ref{sec:setup} or matched-pairs design.\footnote{One can show that the gains from fixed group sizes are not present if one uses a difference-in-means estimator instead. Moreover, in the case of $n$ even and $\pi_i=1/2$, this estimator is equivalent to the Horvitz-Thompson for complete randomization, matched-pairs design, and the cube method. However, it can lead to more precise estimates for coin tosses or stratified designs.}

In sharp contrast with naïve methods, covariate-adaptive randomization uses baseline information to improve balance between treatment and control. Stratified designs are the most used and studied covariate-adaptive method. Stratification has a long tradition in RCTs \citep{fisher_design_1935,higgins_improving_2016}. Stratified designs are the most popular assignment mechanisms used in RCTs as they are simple to grasp and can produce balanced samples. This method consists of using one or several baseline variables to create blocks or strata and then using complete randomization inside each stratum. A common practice in experiments is to block on gender, meaning that randomization is performed independently amongst male and female units, generating the same proportion of men and women in each treatment arm. When using dummy variables to define the strata, stratified or blocked randomization allows almost exact balancing of the variables used to create them. \citet{athey_chapter_2017} recommend balancing on small strata since this method generates substantial precision gains. However, stratified designs do not come without any limitations. Notably, the type and number of covariates that one wants to balance can impose some difficulties. Facing continuous covariates, such as income or grades, makes it impossible to stratify without the researcher deciding how to create the strata. Assumption \ref{hyp:design_bal} imposes continuous covariates. We thus will focus on two ways of generating (possibly-)small strata, discretization and matched pairs.

First, the researcher can discretize continuous variables using $\ell$-quantiles for each covariate, generating thus $\ell^p$ strata. Stratifying will produce balance gains as long as the number of units remains large compared to the number of strata. In particular, we show in Proposition \ref{prop:imbalance} that whenever  $n\ell^{-p}\to \infty$, stratified designs through discretization outperform complete randomization. Discretizing baseline covariates, however, does not ensure fixed sizes for each stratum. In particular, if the number of strata is big compared to the sample size, there is a big chance of having some strata with only one unit. In the limit case where $n\ell^{-p}\to 0$, every non-empty strata has one unit with probability one, and stratifying through discretization performs strictly worse than complete randomization and approximates a coin toss. We also show that, in both limit cases, imbalances grow at a rate of $p/n$. We observe this behavior in Figure \ref{fig:motex} for $\ell=2,4$. In this example, the stratified designs perform better than complete randomization whenever $l^p\leq n/2$ and similarly to a coin toss for $l^p\geq 32n$. Balancing deterioration can thus occur quite rapidly when stratifying is done by discretizing many continuous variables. It is worth noticing that this issue is not exclusive to continuous variables, as it arises when stratifying using many categorical variables. In that case, the number of strata equates to the product of covariate support cardinalities.

To eliminate the issue of single-unit strata, researchers may use a more sophisticated way of creating their strata: matched-pair designs. Following \citet{greevy_optimal_2004,bai_inference_2022,bai_optimality_2022}, the researcher can create $n/2$ strata of two units to minimize the average intra-strata distance. The researcher thus creates pairs of two units that resemble each other. After constructing these strata, the researcher randomly allocates one to treatment. By doing so, she creates control and treatment groups that are very similar. We show in Proposition \ref{prop:imbalance} that this design always outperforms complete randomization. However, this type of strata construction works by trying to have a similar joint distribution of $X$ between treatment and control. This approach to balancing implies that for large $p$, it becomes more difficult to find pairs of units close to each other. In particular, we show that under this design and Assumption \ref{hyp:design_bal}, $\E[||B_{n,p}(X)||^2]\geq \frac{p}{n}\left(\frac{1}{3}-\sqrt{\frac{2\ln(n-1)+4\ln\overline{C}}{p}}\right)$. This result entails that the number of balancing covariates $p$ is large relative to $\ln n$, balance gains shrink, and imbalances increase at the rate of $p/n$. Figure \ref{fig:motex} illustrates this effect. Indeed, when $p$ becomes larger than $\ln 500\approx6$, the matched-pairs design performs better than complete randomization, but its relative gains quickly reduce.

For all these randomization methods but matched pairs, imbalances grow at the rate of $p/n$. For matched pairs, this rate holds whenever $\ln n=o(p).$ We now show that the cube method is less concerned by this curse of dimensionality since imbalances grow at a rate $p^2/n^2$. If $p$ remains smaller than $n$, then this result implies a much slower balance deterioration than for the methods described above. Proposition \ref{prop:pbalance} gives this upper bound for $\E[||B_{n,p}(X)||^2||]$ when using the cube method. This proposition is also stated and proved in Appendix \ref{sec:imbalances}.

\begin{prop}[Imbalance under the Cube Method]~\label{prop:pbalance}\\
    Suppose Assumption \ref{hyp:design_bal} holds.
    Under the cube method using linear programming with positive-definite matrix $M$ for the landing phase, we have
    \begin{align*}
        \E\left[||B_{n,p}(X)||^2\right]\leq 4\frac{(p+1)^2}{n^2}\frac{\lambda_{max}(M)}{\lambda_{min}(M)}
    \end{align*}
for $\lambda_{max}(M)$ and $\lambda_{min}(M)$ the largest and the smallest eigenvalues of $M$. 
\end{prop}
The upper bound depends on the matrix $M$, described in equation Equation (7) in \citet{deville_efficient_2004}, used during the landing phase. Notably, one can take $M$ the identity matrix, and we have $\frac{\lambda_{max}(M)}{\lambda_{min}(M)}=1$. Then, we see that the cube method outperforms other methods that grow at a rate of $p/n$. This is clearly illustrated in Figure \ref{fig:motex}, where we see that imbalances increase only very lightly on the number of covariates when randomizing with the cube method. The main difference between the cube and other designs is that it balances selected moments of the covariates instead of balancing the whole joint distribution of $X$. It thus reduces the burden of balancing a higher number of covariates. Balancing moments can also be achieved through other methods. In particular, we can perform re-randomization such that the stopping criterion requires balancing moments of $X$ or perform a Gram-Schmidt walk design \citep{harshaw_balancing_2024}.

Re-randomization is another method that allows obtaining balance between covariates that has gained focus in the last decades \citep{morgan_rerandomization_2012,li_asymptotic_2018,imbens_experimental_2011}. The main idea of re-randomization is to completely randomize repeatedly until the obtention of balanced groups. Some researchers perform re-randomization without prespecifying it. This repetition affects treatment probabilities in an unknown manner, which induces invalid inference \citep{bruhn_pursuit_2009,athey_chapter_2017}. There are, however, several ways of performing re-randomization that allow valid inferences to some extent. Most of them rely on the simulation of the distribution under the re-randomization procedure used to assign units. This implies that researcher should draw a large number $N$ of balanced samples. One can keep randomizing until $||B_{n,p}(X)||^2\leq 4\frac{(p+1)^2}{n^2}$ to achieve the upper bound of Proposition \ref{prop:pbalance} (with $M=Id$). However, this upper bound is not sharp and to compare re-randomization with the cube method, we counted how many times an researcher should sample with naive randomization to get $N=1000$ samples that are balanced as well as $B^{\ast 2}=\E\left(||B_{n,p}(X)||^2\right)$, where the previous expectation is computed for the cube method through simulations. Under the design used in Figure \ref{fig:motex}, and for complete randomization, $12/4\times n\times||B_{n,p}(X)||^2$  converges in distribution to $\chi^2(p)$. Next, the probability to achieve balancing as good as the cube is $F_{\chi^2(p)}(12/4*n*B^{\ast2})$. To have $N=1000$ samples balanced as well as the cube, researchers thus have to sample approximately $1000/F_{\chi^2(p)}(3*n*B^{\ast2})$. For $p=3$, the researcher have to sample more than $10^6$ samples and for $p=10$ this is more than $9,98\times 10^{11}$.

The probability of getting a sample that has the same properties as the cube method becomes small very quickly, so it becomes demanding computationally, in particular, if one wants several allocations to perform randomization-based inference.

\citet{harshaw_balancing_2024} recently developed the Gram-Schmidt walk design to obtain balanced groups in RCTs. As the authors consider a tradeoff between balance and robustness, they impose a choice of parameter $\phi\in(0,1]$. For $\phi=1$, the algorithm from \cite{harshaw_balancing_2024} reduces to a coin toss and next $\frac{4\underline{C}p}{3n}\leq \E\left(||B_{n,p}(X)||^2\right)\leq  \frac{4\overline{C}p}{3n}$ under Assumption \ref{hyp:design_bal}. For $\phi\in(0;1)$, we conjecture that $\underline{K}_1 \phi \frac{p}{n}+\underline{K}_0 (1-\phi)\frac{p^2}{n^2}\leq\E\left(||B_{n,p}(X)||^2\right)\leq \overline{K}_1 \phi \frac{p}{n}+\overline{K}_0 (1-\phi)\frac{p^2}{n^2}$ for some constant $\underline{K}_0,\underline{K}_1,\overline{K}_0,\overline{K}_1$. The balance of the Gram-Schmidt method increases as $\phi$ tends to zero. However, theoretical results in \citet{harshaw_balancing_2024} and implementation of the Julia package only hold for $\phi$ positive. The choice of $\phi$ is thus critical but difficult to justify. On a side note, the cube method does not require any (subjective) parameter choice.

\subsection{Variance Reduction}

The balance between covariates in the control and treatment groups is also beneficial if these variables are related to the potential outcomes. In this case, using the cube method will also reduce the variance of the Horvitz-Thompson and Hájek estimators. 

\begin{hyp}~\label{hyp:linear}\\
	For $d\in\{0,1\}$,
	$$Y_i(d)=\beta_dZ_{di}+\varepsilon_i(d),  \text{ with } \E[\eps_i(d)|Z_{di}]=0$$.
\end{hyp}

Assumption \ref{hyp:linear} states that potential outcomes are linearly related to observable covariates. However, we allow heterogeneity in treatment effects by specifying different equations for control and treatment groups.

\begin{conj}[Poisson Approximation]~\label{conj:poiss}\\
	For any $k\in \mathbb{N}^{\ast}$ we have with probability one:
	\begin{align*}\lim_{n\rightarrow \infty}\sup_{i_1,...,i_k}\left|\E\left(\prod_{j=1}^{k}\left(D_{i_j}-\pi_{i_j}\right)\big|X_1,...,X_n\right)\right|=0
	\end{align*}
\end{conj}

This conjecture establishes that as $n$ increases, the cube method tends to Poisson sampling. As $n$ goes to infinity, the dependence between the assignment of a finite number of individuals disappears. We draw this conjecture from results in \citet{deville_variance_2005} and simulations that confirm it.

To have a benchmark for the gains in variance decline, we compare the cube method with a coin toss.

\begin{prop}[Asymptotic Normality]~\label{prop:var}\\
	Let $\theta_0$ be the SATE defined in \eqref{eq:SATE}, $\theta_0^\ast$ the PATE in \eqref{eq:PATE} and $\widehat{\theta}$ the HT or H  estimator in \eqref{eq:HT} and \eqref{eq:Hajec}. If Assumptions \ref{hyp:iid+mom}, \ref{hyp:sample} and \ref{hyp:linear}, and Conjecture \ref{conj:poiss} hold, and if $\Pi$ is a balancing sampling using the cube method we have:
	$$\sqrt{n}\left(\widehat{\theta}-\theta_0\right)\convNor{V_0}$$
	and
	$$\sqrt{n}\left(\widehat{\theta}-\theta_0^\ast\right)\convNor{V_0^\ast}.$$
 for $V_0=\E\left(\pi_i(1-\pi_i)\left(\frac{\varepsilon_i(1)}{\pi_i}+\frac{\varepsilon_i(0)}{1-\pi_i}\right)^2\right)$ and $V_0^\ast=\V(Z_{1i}'\beta_1-Z_{0i}'\beta_0) + \E\left[\frac{\varepsilon_i(1)^2}{\pi_i}\right]+ \E\left[\frac{\varepsilon_i(0)^2}{1-\pi_i}\right].$
 If a coin toss is used instead, we have:
 	$$\sqrt{n}\left(\widehat{\theta}-\theta_0\right)\convNor{V_0+\Sigma_0}$$
	and
	$$\sqrt{n}\left(\widehat{\theta}-\theta_0^\ast\right)\convNor{V_0^\ast+\Sigma_0}.$$
 with $\Sigma_0=\E\left[\frac{\pi_i}{1-\pi_i}\left(Z_{i0}'(\beta_1+\beta_0)\right)^2\right]\geq0$.
\end{prop}

Proposition \ref{prop:var} shows the gain in asymptotic variance from balancing covariates using the cube method. The reduction is more substantial when $X$ explains more of the potential outcomes. Estimates of the ATE are thus more precise when using the cube method. This reduction can represent significantly lower costs when conducting an RCT. Notice that under the same set of assumptions, $V_0^\ast$ corresponds to the semiparametric efficiency bound in \citet{hahn_role_1998}. Simulations in Sections \ref{sec:sim} illustrate these gains.

\subsection{Inference}
\label{sec:inf}
This section provides properties of the cube algorithm and methods to perform inference. We elicit two main techniques of conducting inference, one based on the asymptotic properties of the HT estimator and the other based on the randomization mechanism.

\subsubsection{Asymptotics-based Inference}

Some methods, such as re-randomization, alter the inclusion probabilities in a manner that is unclear to the researcher \citep{imbens_experimental_2011}. When the criterion for selection is known and behaves in a known way, such as the Mahalanobis distance, one can perform conservative inference. However, balance is imperfect for numerous covariates.
Since the cube method assigns treatment only once, we can perform asymptotic-based inference. We here give the asymptotic properties and propose an easy way to construct exact confidence intervals.

To construct a confidence interval, one would like to estimate either $V_0$ or $V_0^\ast$. Estimating $V_0/n$ is impossible without making assumptions on the relation between $\varepsilon_i(1)$ and $\varepsilon_i(0)$. This issue is common in RCTs. We can, nonetheless, easily construct an unbiased estimator $\widehat{V}$ for $V_0^\ast/n$. Let $\widehat{\beta}_d$ and $\widehat{\varepsilon}_i(d)$ be the estimated coefficients and residuals, respectively, of a regression of $Y_i(d)$ on $Z_{di}$, for $d\in\{0,1\}$. We then have
\begin{equation}
	\label{eq:estvar}
	\widehat{V}=\frac{1}{n}\left[\widehat{\Omega}+\frac{1}{n}\sum_{i=1}^n\frac{\widehat{\varepsilon}_i(1)^2D_i}{\pi_i^2} +\frac{1}{n}\sum_{i=1}^n\frac{\widehat{\varepsilon}_i(0)^2(1-D_i)}{(1-\pi_i)^2}\right],
\end{equation}\\
with $\widehat{\Omega}=\frac{1}{n-1}\sum_{i=1}^n\left(Z_{1i}\widehat{\beta_1}-Z_{0i}\widehat{\beta_0}-\frac{1}{n}\sum_{i'=1}^n\left(Z_{1i'}\widehat{\beta_1}-Z_{0i'}\widehat{\beta_0}\right)\right)^2.$

Then, we can test the weak hypothesis
\begin{equation}
	\label{eq:weaknull}
	H_0: \theta_0^\ast=0,
\end{equation}
and construct the confidence interval based on
\begin{equation}
	\label{eq:asyCI}
	\widehat{\theta}\pm\Phi^{-1}\left(1-\alpha/2\right)\sqrt{\widehat{V}}
\end{equation}\\
In Section \ref{sec:emp}, we perform simulations that confirm the exact coverage rate of this confidence interval when $n$ is big enough.

\subsubsection{Randomized-based Inference}

We here study the properties of randomization-based inference when permuting treatment status while satisfying balancing constraints. For these tests, we consider the stronger null hypothesis:
\begin{equation}
	\label{eq:strongnull}
	H_0:  (Y_i(1),X_i)\stackrel{d}{=}(Y_i(0),X_i).
\end{equation}
Notice that testing this hypothesis, under Assumptions \ref{hyp:iid+mom} and \ref{hyp:sample} is equivalent to testing $(Y_i)_{i=1}^n\indep (D_i)_{i=1}^n | X_1,\ldots X_n$ (Proof in Appendix \ref{proof:rbinf}).\\
To explain the test, we introduce some new notation. Let $G_n$ be the set of all possible $2^n$ assignments. Then, we can define the set of assignments $G_n^{cube}\subseteq G_n$ satisfying the constraints imposed by the cube method. That is, with Assumptions \ref{hyp:iid+mom} and \ref{hyp:sample},

$$G_n^{cube}=\left\{g\in G_n : \Delta_{j,n}=o_p\left(\frac{q}{\sqrt{n}}\right) \text{ for } 1\leq j\leq p\right\}.$$\\
We note $\mathbf{P_n}=(Y_i,D_i, X_i)_{i=1}^n$ the observed values, and $\mathbf{P_n^{(g)}}=(Y_i,D_i^{(g)}, X_i)_{i=1}^n$,  the new data where we have reassigned treatment according to $g\in G_n^{cube}$. For computational facility, we can replace $G_n^{cube}$ by $G_n^B=\{g_1,\ldots,g_B\}$, such that $g_1$ is the assignment really obtained  and $(g_i)_{i=1}^B$ are drawn independently from a uniform distribution on $G_ n^{cube}$.\\
Then, for a given test statistic $T_n(\mathbf{P_n})$,we consider the test

$$\phi^{rand}(\mathbf{P_n})=\mathbbm{1}\left\{T_n(\mathbf{P_n})>c_n(\mathbf{P_n},1-\alpha)\right\}$$
with
$$c_n(\mathbf{P_n},1-\alpha)=\inf\left\{t\in \mathbb{R} : \frac{1}{B}\sum_{g\in G_n^{B}}\mathbbm{1}\{T_n(\mathbf{P_n^{(g)}}) \leq t\}\geq 1-\alpha\right\}.$$

\begin{prop}[Randomization-based Inference]~\label{prop:randinf}\\
	Under Assumptions \ref{hyp:iid+mom} and \ref{hyp:sample}, and the null hypothesis in \eqref{eq:strongnull},
	$$\E\left[\phi^{rand}_n(\mathbf{P_n})\right]\leq \alpha.$$
\end{prop}
Proposition \ref{prop:randinf} indicates that if $T_n(\mathbf{P_n}) > c_n(\mathbf{P_n},1-\alpha)$, we reject the null hypothesis \eqref{eq:strongnull} at the $\alpha$ level. The proof is similar to previous results on other covariate-adaptive assignment mechanisms \citep{heckman_analyzing_2010,heckman_inference_2011,lee_multiple_2014,bai_inference_2022}, but it is presented for completeness. This proposition ensures that we can compute Fisher's $p$-values by comparing our test statistic with those produced by other assignments made by the Cube method.

\section{Simulations}
\label{sec:sim}
This section compares the cube method to other randomization methods by performing Monte Carlo simulations. We are interested in examining the impact of introducing new covariates in the variance of treatment effect estimates. For this purpose, we evaluate different randomization methods using one example from data following a simple DGP in the spirit of Figure \ref{fig:motex} and another using data-driven methods from an empirical application. We use packages in R for randomization available here:
https://rdrr.io/cran/BalancedSampling/.
Interestingly enough, the cube method is not computationally demanding. While running simulations included in the present paper, we systematically found the cube method to have an order of magnitude faster than the algorithms we used to implement other presented methods.

\subsection{Simple DGP}
\label{sec:simple}
For $k=1,\ldots,K$ the number of iterations, $j=1,\ldots,p$ the number of covariates, and $i=1,\ldots,n$, the number of observations, we independently draw $X_{jik}\sim\mathcal{U}(0,1)$ and $\eps_{ik}(d)\sim\mathcal{N}(0,1) $, for $d=0,1$. We then generate the potential outcomes $Y_{ik}(0)=1+ (X_{ik}-1/2)'\beta_0+\eps_{ik}(0)$ and $Y_{ik}(1)=1+X_{ik}'\beta_1+(X_{ik}-1/2)'A(X_{ik}-1/2)+\eps_{ik}(1),$ with $A=(1/20)\times(\mathbbm{11'}-\operatorname{diag}(1))$. Notice that, in this example, $\theta_0^\ast=0.$

We consider $n=500$, $p=30$, $\beta_0=(1,\boldsymbol{0}')'$, $\beta_1=2\beta_0$, so only one covariate and noise explain variations in the individual treatment effect. We assume that the researcher knows she should always balance this covariate. Still, she does not have previous information about the (un)informativeness of the 29 other covariates. In these simulations, the researcher has to choose which simulation method she uses and how many covariates to include. That choice corresponds to an assignment design $\Pi$ and generates treatment statuses $D_{ik}^\Pi$. We estimate the PATE using the HT estimator $\widehat{\theta}^\Pi_{HT,k}$. To evaluate the precision entailed by the assignment design, we perform $K=5,000$ simulations and compute the standard deviation of the estimator over the simulations. Since the PATE is null, this is equivalent to estimating the root mean square error (RMSE).

Figure \ref{fig:MCsim} shows the RMSE of the HT estimator by number of covariates and randomization method. The simulations show that the cube method is always competitive. Since including more covariates deteriorates balancing only very lightly, precision gains are maintained even when $p=30$. This behavior is not present for other randomization methods. Indeed, stratification using the median (quartile) leads to worse precision than complete randomization as soon as $p>6$ ($p>3)$ and converges to a coin toss for $p>12$ ($p>6)$. Moreover, using a matched-pairs design improves from complete randomization but, for $p>3$, underperforms compared to the cube method: when $p$ increases, precision for matched-pairs design worsens, whereas it remains the same when using the cube. By allowing an abundant set of covariates, the cube method improves the exploitation of balancing gains, even when the researcher chooses to balance covariates that are not explicative of potential outcomes. This behavior could arise if the researcher is interested in several treatment outcomes and collects their pre-treatment values. Then, she would ideally want to balance them all, even if only one covariate is explicative of one outcome. As described through these simulations, the cube method ensures precision gains for a particular outcome variable, even when balancing another 29 baseline variables.

\subsection{Empirical Data}
\label{sec:emp}
We further illustrate the properties of the cube method by using experimental data from \citet{gerber_one_2020}. The authors investigate how informing potential voters about the closeness of an election affects their beliefs and voting behavior. Since the experimental data only represents one of many possible samplings, we proceed by generating a \textit{superpopulation}. We create a large dictionary with baseline outcomes, covariates, and demographics. We consider all possible interactions and second-order polynomials. We thus generate a dataset of 6,424 observations and 7,381 covariates (hereon denoted by $X$), with 3,193 individuals in the treatment group. We consider beliefs about the closeness of the election as the main outcome $Y$. We ran two lasso regressions separately for treated and control units to train two models, $f_1$ and $f_0$. We then estimate $s_1^2=\widehat{\V}(Y-f_1(X)|D=1)$ and $s_0^2=\widehat{\V}(Y-f_0(X)|D=0)$.  To generate the superpopulation we draw $N=50,000$ individuals, with replacement and we generate $Y_i(1)=f_1(X_i)+\eps_i(1)$ and $Y_i(0)=f_0(X_i)+\eps_i(0)$ for $i=1,\ldots,N$ with $(\eps_i(1);\eps_i(0))\sim\mathcal{N}\left((0~,~0) , (s_1^2 ~ ~0.5s_1s_0~,~0.5s_1s_0 ~~ s_0^2
)\right).$ We thus obtain a superpopulation $(X_i, Y_i(1), Y_i(0))_{i})_{i=1,\ldots,N}$.

We then run $K=10,000$ Monte Carlo simulations, where for every iteration, we draw $n\in(100,256,500,1000)$\footnote{We select 256 because exact inference methods for matched pair designs require a sample size divisible by four \citep{bai_inference_2022}.} individuals, allocate them according to five treatment allocation methods: complete randomization, stratified randomization using median values for continuous variables, matched-pairs design using the Mahalanobis distance when balancing multiple covariates, and the cube method with the two first moments per variable. For stratified designs, matched-pairs design, and the cube method, we balance between 1 and 12 covariates. When balancing only one, we use the pre-treatment value of $Y$. For the 12 covariates, we consider five pre-treatment outcomes and seven baseline covariates. We always prioritize pre-treatment outcomes as they are likely the most explicative variable for their post-treatment counterpart. We set $\pi_i=\frac{1}{2}$. For complete randomization, matched pairs, and the cube method, we compute the HT estimator. For these methods, we compute confidence intervals using, respectively, White standard errors, Equation (14) in \citet{bai_optimality_2022}, and Equation \eqref{eq:estvar} in Section \ref{sec:inf} above. For stratification, since  $\pi_i=\frac{1}{2}$, we use an OLS regression with strata fixed-effects, which gives consistent estimators and exact inference as shown by  \citet{bugni_inference_2018}.

Table \ref{tb:empiricalcov} reports estimators of the effective sample size $\operatorname{ESS}=\frac{\V(\widehat{\theta}_{HT}^\Pi)}{\V(\widehat{\theta}_{HT}^{\operatorname{CR}})}\times n$ of each design, based on the variance of the estimators across the $K$ iterations. The ESS indicates, for every allocation design, the experimental sample size required to estimate the treatment effect with the same precision as with complete randomization. We see that almost every covariate-adaptive method does better than complete randomization, as they allow to reduce the sample size, often substantially. The only exception is stratification with many covariates. In general, if $n<2^p$, stratification becomes worse than complete randomization. This phenomenon is due to the small strata issue and worsens with the strata fixed-effects estimator. Across different $n$ sizes, we see that the matched-pairs design does better than the cube method when we balance a few covariates (up to three, in general). However, once balancing more covariates, the cube method becomes more efficient. As expected, these relative gains to the matched-pairs design are more apparent for smaller $n$ since the curse of dimensionality is more stringent. Notice there are clear gains from using the cube method even when allowing for non-linearities in the imputation of $Y(1)$ and $Y(0)$. These results and those in Section \ref{sec:sim}, along with coverage rates displayed in Table \ref{tb:empcoverage} in the Appendix, constitute suggestive evidence for a possible relaxation of Assumption \ref{hyp:linear}.

\begin{table}[htbp]
	\caption{Effect of number of covariates on effective sample size}
	\label{tb:empiricalcov}
	\centering\scriptsize
	\begin{threeparttable}
		\begin{tabularx}{\textwidth}{cY|YYYYY}
			\specialrule{.1em}{.05em}{.05em}\specialrule{.1em}{.05em}{.05em}
&Number of covariates   & Complete Randomization & Stratified Randomization & Matched \quad\quad\quad Pairs & Cube \quad\quad\quad Method  \\  
	&		& (1)         & (2)                      & (3)                     & (4)                   \\\specialrule{.1em}{.05em}{.05em}
&1 & 100.00 & 80.99 & 63.82 & 64.90  \\ 
&  2 & -- & 81.66 & 63.57 & 67.89   \\ 
&  3 & -- & 80.72 & 65.24 & 68.46  \\ 
$n=100$&  5 & -- & 90.57 & 67.28 & 67.37  \\ 
&  7 & -- & 113.60 & 69.10 & 67.70   \\ 
&  9 & -- & 160.25 & 73.14 & 65.31   \\ 
&  12 & -- & 523.57 & 75.08 & 65.40  \\\specialrule{.1em}{.05em}{.05em}
&1 & 256.00 & 215.49 & 162.69 & 167.76  \\ 
&  2 & -- & 212.66 & 162.92 & 170.79  \\ 
&  3 & -- & 215.07 & 163.50 & 169.53  \\ 
$n=256$ &  5 & -- & 215.00 & 174.35 & 168.93  \\ 
&  7 & -- & 240.92 & 180.31 & 171.15  \\ 
 & 9 & -- & 308.55 & 183.49 & 165.50  \\ 
 & 12 & -- & 610.22 & 194.93 & 174.00 \\  
			\specialrule{.1em}{.05em}{.05em}
&1 & 500.00 & 419.11 & 328.17 & 329.99  \\ 
&  2 & -- & 408.31 & 328.15 & 325.72 \\ 
&  3 & -- & 408.02 & 320.39 & 332.88 \\ 
$n=500$&  5 & -- & 418.08 & 326.86 & 331.19 \\ 
& 7 & -- & 445.93 & 329.81 & 332.37\\ 
&  9 & -- & 514.09 & 349.28 & 334.98 \\ 
&  12 & -- & 847.19 & 363.50 & 330.05  \\ 
			\specialrule{.1em}{.05em}{.05em}
&1 & 1000.00 & 828.85 & 646.01 & 669.16  \\ 
&  2 & -- & 826.04 & 640.22 & 647.56  \\ 
&  3 & -- & 830.47 & 641.96 & 638.66 \\ 
$n=1000$&  5 & -- & 819.82 & 651.41 & 666.30  \\ 
&  7 & -- & 845.31 & 657.62 & 663.35  \\ 
 & 9 & -- & 955.18 & 686.53 & 664.19  \\ 
 & 12 & -- & 1329.51 & 717.35 & 656.39  \\ 
 \specialrule{.1em}{.05em}{.05em}\specialrule{.1em}{.05em}{.05em}
		\end{tabularx}
		\begin{tablenotes}[flushleft,para]
			\scriptsize
			This table shows the effective sample size (ESS) for different allocation designs and experimental sample sizes. For each allocation design, the ESS corresponds to the number of observations needed to have the same precision as under complete randomization, equal to $n\times(\widehat{\V}(\widehat{\theta}_{HT}^\Pi)/\widehat{\V}(\widehat{\theta}_{HT}^{\operatorname{CR}}))$, where we compute the variance estimates over $10,000$ simulations. For columns 1, 3, and 4, the estimator used is the Horvitz-Thompson algorithm. Column 1 gives the size $n$, as the design used is complete randomization. For column 3, we assign treatment using the matched-pairs design, pairing individuals to the closest unit and using the Mahalanobis distance whenever more than one covariate is balanced. Column 4 shows the results for the cube method with two moments for each variable. For column 2, we run stratified randomization. We use median values for continuous variables and estimate the PATE using OLS with strata fixed-effects following \citet{bugni_inference_2018}.
		\end{tablenotes}
	\end{threeparttable}
\end{table}

Tables \ref{tb:empsd}-\ref{tb:emppower} in the Appendix show additional results for each design: standard deviation and bias of the HT estimator, confidence coverage rate, and power for testing a null PATE. In particular, Table \ref{tb:empcoverage} verifies that the coverage rate is exact for $n$ large enough.

\section{Discussion}
\label{sec:practical}

Although RCTs are now common in economics, there are still some aspects of randomization that can be further clarified. We here discuss two of these, in light of the contribution of the cube method. We first show that, despite being widely available, pre-treatment information is still underused. The cube method could improve practices. We also discuss how pre-analysis plans are often vague about which covariates should be included. In a second section, we explain how the cube method may help suppress a publication bias arising from imbalances.

\subsection{On the use of balancing methods in RCTs}

\begin{figure}[htb]
	\centering
	\caption{Distribution of Randomization Methods for 104 RCTs in Top-5 + AEA Journals (2019-2023)}
	\label{fig:Top5}
		\centering
		\includegraphics[width=\textwidth]{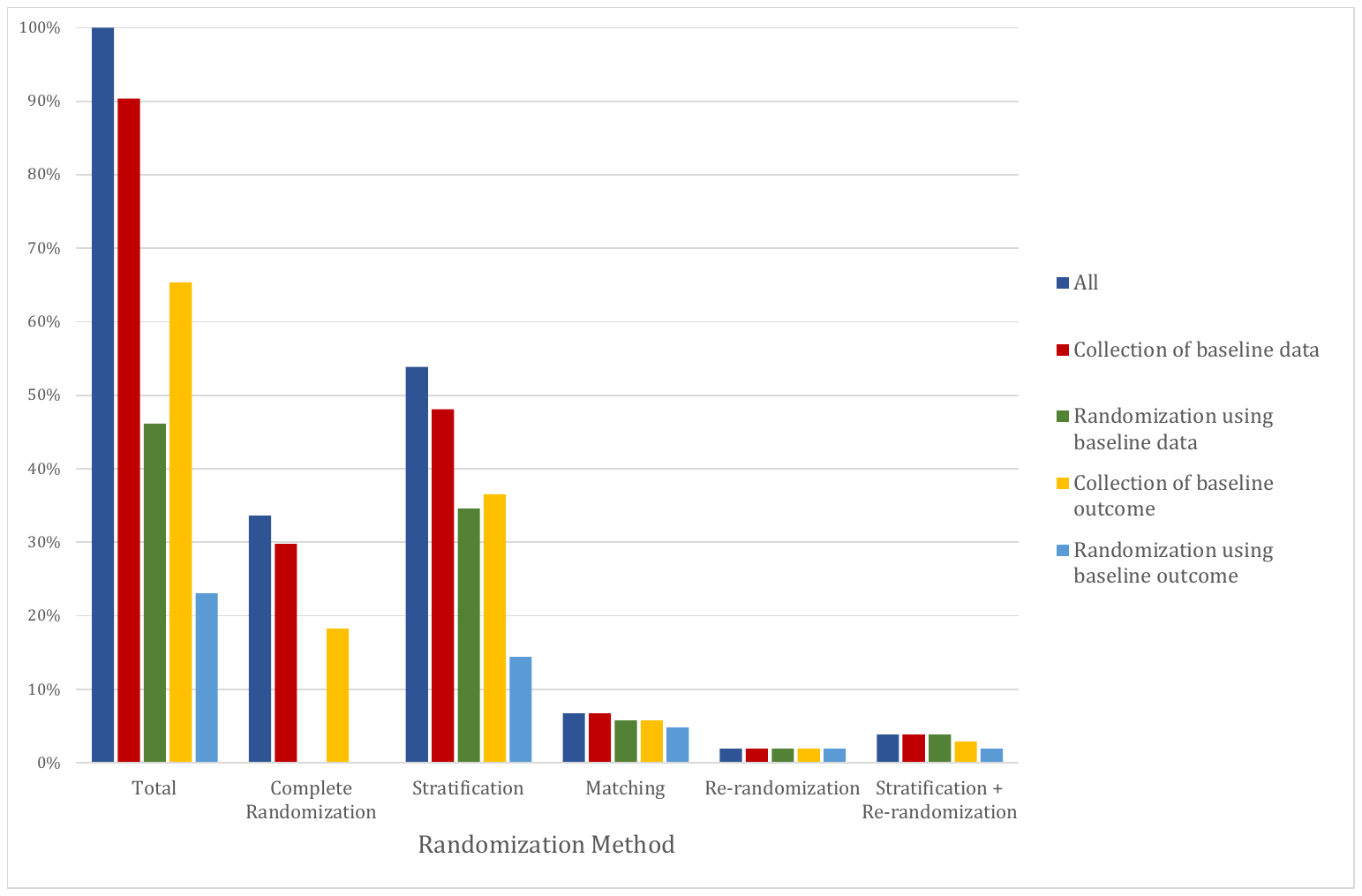}
\end{figure}

\cite{bai_optimality_2022} indicates that among 5,000 RCTs in the AEA RCT Registry, more than 800 are stratified (i.e., about 16\%). We complement this insight by gathering information from 104 randomized controlled trials (RCTs) published in top-5 and AEA journals\footnote{American Economic Review, AEJ: Applied Economics, AEJ: Economic Policy, AEJ: Macroeconomics, AEJ: Microeconomics, Econometrica, Journal of Political Economy, Quarterly Journal of Economics, and Review of Economic Studies} between 2019 and 2023. Specifically, we examined their collection of baseline information, baseline outcomes, and randomization methods. Figure \ref{fig:Top5} summarizes these details. Our findings indicate a lack of consensus among RCTs regarding the method used for allocating individuals to treatment. Most published papers (54\%) employ a stratified design, followed by completely randomized designs (34\%). A minority of researchers utilize alternative methods such as matching or re-randomization. However, there is substantial agreement regarding the collection of baseline data, with 90\% of papers gathering information before treatment allocation. Nevertheless, this information is not always utilized during the randomization process, as only 46\% of the papers leverage it to achieve covariate balance. The remaining studies collect baseline data for balance tests, covariate adjustment in regression, and/or heterogeneity analysis of treatment effects. If outcomes of interest are relatively stable over time, researchers should be interested in balancing their pre-treatment values, as they are highly likely to be correlated with potential outcomes. In our sample, 65\% of researchers collect these variables, yet only 23\% incorporate them into the allocation design, indicating an area for improvement in experimental design and inference. When various outcomes are considered in RCT, the curse of dimensionality arising in stratification, matched pair design, or re-randomization (where computational time could become prohibitive) may prevent researchers from balancing on a large set of pre-treatment outcomes and sociodemographic covariates. In view of results in Section \ref{sec:results}, the cube method could greatly improve experimental randomization by allowing balancing on more variables than other methods.

\subsection{``Balancing checks'': cube randomization and publication bias}
Researchers routinely provide ``balance tables'', i.e., a comparison of the moments of available covariates between control and treatment. Due to bad luck, the researcher can expect a certain proportion of significant imbalances. And the likelihood of imbalances increases with the number of covariates used. There seems to be a gray area around the reporting of balance checks. In particular, pre-analysis plans often report no clear justification regarding the choice of covariates to include in balance tables.

Researchers are not very comfortable with reporting substantial unbalances. \cite{snyder_examining_2024} analyzing a large set of balance checks find that the editorial process removes an ample part of studies reporting imbalances, perhaps as much as 30\%. Imbalances increase the risk of rejection by journals and, even worse, may provide incentives to engage in p-hacking (e.g., removing some covariates from the balance table). Our Proposition \ref{prop:balance} ensures that cube randomization definitively solve these issues making balance checks superfluous.

\subsection{Future research and limitations}

To conclude this discussion, let us mention some limitations of our results and agenda for future research. We use an assumption of linearity in the conditional expectation of potential outcomes to establish some asymptotic results. Based on simulations, we conjecture that our results hold without linearity assumption, but we do not have a formal proof at the current stage.

\newpage
\bibliography{references}

\newpage
\appendix
\numberwithin{equation}{section}
\renewcommand{\theequation}{\thesection.\arabic{equation}}
\setcounter{table}{0}
\renewcommand{\thetable}{\thesection.\arabic{table}}
\renewcommand\thelem{\thesection.\arabic{lem}}
\setcounter{lem}{0}
\renewcommand\theprop{\thesection.\arabic{prop}}
\setcounter{prop}{0}
\renewcommand\thedefin{\thesection.\arabic{defin}}
\setcounter{defin}{0}
\section{Imbalances in Randomization Methods}
\label{sec:imbalances}
In this appendix, we prove the theoretical results discussed in Section \ref{sec:comparison}.\\
~\\
We consider $(D_1,....,D_n)\in \{0;1\}^n$ a random selection of some units among $n$. We will consider the following schemes:
\begin{enumerate}
	\item Coin Toss (CT): for any $(d_1,...,d_n)\in \{0;1\}^n$ $$\Pr\left((D_1,....,D_n)=(d_1,...,d_n)|(X_i)_{1\leq i\leq n}\right)=2^{-n}$$ 
	\item Complete randomization (CR): for any $(d_1,...,d_n)\in \{0;1\}^n$ $$\Pr\left((D_1,....,D_n)=(d_1,...,d_n)|(X_i)_{1\leq i\leq n}\right)=\binom{n}{n/2}^{-1}\mathds{1}\left\{\sum_{i=1}^n d_i=n/2\right\}$$
	\item Stratification (S-$\ell$): we consider $q^k_{p}$ the empirical $p$-quantile of $(X_{ki})_{i=1,...,n}$ and the partition $H_n=\left\{\prod_{k=1}^p[q^k_{j_k/\ell};q^k_{(j_k+1)/\ell}]: \forall k\in \{1,...,p\}, j_k\in\{0,1,...,\ell-1\} \right\}$ of $[0;1]^k$ for some $\ell\geq 2$ and $\mathcal{S}_n=\left\{\{i: X_i\in h\}:h\in H_n\text{ such that }\exists i: X_i\in h\right\}$ a partition of $\{1,...,n\}$.	For any $(d_1,...,d_n)\in \{0;1\}^n$
	$$\Pr\left((D_1,....,D_n)=(d_1,...,d_n)|(X_i)_{1\leq i\leq n}\right)=\prod_{s\in \mathcal{S}_n}\binom{|s|}{\lfloor|s|/2\rfloor}^{-1}\mathds{1}\{\lfloor |s|/2\rfloor\leq \sum_{i\in s} d_i\leq \lfloor (|s|+1)/2\rfloor\}.$$  

	\item Matched-pairs design (MP): we consider a partition $\mathcal{S}_n$ of $\{1,...,n\}$ such that \begin{align}\mathcal{S}_n&:=\arg \min_{\mathcal{S}: |s|=2\text{ for any }s\in \mathcal{S}}\sum_{s\in \mathcal{S}}\left|\left|\sum_{i,j\in s^2, i\neq j}X_i-X_j\right|\right|^{2}
		\label{opti:MP} \end{align}
  \item Cube method with landing phase using linear programming (CM) as described in Section \ref{sec:cube}.
\end{enumerate}
~\\
In all the previous designs, we have $\Pr(D_i=1|(X_i)_{i=1,...,n})=1/2$ and next, the estimator $B_{n,p}(X)=\frac{2}{n}\sum_{i=1}^nX_iD_i-\frac{2}{n}\sum_{i=1}^nX_i(1-D_i)$ has expectation $\E(X|D=1)-\E(X|D=0)=0$, meaning that these assignments  balance $X$ on average. To compare the balancing of these assignments, we will bound $\E(||B_{n,p}(X)||^2)$ for each design.\\
~\\
\begin{prop}~\label{prop:imbalance}\\
	Suppose that Assumption \ref{hyp:design_bal} holds. 
	\begin{enumerate}
	\item Under assignment design (CT) we have:
	\begin{align}&\E\left(||B_{n,p}(X)||^2\right)=\frac{4}{n}\sum_{k=1}^p\E\left(X_{k1}^2\right)\label{bal:Bernoulli1}\\
 \text{ and~~ }&\frac{4\underline{C}}{3}\frac{p}{n}\leq \E\left(||B_{n,p}(X)||^2\right)\leq \frac{4\overline{C}}{3}\frac{p}{n}.\label{bal:Bernoulli2}
	\end{align}
	\item Under assignment design (CR) we have:
	\begin{align}&\E\left(||B_{n,p}(X)||^2\right)=\frac{4}{n}\sum_{k=1}^p\V\left(X_{k1}^2\right)\label{bal:CR1}\\
 \text{ and~~ }&\frac{\underline{C}}{3}\frac{p}{n}\leq \E\left(||B_{n,p}(X)||^2\right)\leq \frac{\overline{C}}{3}\frac{p}{n}.\label{bal:CR2}
\end{align}
		\item Under assignment design (S-$\ell$), we have:
		\begin{enumerate}
			\item if $n\ell^{-p}\rightarrow \infty$: 
		\begin{align}||B_{n,p}(X)||^2&=B_1^2+o_p\left(\frac{p}{n}\right)\label{bal:Strata1}\end{align}
		with  
	$ \frac{\underline{C}}{6\ell^2\overline{C}}\frac{p}{n}\left(1-o\left(1\right)\right)\leq \E\left(B_1^2\right)\leq \frac{4}{n}\sum_{k=1}^p\V(X_{k1})$
		\item if $n\ell^{-p}\rightarrow 0$:
	\begin{align}||B_{n,p}(X)||^2&=B_2^2+o_p\left(\frac{p}{n}\right).\label{bal:Strata2}
\end{align}
with $\E(B_2^2)=\frac{4}{n}\sum_{k=1}^p\E\left(X_{k1}^2\right)$.\end{enumerate}
\item Under assignment design (MP), we have:
\begin{align}\frac{p}{n}\left(\frac{1}{3}-\sqrt{\frac{2\ln(n-1)+4\ln{\overline{C}}}{p}}\right)\leq \E\left(||B_{n,p}(X)||^2\right)\leq \frac{4}{n}\sum_{k=1}^p\V(X_{1k})\label{bal:MP1}\end{align}
\item Under assignment design (CM), w have:
\begin{align}
	\E\left(||B_{n,p}(X)||^2\right)\leq 4\frac{(p+1)^2}{n^2}\frac{\lambda_{max}(M)}{\lambda_{min}(M)},\label{bal:CUBE}
	\end{align}
for $\lambda_{max}(M)$ and $\lambda_{min}(M)$ the largest and the smallest eigenvalues of $M$. 
\end{enumerate}
\end{prop}
\underline{Proof:}\\
For all the assignment designs considered, we have
\begin{align*}
\E\left(||B_{n,p}(X)||^2\right)&=4\sum_{k=1}^p\E\left[\left(\frac{1}{n}\sum_{i=1}^n X_{ki}(2D_i-1)\right)^2\right]\\
&=4\sum_{k=1}^p\E\left[\left(\frac{1}{n}\sum_{i=1}^n\frac{X_{ki}D_i}{\Pr(D_i=1|(X_j)_{j=1,...,n})}-\overline{X}_k\right)^2\right]\\
&=4\sum_{k=1}^p\E\left[\V\left(\frac{1}{n}\sum_{i=1}^n\frac{X_{ki}D_i}{\Pr(D_i=1|(X_j)_{j=1,...,n})}\bigg|(X_j)_{j=1,...,n}\right)\right].
	\end{align*}
~\\
Under assignment design (CT), we have $D_i\indep D_j|(X_j)_{j=1,...,n}$ and $\E\left(2D_i-1|(X_j)_{j=1,...,n}\right)=0$. Next, expanding the square we have
\begin{align*}
\E\left[\left(\frac{1}{n}\sum_{i=1}^n X_{ki}(2D_i-1)\right)^2\bigg| (X_{j})_{j=1,...,n}\right]&=\frac{1}{n^2}\sum_{i=1}^nX_{ki}^2.	\end{align*}
Identical distribution across $i$ ensures \eqref{bal:Bernoulli1}.\\
For any $k\in\{1,...,p\}$ we have:
\begin{align*}&\frac{\underline{C}}{3}\leq \frac{1}{3}\left(\underline{C}+\frac{(1-\underline{C})^3}{(\overline{C}-\underline{C})^2}\right)=\int_0^1 u^2\left(\overline{C}\mathds{1}\{u\leq \frac{1-\underline{C}}{\overline{C}-\underline{C}}\}+\underline{C}\mathds{1}\{u\geq \frac{1-\underline{C}}{\overline{C}-\underline{C}}\}\right)du\\
	&\leq \E(X_{k1}^2)\leq
	\int_0^1 u^2\left(\underline{C}\mathds{1}\{u\leq \frac{\overline{C}-1}{\overline{C}-\underline{C}}\}+\overline{C}\mathds{1}\{u\geq \frac{\overline{C}-1}{\overline{C}-\underline{C}}\}\right)du= \frac{1}{3}\left(\overline{C}-\frac{(\overline{C}-1)^3}{(\overline{C}-\underline{C})^2}\right)\leq \frac{\overline{C}}{3},\end{align*}
and next \eqref{bal:Bernoulli2} follows.\\
~\\
Under assignment design (CR), we have:
\begin{align*}
	\V\left(\frac{1}{n}\sum_{i=1}^n\frac{X_{ki}D_i}{\Pr(D_i=1|(X_j)_{j=1,...,n})}\bigg|(X_j)_{j=1,...,n}\right)&=\left(1-\frac{1}{2}\right)\frac{2}{n}S^2_k
	\end{align*}
for $S^2_k=\frac{1}{n-1}\sum_{i=1}^n\left(X_{ki}-\overline{X}_k\right)^2$ and $\overline{X}_k=\frac{1}{n}\sum_{i=1}^nX_{ki}$. Because $\E(S^2_k)=\V(X_{k1})$, \ref{bal:CR1} follows. \refeq{bal:CR2} is ensured by \begin{align*}&\frac{\underline{C}}{12}\leq \frac{1}{12}\left(\underline{C}+\frac{(1-\underline{C})^3}{(\overline{C}-\underline{C})^2}\right)\\
	&=\int_0^1 \left(u-\frac{1}{2}\right)^2\left(\overline{C}\mathds{1}\left\{\left|u-\frac{1}{2}\right|\leq \frac{1-\underline{C}}{2\left(\overline{C}-\underline{C}\right)}\right\}+\underline{C}\mathds{1}\left\{\left|u-\frac{1}{2}\right|\geq \frac{1-\underline{C}}{2\left(\overline{C}-\underline{C}\right)}\right\}\right)du\\
	&\leq \V(X_{k1})\leq
	\int_0^1 \left(u-\frac{1}{2}\right)^2\left(\underline{C}\mathds{1}\left\{\left|u-\frac{1}{2}\right|\leq \frac{\overline{C}-1}{2\left(\overline{C}-\underline{C}\right)}\right\}+\overline{C}\mathds{1}\left\{\left|u-\frac{1}{2}\right|\geq \frac{\overline{C}-1}{2\left(\overline{C}-\underline{C}\right)}\right\}\right)du\\&= \frac{1}{12}\left(\overline{C}-\frac{(\overline{C}-1)^3}{(\overline{C}-\underline{C})^2}\right)\leq \frac{\overline{C}}{12}.\end{align*}
 ~\\
Under assignment design (S-$\ell$), for $s\in \mathcal{S}_n$ let $m_s$ the number of units assigned to the treatment in strata $s$. Similarly, for $h\in \mathcal{H}_n$ let $m_h$ the number of units assigned to the treatment in strata $\{i: X_i\in h\}$ and $m=\sum_{s\in \mathcal{S}_n}m_s=\sum_{h\in \mathcal{H}_n}m_h$. Last, $n_h$ denotes $|\{i:X_i\in h\}|$ for $h\in \mathcal{H}_n$. We have
\begin{align*}
	&\V\left(\frac{1}{n}\sum_{i=1}^n\frac{X_{ki}D_i}{\Pr(D_i=1|(X_j)_{j=1,...,n})}\bigg|(X_j)_{j=1,...,n},(m_s)_{s\in \mathcal{S}}\right)\\=&\sum_{s\in \mathcal{S}_n}\left[\left(\frac{m_s}{m}\right)^2\left(1-\frac{m_s}{|s|}\right)\frac{1}{m_s}S^2_{ks}\mathds{1}\{|s|\geq 2\}+\frac{1}{n^2}\sum_{i\in s}X^2_{ki}\mathds{1}\{|s|=1\}\right]\\
	=& \sum_{h\in \mathcal{H}_n}\left[\left(\frac{m_h}{m}\right)^2\left(1-\frac{m_h}{n_h}\right)\frac{1}{m_h}S^2_{kh}\mathds{1}\{n_h\geq 2\}+\frac{1}{n^2}\left(\sum_{i\in \{j: X_j\in h\}}X^2_{ki}\right)\mathds{1}\{n_h=1\}\right]
\end{align*}
with $S^2_{ks}=\frac{1}{|s|-1}\sum_{i\in s}(X_{ki}-\overline{X}_{ks})^2$ for $\overline{X}_{ks}=\frac{1}{|s|}\sum_{i\in s}X_{ki}$ and $S^2_{kh}=\frac{1}{n_h-1}\sum_{i\in \{j:X_j\in h\}}(X_{ki}-\overline{X}_{ks})^2$ for $\overline{X}_{kh}=\frac{1}{n_h}\sum_{i\in \{j: X_j\in h\}}X_{ki}$. $\left(n_h\right)_{h\in \mathcal{H}_n}$ follows a multinomial distribution of parameters $(n,\ell^{p}, (p_h)_{h\in H})$ with $\frac{\underline{C}}{\overline{C}}\ell^{-p}\leq\frac{\underline{C}}{\underline{C}+\overline{C}(\ell^{p}-1)}\leq p_h\leq \frac{\overline{C}}{\overline{C}+\underline{C}(\ell^p-1)}\leq \frac{\overline{C}}{\underline{C}}\ell^{-p}$.  Then for any $h\in \mathcal{H}_n$, we have $\Pr\left(n_h= 1\right)=np_h(1-p_h)^{n-1}$ and $\Pr\left(n_h\geq 2\right)=1-(1-p_h)^n-np_h(1-p_h)^{n-1}$. Moreover, $(m_h)_{h\in \mathcal{H}_n}\indep (S_{kh'})_{h'\in \mathcal{H}_n,k=1,...,p}|(n_{h''})_{h''\in \mathcal{H}_n}$, $\E(S_{kh}^2|(n_{h'})_{h'\in \mathcal{H}_n})=\V(X_{k1}|X_1\in h)\in \left[\frac{\underline{C}}{12\overline{C}\ell^2};\frac{\overline{C}}{12\underline{C}\ell^2}\right]$ and, conditional on $(n_h)_{h\in \mathcal{H}_n}$, $m_h$ are independent across $h$ with probability distribution $\frac{1}{2}\delta_{\lfloor\frac{n_h}{2}\rfloor}+\frac{1}{2}\delta_{\lfloor\frac{n_h+1}{2}\rfloor}$. It follows that $\E(m_h|(n_{h'})_{h'\in \mathcal{H}_n})=n_h/2$, $\V(m_h|(n_{h'})_{h'\in \mathcal{H}_n})=(\lfloor (n_h+1)/2\rfloor-\lfloor n_h/2\rfloor)^2/4\leq 1/4$, $\E(m|(n_{h'})_{h'\in \mathcal{H}_n})=n/2$, $\V(m|(n_{h'})_{h'\in \mathcal{H}_n})\leq \ell^p/4$.
Chebyshev's inequality implies $\Pr\left(|m-n/2|\geq M\right)\leq \frac{\ell^p}{4M^2}$. Next, $m=\frac{n}{2}\left(1+O_p(n^{-1}\ell^{p/2})\right)=\frac{n}{2}\left(1+o_p(1)\right)$ and  
\begin{align*}
	\sum_{h\in \mathcal{H}_n}\left(\frac{m_h}{m}\right)^2\left(1-\frac{m_h}{n_h}\right)\frac{1}{m_h}S^2_{kh}\mathds{1}\{n_h\geq 2\}&=\frac{4}{n^2}\left(1+o_p(1)\right)\sum_{h\in \mathcal{H}_n}m_h\left(1-\frac{m_h}{n_h}\right)S^2_{kh}\mathds{1}\{n_h\geq 2\}
\end{align*}
and $\frac{n_h}{8}\mathds{1}\{n_h\geq 2\}\leq\frac{n_h^2-1}{4n_h}\mathds{1}\{n_h\geq 2\}\leq m_h\left(1-\frac{m_h}{n_h}\right)\mathds{1}\{n_h\geq 2\}\leq\frac{n_h}{4}\mathds{1}\{n_h\geq 2\}$.
\\If $n \ell^{-p}\rightarrow \infty$ as $n$ and $p$ increase, we have $n p_h\geq \frac{\underline{C}}{\overline{C}}n\ell^{-p}\rightarrow \infty$ and next $\sup_{h\in \mathcal{H}_n}\Pr(n_h=1)\leq\sup_{h\in \mathcal{H}_n}np_h e^{(n-1)\ln(1-p_h)}\leq\sup_{h\in \mathcal{H}_n}np_h e^{-(n-1)p_h}\leq n\ell^{-p}\frac{\overline{C}}{\underline{C}} \exp(-(n-1)\ell^{-p}\underline{C}/\overline{C})$ and because $|\mathcal{H}_n|=\ell^p$ and $|X_{ki}^2|\leq 1$, we have:
\begin{align*}\E\left[\sum_{k=1}^p\sum_{h\in \mathcal{H}_n}\frac{1}{n^2}\sum_{i\in \{j: X_j\in h\}}X^2_{ki}\mathds{1}\{n_h=1\}\right]&\leq\frac{p}{n}\frac{|\mathcal{H}_n|}{n}\sup_{h\in \mathcal{H}_n}\Pr\left(\mathds{1}\{n_h=1\}\right)\\&\leq\frac{p}{n} \frac{\overline{C}}{\underline{C}}\exp\left(\frac{\underline{C}}{\ell^p\overline{C}}\right)\exp\left(-n\ell^{-p}\frac{\underline{C}}{\overline{C}}\right)=o\left(\frac{p}{n}\right).
\end{align*}\\
We have $\Pr\left(|m-n/2|\geq M\right)\leq \frac{\ell^p}{4M^2}$, next $m=\frac{n}{2}\left(1+O_p(1/(n\ell^{p/2})\right)$ and  
\begin{align*}
	\sum_{h\in \mathcal{H}_n}\left(\frac{m_h}{m}\right)^2\left(1-\frac{m_h}{n_h}\right)\frac{1}{m_h}S^2_{kh}\mathds{1}\{n_h\geq 2\}&=\frac{4}{n^2}\left(1+o_p(1)\right)\sum_{h\in \mathcal{H}_n}m_h\left(1-\frac{m_h}{n_h}\right)S^2_{kh}\mathds{1}\{n_h\geq 2\}\\
	&\leq \frac{1+o_p(1)}{n}\sum_{h\in \mathcal{H}_n}\frac{n_h}{n}S^2_{kh}\mathds{1}\{n_h\geq 2\}\\
	&\leq \frac{1+o_p(1)}{n}\sum_{h\in \mathcal{H}_n}\frac{n_h}{n}S^2_{kh}
	\end{align*}
Last, note that $\E(S_{kh}^2|(n_{h'})_{h'\in \mathcal{H}_n})=\V(X_k|X\in h)$ and $\E(n_h/n)=p_h$ ensuring that $\E\left(\sum_{h\in \mathcal{H}_n}\frac{n_h}{n}S^2_{kh}\right)=\E\left(\V(X_k|(\mathds{1}\{X\in h\})_{h\in \mathcal{H}_n})\right)\leq \V(X_k)$. Next for $B_1^2=\frac{16}{n^2}\sum_{k=1}^p \sum_{h\in \mathcal{H}_n}m_h\left(1-\frac{m_h}{n_h}\right)S_{kh}^2\mathds{1}\{n_h\geq 2\}$, we have $\E\left(B_1^2\right)\leq \frac{4}{n}\sum_{k=1}^p\E\left(\V\left(X_{k1}|\mathds{1}\{X_{1}\in h\}\right)_{h\in \mathcal{H}_n}\right)=O\left(\frac{p}{n}\right)$. It follows that $o_p(B_1^2)=o_p\left(\frac{p}{n}\right)$ and this prove \eqref{bal:Strata1} and the upper bound of $\E\left(B_1^2\right)$. It remains to show the lower bound for $\E\left(B_1^2\right)$. \begin{align*}\E\left(B_1^2\right)&\geq \frac{2p}{n^2}\frac{\underline{C}}{12\ell^2\overline{C}}\E\left(\sum_{h\in \mathcal{H}_n}n_h\mathds{1}\{n_h\geq 2\}\right)\\
	&\geq \frac{p}{n^2}\frac{\underline{C}}{6\ell^2\overline{C}}\E\left(\sum_{h\in \mathcal{H}_n}n_h\left(1-\mathds{1}\{n_h=1 \}\right)\right)\\
	&\geq \frac{p}{n}\frac{\underline{C}}{6\ell^2\overline{C}}\left(1-\frac{\ell^p}{n}\sup_{h\in \mathcal{H}_n}\Pr(n_h=1)\right)\\
	&\geq \frac{p}{n}\frac{\underline{C}}{6\ell^2\overline{C}}\left(1-\frac{\overline{C}}{\underline{C}}\exp\left(-(n-1)\ell^{-p}\frac{\underline{C}}{\overline{C}}\right)\right).
\end{align*}
If $n\ell^{-p}\rightarrow 0$, for any $h\in \mathcal{H}_n$ we have $\Pr\left(n_h\geq 2\right)\leq 1-(1-p_h)^{n}-np_h(1-p_h)^{n-1}\leq 1-(1-np_h)-np_h(1-(n-1)p_h)\leq n^2p_h^2\leq \frac{\overline{C}^2}{\underline{C}^2}(n\ell^{-p})^2$. This ensures that $\sup_{h\in \mathcal{H}_n}\Pr(n_h\geq 2)=o(1)$. Next, we have:
\begin{align*}
	\E\left[\frac{1}{n}\sum_{h\in \mathcal{H}_n}n_h\mathds{1}\{n_h\geq 2\}\right]&\leq n^{-1}\ell^p\sup_{h\in \mathcal{H}_n}\E\left(n_h\mathds{1}\{n_h\geq 2\}\right)\\
	&=n^{-1}\ell^p\sup_{h\in \mathcal{H}_n}\left(\E(n_h)-\Pr(n_h=1)\right)\\
	&=n^{-1}\ell^p\sup_{h\in \mathcal{H}_n} \left(np_h\left(1-(1-p_h)^{n-1}\right)\right)\\
	&\leq \ell^p (n-1)\sup_{h\in \mathcal{H}_n}p_h^2\\
	&\leq \frac{\overline{C}^2}{\underline{C}^2}n\ell^{-p}=o(1).
	\end{align*}
We have $S_{kh}^2\mathds{1}\{n_h\geq 2\}\leq 2\mathds{1}\{n_h\geq 2\}$, $m=\frac{n}{2}\left(1+o_p(1)\right)$ and $m_h\left(1-\frac{m_h}{n_h}\right)\mathds{1}\{n_h\geq 2\}\leq \frac{n_h}{4}\mathds{1}\{n_h\geq 2\}$ and next
\begin{align*}
	\sup_{k=1,...,p}\sum_{h\in \mathcal{H}_n}\left(\frac{m_h}{m}\right)^2\left(1-\frac{m_h}{n_h}\right)\frac{1}{m_h}S^2_{kh}\mathds{1}\{n_h\geq 2\}&\leq\frac{4}{n^2}\left(1+o_p(1)\right)\sum_{h\in \mathcal{H}_n}\frac{n_h}{2}\mathds{1}\{n_h\geq 2\}.
	\end{align*}
Then, $4\sum_{k=1}^p\sum_{h\in \mathcal{H}_n}\frac{m_h^2}{m^2}\left(1-\frac{m_h}{n_h}\right)\frac{1}{m_h}S^2_{kh}\mathds{1}\{n_h\geq 2\}=o_p\left(\frac{p}{n}\right)$. On the other hand,
\begin{align*}
\left|\frac{1}{n}
\sum_{h\in \mathcal{H}_n} \left(\sum_{i\in \{j:X_j\in h\}}X_{ki}^2\right)\mathds{1}\{n_h=1\}-\frac{1}{n}\sum_{i=1}^nX_{ki}^2\right|&=\frac{1}{n}
\sum_{h\in \mathcal{H}_n} \left(\sum_{i\in \{j:X_j\in h\}}X_{ki}^2\right)\mathds{1}\{n_h\geq2\}\\
&\leq \frac{1}{n}
\sum_{h\in \mathcal{H}_n} n_h\mathds{1}\{n_h\geq2\}.
\end{align*}
It follows that $4\sum_{k=1}^p\E\left(\frac{1}{n^2}
\sum_{h\in \mathcal{H}_n} \left(\sum_{i\in \{j:X_j\in h\}}X_{ki}^2\right)\mathds{1}\{n_h=1\}\right)=\frac{4}{n}\sum_{p=1}^k\E(X^2_{k1})+o\left(\frac{p}{n}\right)$. This proves \eqref{bal:Strata2}.\\
~\\
Under assignment design (MP), partition $\mathcal{S}_n$ is adaptive to the sample and $\mathcal{S_n}$ is $(X_{i})_{i=1,...,n}$-measurable. All element of $\mathcal{S}_n$ are of size 2 and in each strata $s$ one unit over two is randomly assigned to the treatment. Next, the formula of the variance of the Horwitz-Thompson estimator of an average for a stratified sampling ensures
\begin{align*}
	\E\left(||B_{n,p}(X)||^2|(X_{j})_{j=1,...,n}\right)&=4\sum_{k=1}^p \sum_{s\in \mathcal{S}_n}\left(\frac{2}{n}\right)^2\left(1-\frac{1}{2}\right)S^2_{ks}\\
	&=\frac{4}{n}\sum_{k=1}^p\frac{2}{n}\sum_{s\in\mathcal{S}_n}S^2_{ks}
\end{align*}
with $S^2_{ks}=\left(X_{ki}-\frac{X_{ki}+X_{ki'}}{2}\right)^2+\left(X_{ki'}-\frac{X_{ki}+X_{ki'}}{2}\right)^2=\frac{1}{2}\left(X_{ki}-X_{ki'}\right)^2$ for $i$ and $i'$ such that $s=\{i,i'\}$. According to Lemma 2 in \cite{bai_optimality_2022}, complete randomization could be implemented as a two-stage random process. In a first step, a random partition $\mathcal{S}^{\ast}_n$ is selected with uniform probability among all the partitions such that $|s|=2$ for any $s\in \mathcal{S}^{\ast}_n$. In a second stage, stratified sampling is used in each strata $s\in \mathcal{S}^{\ast}_n$. According to \eqref{bal:CR1}, we have $\frac{4}{n}\sum_{k=1}^p\V(X_{k1})=\E\left(\frac{8}{n^2}\sum_{s\in\mathcal{S}_n^{\ast}}\sum_{k=1}^pS^2_{ks}\right)$. But program \eqref{opti:MP} is equivalent to $\sum_{s\in\mathcal{S}_n}\sum_{k=1}^pS^2_{ks}\leq \sum_{s\in\mathcal{S}_n^{\ast}}\sum_{k=1}^pS^2_{ks}$. Next, the right hand side of \eqref{bal:MP1} follows. 
To prove the left hand side of \eqref{bal:MP1}, note that $\E\left(||B_{n,p}(X)||^2|(X_{j})_{j=1,...,n}\right)=\frac{4}{n^2} \sum_{s\in \mathcal{S}_n}\text{diam}^2(s,\mathbb{R}^p)$ for $\text{diam}^2(s,\mathbb{R}^p)=\sup_{(i,j)\in s^2}||X_{i}-X_{j}||^2$.  Let $d_i$ the distance in $\mathbb{R}^p$ of the unit $i$ to its nearest neighbor $d_i=\min_{j\in\{1,...,n\}\backslash\{i\}}||X_j-X_i||$. We have $\text{diam}^2(\{i,i'\},\mathbb{R}^p)\geq \max (d_i^2,d_{i'}^2)\geq \frac{1}{2}(d_i^2+d_{i'}^2)$ and next $\E\left(||B_{n,p}(X)||^2|(X_{j})_{j=1,...,n}\right)\geq \frac{2}{n^2}\sum_{i=1}^n d_i^2=\frac{p}{3n}+\frac{2}{n^2}\sum_{i=1}^n\min_{j\in\{1,...,n\}\backslash\{i\}}\sum_{k=1}^pZ_{kij}$ for $Z_{kij}=\left(X_{ki}-X_{kj}\right)^2-\frac{1}{6}$. 
 Let $\widetilde{Z}_{kij}=\left(U_{ki}-U_{kj}\right)^2-\frac{1}{6}$ for $(U_{ki})_{k=1,...,p,i=1,...,n}$ some independent uniform variables on $[0;1]$. For any $k$ and any $i\neq j$, the $\widetilde{Z}_{kij}$ are zero mean variables with bounded support $[-1/6; 5/6]$. A zero mean variable $Z$ is sub-Gaussian with parameter $\nu>0$ if for any $\lambda\in\mathbb{R}$, $\E\left(e^{\lambda Z}\right)\leq e^{\lambda^2\nu^2/2}$. Hoeffding's lemma ensures that a zero mean variable with bounded support included in $[a;b]$ is sub-Gaussian with parameter $(b-a)/2$. It follows that $\widetilde{Z}_{kij}$ and $-\widetilde{Z}_{kij}$ are sub-Gaussian with parameter 1/2. Moreover, because $f_X(x_{1i},...,x_{pi})f_X(x_{1j},...,x_{pj})\leq \overline{C}^2$ and because $(\widetilde{Z}_{kij})_{k=1,...,p}$ are independent across $k$, we have for any $i\neq j$: \begin{align*}\E\left(e^{-\lambda \sum_{k=1}^pZ_{kij}}\right)&\leq \overline{C}^2\E\left(e^{-\lambda \sum_{k=1}^p\widetilde{Z}_{kij}}\right)\\
	&=\overline{C}^2\prod_{k=1}^p\E\left(e^{-\lambda \widetilde{Z}_{kij}}\right)\\
	&\leq \overline{C}^2e^{\lambda^2p/8}.\end{align*} For any $i\in \{1,...,n\}$, by Jensen inequality, we have
\begin{align*}
	e^{\lambda \E\left(\max_{j\neq i}-\sum_{k=1}^pZ_{kij}\right)}&\leq\E\left( e^{\lambda \max_{j\neq i}-\sum_{k=1}^pZ_{kij}}\right)\\
	&\leq\sum_{j\neq i}\E\left( e^{-\lambda \sum_{k=1}^pZ_{kij}}\right)\\
	&\leq (n-1)\overline{C}^2 e^{\lambda^2p/8}.
	\end{align*}
Next, $\E\left(\max_{j\neq i}(-\sum_{k=1}^pZ_{kij})\right)\leq \frac{\ln(\overline{C}^2(n-1))}{\lambda}+\lambda\frac{p}{8}$ for any $\lambda>0$. Choosing $\lambda=\sqrt{\frac{8\ln(\overline{C}^2(n-1))}{p}}$, we have $\E\left(\max_{j\neq i}(-\sum_{k=1}^pZ_{kij})\right)\leq\sqrt{p\ln(\overline{C}^2(n-1))/2}$ or equivalently $\E\left(\min_{j\neq i}\sum_{k=1}^pZ_{kij}\right)\geq-\sqrt{p\ln(\overline{C}^2(n-1))/2}$.
From what precedes, we have
\begin{align*}
	\E\left(||B_{n,p}(X)||^2\right)\geq& \frac{p}{3n}+\frac{2}{n^2}\sum_{i=1}^n\E\left(\min_{j\in\{1,...,n\}\backslash \{i\}}\sum_{k=1}^pZ_{kij}\right)\\
	\geq &\frac{p}{3n}-\frac{2}{n}\sqrt{p\ln\left(\overline{C}^2(n-1)\right)/2}\\
 =&\frac{p}{n}\left(\frac{1}{3}-\sqrt{\frac{2\ln(n-1)+4\ln{\overline{C}}}{p}}\right).
\end{align*}
This achieves the proof of \eqref{bal:MP1}.\\
~\\
Concerning assignment (CM), let $Z_i=(1,X_i')'$. Then, we have $||B_{n,p}(X)||^2\leq ||B_{n,p}(Z)||^2$. At the end of the flying phase of the cube method, all units of a $A\subset\{1,...,n\}$ such that $|A|=n-\dim(Z)=n-p-1$ have been allocated. This means that $D_i$ has been drawn for any $i\in A$. For $i\notin A$, some random variables $\pi_i^{\star}$ have been generated such that $\E(\pi_i^{\star}|(Z_i)_{i=1,...,n})=1/2$ and
\begin{align*}
	\sum_{i\in A}Z_i(2D_i-1)+\sum_{i\notin A}Z_i(2\pi_i^{\ast}-1)&=0,
	\end{align*}
or equivalently:
\begin{align*}B_{n,p}(Z)&=\frac{4}{n}\sum_{i\notin A} Z_i(D_i-\pi_i^{\ast}).
	\end{align*}
In the landing phase of the cube algorithm, given $W=(A, (D_i)_{i\in A}, (\pi_i^{\ast})_{i\notin A}, (Z_{i})_{i=1,...,n})$, the cube method samples $(D_i)_{i\notin A}$ such that for any $i\notin A$, $D_j|W$ follows a Bernoulli of mean $\pi^{\ast}_j$. But the sampling of $(D_i)_{i\notin A}$ is not independent across $i\notin A$. Indeed, sampling probabilities are correlated in view to minimize
\begin{align*}
	\E\left(\sum_{i\notin A}Z'_i(D_i-\pi_i^{\ast})M\sum_{i\notin A}Z_i(D_i-\pi_i^{\ast})\big|W\right)
	\end{align*}  
where $M$ is a symmetric-positive matrix $dim(Z)\times dim(Z)$. When $M=Id$, this means that sampling probabilities of $(D_i)_{i\notin A}$ during the landing phase are coordinated to minimize $\E\left(||\sum_{i\notin A} Z_i(D_i-\pi_i^{\ast})||^2|W\right)$. Next, this means that
\begin{align*}\E\left(||\sum_{i\notin A} Z_i(D_i-\pi_i^{\ast})||^2|W\right)&\leq \E\left(||\sum_{i\notin A} Z_i(\tilde{D}_i-\pi_i^{\ast})||^2|W\right)\end{align*}
where $(\tilde{D}_i)_{i\notin A}|W$ are sampled as some independent Bernoulli of mean $\pi^{\ast}_i$.
If $M\neq Id$, we have
\begin{align*}\E\left(||\sum_{i\notin A} Z_i(D_i-\pi_i^{\ast})||^2|W\right)&\leq \frac{1}{\lambda_{min}(M)} \E\left(||M^{1/2}\sum_{i\notin A} Z_i(D_i-\pi_i^{\ast})||^2|W\right)\\
	&\leq \frac{1}{\lambda_{min}(M)} \E\left(||M^{1/2}\sum_{i\notin A} Z_i(\tilde{D}_i-\pi_i^{\ast})||^2|W\right)\\
	&\leq \frac{\lambda_{max}(M)}{\lambda_{min}(M)} \E\left(||\sum_{i\notin A} Z_i(\tilde{D}_i-\pi_i^{\ast})||^2|W\right).\end{align*}
To conclude, note that
\begin{align*}
	\E\left(||\sum_{i\notin A} Z_i(\tilde{D}_i-\pi_i^{\ast})||^2\big|W\right)&=\sum_{k=1}^{p+1}\E\left[\left(\sum_{i\notin A}Z_{ki}(\tilde{D}_i-\pi^{\ast}_i)\right)^2\big|W\right]\\
	&=\sum_{k=1}^{p+1}\E\left[\sum_{i\notin A}Z_{ki}^2\pi^{\ast}_i(1-\pi^{\ast}_i)\big|W\right]\\
	&\leq \frac{(p+1)|A|}{4}=\frac{(p+1)^2}{4}.
	\end{align*}
\newpage
\section{Proofs of Propositions}
\subsection{Proof of Balancing Approximations for the Cube Method (Proposition \ref{prop:balance})}
\label{proof:balance}
From Assumption \ref{hyp:sample} and Proposition 4 in \citet{deville_efficient_2004}, we have: $$\left|\frac{1}{n}\sum_{i=1}^n\frac{X_{ji} D_i}{\pi_i}-\frac{1}{n}\sum_{i=1}^n X_{ji}\right|\leq \frac{q}{n} \max_{i=1,...,n}\left|\frac{X_{ji}}{\pi_i}\right|\leq \frac{q \max_{i=1,...,n}\left|X_{ji}\right|}{cn}.$$ 
If Assumption \ref{hyp:iid+mom} holds and if moments of order $r$ exist for $X_{j1}$, from Proposition 1.5 and Theorem 2.1 in Chapter 6 in \cite{gut_probability_2013}, we have $\max_{i=1,...,n}\left|X_{ji}\right|=o_p(n^{1/r})$. If $X_{ji}$ sub-Gaussian  $\max_{i=1,...,n}\left|X_{ji}\right|=O_p\left(\sqrt{\ln(n)}\right)$ and if $X_{ji}$ is bounded by $K$, $\max_{i=1,...,n}\left|X_{ji}\right|\leq K$.

\subsection{Proof of Asymptotic Normality for the SATE (First part of Proposition \ref{prop:var})}
\label{proof:SATE}
We want to prove $$\sqrt{n}\left(\widehat{\theta}_{HT}-\theta_0\right)\convNor{V_0}$$ and  $$\sqrt{n}\left(\widehat{\theta}_{H}-\theta_0\right)\convNor{V_0}$$ 
where $V_0=\E\left[\pi_i\left(1-\pi_i\right)\left(\frac{\eps_i(1)}{\pi_i}+\frac{\eps_i(0)}{1-\pi_i}\right)^2\right].$\\\\
We first show that it is sufficient to prove asymptotic normality for one of the two estimators. Reasoning similar to the proof of Proposition \ref{prop:balance} ensures that if the researcher includes a constant in the set of covariates to balance, one has $\frac{1}{n}\sum_{i=1}^n\frac{D_i}{\pi_i}=1+o_p\left(\frac{1}{\sqrt{n}}\right)$ and $\frac{1}{n}\sum_{i=1}^n\frac{1-D_i}{1-\pi_i}=1+o_p\left(\frac{1}{\sqrt{n}}\right)$. Then, the Hájek estimator in \eqref{eq:Hajec} is given by
\begin{align*}
	\widehat{\theta}_H&=\frac{1}{n}\left(\sum_{i=1}^n\frac{Y_i D_i}{\pi_i}-\frac{Y_i (1-D_i)}{1-\pi_i}\right)\left(1+o_p\left(n^{-1/2}\right)\right)\\
    &=\widehat{\theta}_{HT}\left(1+o_p\left(n^{-1/2}\right)\right).
\end{align*}
Then,
\begin{align*}
	\sqrt{n}\left(\widehat{\theta}_{H}-\theta_0\right)&=\sqrt{n}\left(\widehat{\theta}_{HT}-\theta_0\right)+o_p\left(1\right)\left(\widehat{\theta}_{HT}-\theta_0\right).
\end{align*}
By Slutsky's theorem, if $\sqrt{n}\left(\widehat{\theta}_{HT}-\theta_0\right)\convNor{V_0}$, then $\sqrt{n}\left(\widehat{\theta}_{H}-\theta_0\right)\convNor{V_0}$. It is thus sufficient to prove asymptotic normality of the Horvitz-Thompson estimator, i.e.,
$$\sqrt{n}\left(\widehat{\theta}_{HT}-\theta_0\right)\convNor{V_0}.$$\\
Under Assumptions \ref{hyp:iid+mom} and \ref{hyp:sample}, Proposition \ref{prop:balance} ensures
\begin{align*}
	\sqrt{n}\left(\widehat{\theta}_{HT}-\theta_0\right)=&\frac{1}{\sqrt{n}}\sum_{i=1}^n(D_i-\pi_i)\left(\frac{\eps_i(1)}{\pi_i}+\frac{\eps_i(0)}{1-\pi_i}\right) +\sqrt{n}\left(\frac{1}{n}\sum_{i=1}^n\frac{D_iZ_{1i}'}{\pi_i}-\sum_{i=1}^nZ_{1i}'\right)\beta_1\\&-\sqrt{n}\left(\frac{1}{n}\sum_{i=1}^n\frac{(1-D_i)Z_{0i}'}{1-\pi_i}-\sum_{i=1}^nZ_{0i}'\right)\beta_0\\
	=&\frac{1}{\sqrt{n}}\sum_{i=1}^n-\left(\frac{\eps_i(1)}{\pi_i}+\frac{\eps_i(0)}{1-\pi_i}\right)\pi_i+\left(\frac{\eps_i(1)}{\pi_i}+\frac{\eps_i(0)}{1-\pi_i}\right)D_i+o_p(1)
\end{align*}
Then, we have
\begin{equation*}
    \sqrt{n}\left(\widehat{\theta}_{HT}-\theta_0\right)=\frac{1}{\sqrt{n}}\sum_{i=1}^nf_i+g_iD_i+o_p(1)
\end{equation*}
with $f_i\coloneqq f\left(X_i,\eps_i(0),\eps_i(1)\right)= -\left(\frac{\eps_i(1)}{\pi_i}+\frac{\eps_i(0)}{1-\pi_i}\right)\pi_i$ and $g\left(X_i,\eps_i(0),\eps_i(1)\right)= \frac{\eps_i(1)}{\pi_i}+\frac{\eps_i(0)}{1-\pi_i}$.
Slutsky's theorem ensures that we have to prove
\begin{align*}\frac{1}{\sqrt{n}}\sum_{i=1}^nf_i+g_iD_i\convNor{V_0}.
\end{align*}
By Assumption \ref{hyp:linear} $\E[f|X]= \E[g|X]=0$. Then, Conjecture \ref{conj:poiss} and Lemma \ref{lem:an} give that, conditional on $(X_i)_{i\geq1}$, $\frac{1}{\sqrt{n}}\sum_{i=1}^nf_i+g_iD_i\convNor{V_0}$,
with $V_0=\E[f_i^2+(2g_if_i+g_i^2)\pi_i]=\E\left[\pi_i(1-\pi_i)\left(\frac{\eps_i(1)}{\pi_i}+\frac{\eps_i(0)}{1-\pi_i}\right)\right]$, in the sense of Definition \ref{def:condconv}. Notice that $V_0$ does not depend on  $(X_1)_{i\geq1}$, so convergence in distribution is unconditional. This concludes the proof.
\\\\
\underline{Comparison with Poisson randomization}
\\
If treated units are selected through Poisson randomization and Assumptions \ref{hyp:iid+mom} and \ref{hyp:sample} hold, then $(Y_i(1),Y_i(0),X_i,D_i)_{i=1,\ldots,n}$ are i.i.d., and by the CLT,
$$\sqrt{n}(\widehat{\theta}_{HT}-\theta_0)\convNor{W_0}$$
with \begin{align*}
    W_0&=\E\left[\left(D_i-\pi_i\right)^2\left(\frac{Y_i(1)}{\pi_i}+\frac{Y_i(0)}{1-\pi_i}\right)^2\right]\\
    &=\E\left[\pi_i(1-\pi_i)\left(\frac{Y_i(1)}{\pi_i}+\frac{Y_i(0)}{1-\pi_i}\right)^2\right]\\
    &=\E\left[\pi_i(1-\pi_i)\left(\frac{Z_{1i}'\beta_1}{\pi_i}+\frac{Z_{0i}'\beta_0}{1-\pi_i}\right)^2\right]+\E\left[\pi_i(1-\pi_i)\left(\frac{\eps_i(1)}{\pi_i}+\frac{\eps_i(0)}{1-\pi_i}\right)^2\right]\\
    &=\E\left[\pi_i(1-\pi_i)\left(Z_{0i}'\frac{\beta_1+\beta_0}{1-\pi_i}\right)^2\right]+V_0 =\Sigma_0+V_0.
\end{align*}

\subsection{Proof of Asymptotic Normality for the PATE (Second part of Propositions \ref{prop:var})}
\label{proof:PATE}
As shown in Proof \ref{proof:SATE}, if $\sqrt{n}\left(\widehat{\theta}_{HT}-\theta^\ast_0\right)\convNor{V_0^\ast}$, we have $\sqrt{n}\left(\widehat{\theta}_{H}-\theta^\ast_0\right)\convNor{V_0^\ast}$, we thus restrict ourselves to proving asymptotic normality for the Horvitz-Thompson estimator, i.e,
$$\sqrt{n}\left(\widehat{\theta}_{HT}-\theta^\ast_0\right)\convNor{V_0^\ast}$$
where $V_0^\ast=\V(Z_{1i}'\beta_1-Z_{0i}'\beta_0)+\E\left[\frac{\varepsilon_i(1)^2}{\pi_i}\right] +\E\left[\frac{\varepsilon_i(0)^2}{1-\pi_i}\right].$\\\\
Let us consider $f_i\coloneqq f(X_i,\varepsilon_i(1),\varepsilon_i(0))= -\frac{\varepsilon_i(0)}{1-\pi_i}$, $g_i\coloneqq g(X_i,\varepsilon_i(1),\varepsilon_i(0))=\frac{\varepsilon_i(1)}{\pi_i}+\frac{\varepsilon_i(0)}{1-\pi_i}$, and $h_i\coloneqq h(X_i)=(Z_{1i}-\E[Z_{1i}])'\beta_1-(Z_{0i}-\E[Z_{0i}])'\beta_0$.\\
Under Assumptions \ref{hyp:iid+mom} and \ref{hyp:sample}, Proposition \ref{prop:balance} ensures
\begin{align*}
	\sqrt{n}\left(\widehat{\theta}_{HT}-\theta^\ast_0\right)=&\frac{1}{\sqrt{n}}\sum_{i=1}^n\left(\frac{Z_{1i}'D_i}{\pi_i}-E[Z_{1i}]'\right)\beta_1\\
	&-\frac{1}{\sqrt{n}}\sum_{i=1}^n\left(\frac{Z_{0i}'(1-D_i)}{1-\pi_i}-E[Z_{0i}]'\right)\beta_0\\
	&+\frac{1}{\sqrt{n}}\sum_{i=1}^n\frac{\varepsilon_i(1)D_i}{\pi_i}-\frac{\varepsilon_i(0)(1-D_i)}{1-\pi_i}\\
	=& \frac{1}{\sqrt{n}}\sum_{i=1}^n f_i + g_iD_i + \frac{1}{\sqrt{n}}\sum_{i=1}^nh_i+ o_p(1).
\end{align*}
Slutsky's theorem ensures that we have to prove
\begin{equation}
	\label{eq:CLTsuper}
	\frac{1}{\sqrt{n}}\sum_{i=1}^n f_i + g_iD_i+\frac{1}{\sqrt{n}}\sum_{i=1}^n h_i\convNor{V_0^\ast}.
\end{equation}
By Assumption \ref{hyp:linear}, $\E[f|X]= \E[g|X]=0$. Then, Conjecture \ref{conj:poiss} and Lemma \ref{lem:an} give that, conditional on $(X_i)_{i\geq1}$, $\sqrt{n}\left(\widehat{\theta}_{HT}-\theta_0^\ast\right) \convNor{V_{01}^\ast}$ with $V_{01}^\ast=\E[f_i^2+(2g_if_i+g_i^2)\pi_i]=\E\left[\frac{\eps_i(1)^2}{\pi_i}\right]+\E\left[\frac{\eps_i(0)^2}{1-\pi_i}\right].$ Moreover, by the central limit theorem, $\frac{1}{\sqrt{n}}\sum_{i=1}^n h_i\convNor{V_{02}^\ast}$ with $V_{02}^\ast=\V(Z_{1i}'\beta_1-Z_{0i}'\beta_0)$. Theorem 2 in \citet{chen_asymptotic_2007} ensures that $\frac{1}{\sqrt{n}}\sum_{i=1}^n f_i + g_iD_i+\frac{1}{\sqrt{n}}\sum_{i=1}^n h_i\convNor{V_{01}^\ast+V_{02}^\ast}.$ This concludes the proof.
\\\\
\underline{Comparison with Poisson randomization}
\\
If treated units are selected through Poisson randomization and Assumptions \ref{hyp:iid+mom} and \ref{hyp:sample} hold, then $(Y_i(1),Y_i(0),X_i,D_i)_{i=1,\ldots,n}$ are i.i.d., and by the CLT,
$$\sqrt{n}(\widehat{\theta}_{HT}-\theta_0^\ast)\convNor{W_0^\ast}$$
with \begin{align*}
    W_0^\ast&=\E\left[\left(\frac{Y_i(1)D_i}{\pi_i}-\frac{Y_i(0)(1-D_i)}{1-\pi_i}-\E\left[Y_i(1)-Y_i(0)\right]\right)^2\right]\\
    &=\E\left[\frac{Y_i(1)^2}{\pi_i}\right]+\E\left[\frac{Y_i(0)^2}{(1-\pi_i)}\right]-\E\left[Y_i(1)-Y_i(0)\right]^2\\
    &=\E\left[\frac{(Z_{1i}'\beta_1)^2}{\pi_i}\right]+\E\left[\frac{\eps_i(1)^2}{\pi_i}\right]+\E\left[\frac{(Z_{0i}'\beta_0)^2}{(1-\pi_i)}\right]+\E\left[\frac{\eps_i(0)^2}{1-\pi_i}\right]-\E\left[Z_{1i}'\beta_1-Z_{0i}'\beta_0\right]^2\\
    &=\E\left[\frac{(Z_{1i}'\beta_1)^2}{\pi_i}\right]+\E\left[\frac{(Z_{0i}'\beta_0)^2}{(1-\pi_i)}\right]-\E\left[(Z_{1i}'\beta_1)^2\right]-\E\left[(Z_{0i}'\beta_0)^2\right]+2\E\left[(Z_{1i}'\beta_1)(Z_{0i}'\beta_0)\right]+V_0^\ast\\
    &=\E\left[\frac{(Z_{1i}'\beta_1)^2(1-\pi_i)^2+(Z_{0i}'\beta_0)^2\pi_i^2+2(Z_{1i}'\beta_1)(Z_{0i}'\beta_0)\pi_i(1-\pi_i)}{\pi_i(1-\pi_i)}\right]+V_0^\ast\\
    &=\E\left[\frac{\pi_i}{1-\pi_i}(Z_{0i}'(\beta_1+\beta_0))^2\right]+V_0^\ast=\Sigma_0+V_0^\ast.\\
\end{align*}

\subsection{Proof of Randomized-based Inference (Proposition \ref{prop:randinf})}
\label{proof:rbinf}

For completeness, we show first, as in \citet{bai_inference_2022}, that the strong null hypothesis \eqref{eq:strongnull} $(Y_i(1),X_i)\stackrel{d}{=}(Y_i(0),X_i)$ is equivalent to stating $Y_1,\ldots,Y_n\indep D_1,\ldots,D_n|X_1,\ldots,X_n$.
\\Let us consider random allocations  generated by the cube method $d$ and $d'$ in the support of $D_1,\ldots,D_n|X_1,\ldots,X_n$ and any set $A$. Then we have,
\begin{align*}
	&\Pr\left[(Y_1,\ldots,Y_n)\in A | (D_1,\ldots,D_n)=(d_1,\ldots,d_n), X_1,\ldots,X_n)\right]\\
	=&\Pr\left[(Y_1(d_1),\ldots,Y_n(d_n))\in A | (D_1,\ldots,D_n)=(d_1,\ldots,d_n), X_1,\ldots,X_n)\right]\\
	=&\Pr\left[(Y_1(d_1),\ldots,Y_n(d_n))\in A | (D_1,\ldots,D_n)=(d_1,\ldots,d_n), X_1,\ldots,X_n)\right]\\
	=&\Pr\left[(Y_1(d_1),\ldots,Y_n(d_n))\in A |  X_1,\ldots,X_n)\right]\\
	=&\Pr\left[(Y_1(d'_1),\ldots,Y_n(d'_n))\in A |  X_1,\ldots,X_n)\right]\\
	=&\Pr\left[(Y_1(d'_1),\ldots,Y_n(d'_n))\in A | (D_1,\ldots,D_n)=(d'_1,\ldots,d'_n), X_1,\ldots,X_n)\right]\\
	=&\Pr\left[(Y_1,\ldots,Y_n)\in A | (D_1,\ldots,D_n)=(d'_1,\ldots,d'_n), X_1,\ldots,X_n)\right],\\
\end{align*}so both hypothesis are equivalent.
Then, under Assumptions \ref{hyp:iid+mom} and \ref{hyp:sample}, and the strong null hypothesis \eqref{eq:strongnull}, $$(Y_i,D_i,X_i)_{i\geq1}\stackrel{d}{=}(Y_i,D_i^{(g)},X_i)_{i\geq1}.$$
We thus have
\begin{align}
	\E\left[\sum_{g\in G_n^{B}}\phi^{rand}_n\left(\mathbf{P_n^{(g)}}\right)\right] &= 	\sum_{g\in \nonumber G_n^{B}}\E\left[\E\left[\phi^{rand}_n\left(\mathbf{P_n^{(g)}}\right)\big|X_1,\ldots,X_n\right]\right] \nonumber\\
	&= \sum_{g\in G_n^{B}}\E\left[\E\left[\phi^{rand}_n\left(\mathbf{P_n}\right)\big|X_1,\ldots ,X_n\right]\right] \nonumber \\
	&=B\E\left[\phi^{rand}_n\left(\mathbf{P_n}\right)\right] \label{eq:RBinf1}
\end{align}
Moreover, $c_n\left(\mathbf{P_n},1-\alpha\right)=c_n\left(\mathbf{P_n^{(g)}},1-\alpha\right)$ for any $g\in G_n^B$ ensures by definition of $c_n\left(\mathbf{P_n},1-\alpha\right)$
\begin{align}
&\sum_{g\in G_n^{B}}\phi^{rand}_n\left(\mathbf{P_n^{(g)}}\right)\leq B\alpha	\label{eq:RBinf2}
\end{align}
Combining equations \eqref{eq:RBinf1} and \eqref{eq:RBinf2} we get 
$\E\left[\phi^{rand}_n\left(\mathbf{P_n}\right)\right]\leq \alpha$, which concludes the proof.
\newpage
\section{Lemmas for the Cube Method}

\begin{lem}[Exchangeability]~\label{lem:ex}\\
	For any permutation $\sigma$ of $\{1,...,n\}$ we have:
	$$(D_{\sigma(i)},\pi^{\ast}_{\sigma(i)}X_{\sigma(i)})_{i=1,...,n}\stackrel{d}{=}(D_{i},X_{i})_{i=1,...,n}$$
\end{lem}
\underline{Proof:}\\ For any value of $n$, the Cube algorithm ensures there exists a finite collection of independent uniform random variables $(U_1,...,U_K)$ independent of $(X_1,...,X_n)$ such that $(D_1,...,D_n,\pi^{\ast}_1,...,\pi^{\ast}_n)=f(X_1,...,X_n,U_1,...,U_K)$. Because the $X$ are iid and independent of the $U$, we have: $$(X_{\sigma(1)},...,X_{\sigma(n)},U_1,...,U_K)\stackrel{d}{=}(X_{1},...,X_{n},U_1,...,U_K).$$ The result follows.

\begin{defin}~\label{def:condconv}\\
$W_n\convNor{\sigma^2}$ conditional on $(X_i)_{i\geq1}$ if and only if for any $h$ bounded Lipschitz
$\E\left(h(W_n)|(X_{i})_{i\geq 1}\right)$ converges almost surely to $\int h(u) \frac{1}{\sqrt{2\pi \sigma^2}}\exp\left(-\frac{u^2}{2\sigma^2}\right)du$.    
\end{defin}
Usual criteria (e.g, Portmanteau's lemma or Lévy's continuity theorem) to prove convergence in distribution could be adapted to prove the convergence in distribution conditional on $(X_i)_{i\geq 1}$ apply if the usual expectations and probabilities are replaced by conditional expectations and probabilities and usual convergence of sequences is replaced by almost sure convergence of random variables. More concretely, we will use the fact that if for any $k\geq 1$, $\E((W_n)^k|(X_{i})_{i\geq 1})$ converges almost surely to the $k$th-raw moment of a Gaussian distribution of variance $\sigma^2$ then $W_n\convNor{\sigma^2}$ conditional on $(X_i)_{i\geq1}$. This is an adaptation of the theorem of \cite{takacs_moment_1991} that states that if for any $k\geq 1$, $\E((W_n)^k)$ converges to the $k$th-raw moment of a Gaussian distribution of variance $\sigma^2$ then $W_n\convNor{\sigma^2}$. Moreover, $W_n\convNor{\sigma^2}$ conditional on $(X_i)_{i\geq1}$ if and only if $\forall t\in \mathbb{R}$, $\Pr\left(W_n\leq t|(X_i)_{i\geq 1}\right)$ converges almost surely to $\Phi(\frac{t}{\sqrt{\sigma^2}})$ for $\Phi$ the c.d.f. of the standard Gaussian.

\begin{lem}[Asymptotic normality]~\label{lem:an}\\
Let $f$ and $g$ be two functions such that for $f_i=f(\delta_i(1),\delta_i(0),X_i)$ and $g_i=g(\delta_i(1),\delta_i(0),X_i)$ we have $\E(f_i^2+g_i^2)<\infty$ and $\E[f_i|X_i]=\E[g_i|X_i]=0$.\\
If Assumptions \ref{hyp:iid+mom} and \ref{hyp:sample} and Conjecture \ref{conj:poiss} hold. Then, conditional on $(X_i)_{i\geq1}$,
\begin{equation}
\label{eq:an}
\frac{1}{\sqrt{n}}\sum_{i=1}^nf_i+g_iD_i\convNor{V_0}
\end{equation}
with $V_0=\E\left[f_1^2+(2g_1f_1+g_1^2)\pi_1\right].$
\end{lem}
\underline{Proof:}\\
\textbf{First step: $|f_i| + |g_i|$ bounded implies \eqref{eq:an}}\\
Let us assume it exists $K>0$ such that $|f_1|+|g_1|<K$ for any $k \in \N$. This ensures that all the moments of $f_i+g_i D_i$ exist. Let $M_{n,k}=\E\left[\left(\frac{1}{\sqrt{n}}\sum_{i=1}^nf_i+g_iD_i\right)^k \left| (X_i)_{i\geq 1}\right.\right].$\\
We have $$M_{n,k}=\E\left[n^{-k/2}\sum_{1\leq i_1,\ldots,i_k\leq n} \prod_{\ell=1}^k\left(f_{i_\ell}+g_{i_\ell}D_{i_\ell}\right)\left| (X_i)_{i\geq 1}\right.\right].$$
Let us order the indices $i_1,\ldots,i_k$ as $j_1,\ldots,j_m$ for some $1\leq m \leq k$ with each $j_\ell$ occurring with multiplicity $a_\ell$. Let $A_{k,m} \coloneqq \left\{ a=(a_1,\ldots,a_m)\in\N^{\ast m} : \sum_{\ell=1}^ma_\ell=k\right\}$ and for $a\in A_{k,m}$, $c_{k,a}=\frac{k!}{\prod_{\ell=1}^m a_{\ell}!}$. We have:
\begin{align*}
    M_{n,k} &=\sum_{m=1}^kn^{-k/2}\sum_{1\leq j_1<\ldots<j_m\leq n} \sum_{a\in A_{k,m}}c_{k,a}\E\left[\prod_{\ell=1}^m\left(f_{j_\ell}+g_{j_\ell}D_{j_\ell}\right)^{a_\ell}\left| (X_i)_{i\geq 1}\right.\right].
\end{align*}
In order to prove the convergence of moments, we will focus on the summands
$$B_{n,k,m}=n^{-k/2}\sum_{1\leq j_1<\ldots<j_m\leq n} \sum_{a\in A_{k,m}}c_{k,a}\E\left[\prod_{\ell=1}^m\left(f_{j_\ell}+g_{j_\ell}D_{j_\ell}\right)^{a_\ell}\left| (X_i)_{i\geq 1}\right.\right].$$
Notice that $|B_{n,k,m}|\leq n^{-k/2}\binom{n}{m}\sum_{m=1}^k \sum_{a\in A_{k,m}}c_{k,a}K^k=O\left(n^{m-k/2}\right)$. For $m<k/2$, we thus have $\lim_nB_{n,k,m}=0$.\\
We focus now in the case $m> k/2$.
For $\mathcal{K}\subseteq \{1,\ldots,m\}$, we note $\mathcal{K}^c=\{1,\ldots,m\}\setminus\mathcal{K}$. Then, the binomial theorem and the expansion $\prod_{\ell=1}^m(x_{\ell}+y_{\ell})=\sum_{\mathcal{K}\subseteq\{1,\ldots,m\}}\prod_{\ell \in \mathcal{K}}x_{\ell}\prod_{\ell' \in \mathcal{K}^{c}}y_{\ell'}$ and identity $D^{a}=D$ for $a\geq 1$ ensure
\begin{align*}
   &B_{n,k,m}\\ &=n^{-k/2}\sum_{1\leq j_1<\ldots<j_{m}\leq n} \sum_{a\in A_{k,m}}c_{k,a}\E\left[\prod_{\ell=1}^k\left(f_{j_\ell}+g_{j_\ell}D_{j_\ell}\right)^{a_\ell}\Big| (X_i)_{i\geq 1}\right]\\
    &=n^{-k/2}\sum_{1\leq j_1<\ldots<j_m\leq n} \sum_{a\in A_{k,m}}c_{k,a}\E\left[\prod_{\ell=1}^m\left[ f_{j_\ell}^{a_\ell}+\left(\sum_{r=1}^{a_\ell}\binom{a_\ell}{r}f_{j_\ell}^{a_\ell-r}g_{j_\ell}^r\right)D_{j_\ell}\right]\Big| (X_i)_{i\geq 1}\right]\\
    &=n^{-k/2}\sum_{1\leq j_1<\ldots<j_m\leq n} \sum_{a\in A_{k,m}}c_{k,a}\sum_{\mathcal{K}\subseteq\{1,\ldots,m\}}\E\left[\prod_{\ell\in\mathcal{K}}f_{j_\ell}^{a_\ell}\prod_{\ell'\in\mathcal{K}^c}\left(\sum_{r=1}^{a_{\ell'}}\binom{a_{\ell'}}{r}f_{j_{\ell'}}^{a_{\ell'}-r}g_{j_{\ell'}}^j\right)\prod_{{\ell''}\in\mathcal{K}^c}D_{j_{\ell''}}\Big| (X_i)_{i\geq 1}\right]
\end{align*}
Then, independence of $(f_i,g_i)_{i\geq1}$ across $i$ and conditional independence $(f_i,g_i)\indep D_i |(X_{i'})_{i'\geq 1}$ ensure
\begin{align*}
   & B_{n,k,m}\\
    &=n^{-k/2}\sum_{1\leq j_1<\ldots<j_m\leq n} \sum_{a\in A_{k,m}}c_{k,a}\sum_{\mathcal{K}\subseteq\{1,\ldots,m\}}\prod_{\ell\in\mathcal{K}}\E\left[f_{j_\ell}^{a_\ell}|X_{j_\ell}\right]\prod_{{\ell'}\in\mathcal{K}^c}\left(\sum_{r=1}^{a_{\ell'}}\binom{a_{\ell'}}{r}\E\left[f_{j_{\ell'}}^{a_{\ell'}-r}g_{j_{\ell'}}^r|X_{j_{\ell'}}\right]\right)\\
    &~~~~~~~~~~~~~~~~~~~~~~~~~~~~~~~~~~~~~~~~~~~~~~~~~~~~~~\E\left[\prod_{{\ell''}\in\mathcal{K}^c}D_{j_{\ell''}}\Big| (X_i)_{i\geq 1}\right].
\end{align*}
Because $m>k/2$, for any $a\in A_{k,m}$ there exists $s$ such that $a_s=1$. For any $\mathcal{K}$, if $s\in\mathcal{K}$, then $\prod_{\ell\in\mathcal{K}}\E\left[f_{j_\ell}^{a_\ell}|X_{j_\ell}\right]=\E\left[f_{j_s}|X_{j_s}\right]\prod_{\ell\in\mathcal{K}\backslash\{s\}}\E\left[f_{j_\ell}^{a_\ell}|X_{j_\ell}\right]=0$, else $s\in \mathcal{K}^c$ and $\prod_{\ell'\in\mathcal{K}^c}\left(\sum_{r=1}^{a_{\ell'}}\binom{a_{\ell'}}{r}f_{j_{\ell'}}^{a_{\ell'}-r}g_{j_{\ell'}}^j\right)=\E(g_{j_s}|X_{j_s})\prod_{\ell'\in\mathcal{K}^c\backslash\{s\}}\left(\sum_{r=1}^{a_{\ell'}}\binom{a_{\ell'}}{r}f_{j_{\ell'}}^{a_{\ell'}-r}g_{j_{\ell'}}^j\right)=0$.
It follows that if $m>k/2$ we have $B_{n,k,m}=0$.\\ 
Let now consider the last case $m=k/2$.
For $a\in A_{k,k/2}$ either there exists $s$ such that $a_s=1$ and by the previous reasoning, we have $\prod_{\ell\in\mathcal{K}}\E\left[f_{j_\ell}^{a_\ell}|X_{j_\ell}\right]\prod_{{\ell'}\in\mathcal{K}^c}\left(\sum_{r=1}^{a_{\ell'}}\binom{a_{\ell'}}{r}\E\left[f_{j_{\ell'}}^{a_{\ell'}-r}g_{j_{\ell'}}^r|X_{j_{\ell'}}\right]\right)=0$ for any $\mathcal{K}$, either $a=(2,...,2)$ and it follows
\begin{align*}
    B_{n,k,k/2}&=n^{-k/2}\sum_{1\leq j_1<\ldots<j_{k/2}\leq n}\frac{k!}{2^{k/2}}\sum_{\mathcal{K}\subseteq\{1,\ldots,{k/2}\}}\prod_{\ell\in\mathcal{K}}\E\left[f_{j_\ell}^{2}|X_{j_\ell}\right]\prod_{{\ell'}\in\mathcal{K}^c}\E\left[2f_{j_{\ell'}}g_{j_{\ell'}}+g_{j_{\ell'}}^2|X_{j_{\ell'}}\right]\\
    &~~~~~~~~~~~~~~~~~~~~~~~~~~~~~~~~~~~~~~~~~~~~~~~~~~~~~~\E\left[\prod_{{\ell''}\in\mathcal{K}^c}D_{j_{\ell''}}\Big| (X_i)_{i\geq 1}\right]
\end{align*}
Conjecture \ref{conj:poiss} and the fact that $\max(|f_i|^2,|2f_ig_i+g_i^2|)\leq 3K^2$ ensure
\begin{align*}
    B_{n,k,k/2}
    =&n^{-k/2}\sum_{1\leq j_1<\ldots<j_{k/2}\leq n}\frac{k!}{2^{k/2}}\sum_{\mathcal{K}\subseteq\{1,\ldots,{k/2}\}}\prod_{\ell\in\mathcal{K}}\E\left[f_{j_\ell}^{2}|X_{j_\ell}\right]\prod_{{\ell'}\in\mathcal{K}^c}\E\left[2f_{j_{\ell'}}g_{j_{\ell'}}+g_{j_{\ell'}}^2|X_{j_{\ell'}}\right]\prod_{{\ell''}\in\mathcal{K}^c}\pi_{j_{\ell''}}\\
    &+n^{-k/2}\binom{n}{k/2}\frac{k!}{2^{k/2}}2^{k/2} (3K^2)^{k/2} o(1)\\
        =&n^{-k/2}\sum_{1\leq j_1<\ldots<j_{k/2}\leq n}\frac{k!}{2^{k/2}}\sum_{\mathcal{K}\subseteq\{1,\ldots,{k/2}\}}\prod_{\ell\in\mathcal{K}}\E\left[f_{j_\ell}^{2}|X_{j_\ell}\right]\prod_{{\ell'}\in\mathcal{K}^c}\E\left[\left(2f_{j_{\ell'}}g_{j_{\ell'}}+g_{j_{\ell'}}^2\right)\pi_{j_{\ell'}}|X_{j_{\ell'}}\right]\\
            &+ o(1)\end{align*}
            Factorization formula $\sum_{\mathcal{K}\subseteq\{1,\ldots,m\}}\prod_{\ell \in \mathcal{K}}x_{\ell}\prod_{\ell' \in \mathcal{K}^{c}}y_{\ell'}=\prod_{\ell=1}^m(x_{\ell}+y_{\ell})$ ensures
            \begin{align*}
                B_{n,k,k/2}
    &=\frac{k!}{2^{k/2}}n^{-k/2}\sum_{1\leq j_1<\ldots<j_{k/2}\leq n}\prod_{\ell=1}^{k/2}\E\left[ f_{j_\ell}^2+\left(2f_{j_\ell}g_{j_\ell}+g_{j_\ell}^2\right)\pi_{j_\ell}\Big| X_{j_{\ell}}\right]+o(1)\\
        &=\frac{k!}{2^{k/2}} n^{-k/2}\binom{n}{k/2}\binom{n}{k/2}^{-1}\sum_{1\leq j_1<\ldots<j_{k/2}\leq n}h(X_{j_1},...,X_{j_{k/2}})+o(1)
\end{align*}
for $h(u_1,...,u_{k/2})=\prod_{i=1}^{k/2}\E(f^2+(2fg+g^2)\pi|X=u_i)$. Strong law of large numbers for U-statistics \citep{aaronson_strong_1996} ensures that $\binom{n}{k/2}^{-1}\sum_{1\leq j_1<\ldots<j_{k/2}\leq n}h(X_{j_1},...,X_{j_{k/2}})$ converges almost surely to $\E(h(X_1,...,X_{k/2}))=(V_0)^{k/2}$ and $\lim_n n^{-k/2}\binom{n}{k/2}=\frac{1}{(k/2)!}$.  Then, $\lim_n M_{n,k}=0$ for $k$ odd, and $\lim_n M_{n,k}=\frac{k!}{2^{k/2}(k/2)!}V_0^{k/2}$ for $k$ even. By the adapted form of the theorem in \citet{takacs_moment_1991}, if $f_i$ and $g_i$, are bounded, we have that, conditional on $(X_i)_{i\geq 1}$ $\frac{1}{\sqrt{n}}\sum_{i=1}^nf_i+g_iD_i$ converges almost surely to a Gaussian of variance $V_0$.\\~\\
\textbf{Second step: $\E(Y(0)^2+Y(1)^2+||X||^2)<\infty$ implies \eqref{eq:an}}\\
Assumption \ref{hyp:iid+mom} ensures only that $f_i$ and $g_i$ admit moments of order 2. Then, for $M>0$, let $f_{\leq M,i}$, $f_{> M,i}$, $g_{\leq M,i}$ and $g_{> M,i}$ the truncated variables $f_{\leq M,i}=f_i\mathbbm{1}\{|f_i|\leq M\}$, $f_{> M,i}=f_i\mathbbm{1}\{|f_i|> M\}$, $g_{\leq M,i}=g_i\mathbbm{1}\{|g_i|\leq M\}$ and $g_{> M,i}=g_i\mathbbm{1}\{|g_i|> M\}$. We define $\Tilde{f}_{\leq M,i}=f_{\leq M,i}-\E[f_{\leq M,i}|X_i]$, $\Tilde{f}_{> M,i}=f_{> M,i}-\E[f_{> M,i}|X_i]$, $\Tilde{g}_{\leq M,i}=g_{\leq M,i}-\E[g_{\leq M,i}|X_i]$ and $\Tilde{g}_{> M,i}=g_{> M,i}-\E[g_{> M,i}|X_i]$. We have:
\begin{align*}
	&\E\left[\left.\left|\frac{1}{\sqrt{n}}\sum_{i=1}^n(\Tilde{f}_{> M,i}+\Tilde{g}_{> M,i}D_i)\right|^2 \right|  (X_\ell)_{\ell\geq1}\right]\\
   =&\frac{1}{n}\sum_{i=1}^n \E\left[(\Tilde{f}_{> M,i}+\Tilde{g}_{> M,i}D_i)^2|(X_\ell)_{\ell\geq1}\right]+\frac{1}{n}\sum_{\substack{1\leq i,j \leq n \\ i\neq j}}\E\left[\left(\Tilde{f}_{> M,i}+\Tilde{g}_{> M,i}D_i\right)\left(\Tilde{f}_{> M,j}+\Tilde{g}_{> M,j}D_j\right)|(X_\ell)_{\ell\geq1}\right]\\
   =&\frac{1}{n}\sum_{i=1}^n \E\left[\left.\Tilde{f}_{> M,i}^2\right|(X_\ell)_{\ell\geq1}\right]+\E\left[\left.\left(2\Tilde{f}_{> M,i}\Tilde{g}_{> M,i}+\Tilde{g}_{> M,i}^2\right)\right|(X_\ell)_{\ell\geq1}\right]\E\left[\left.D_i\right|(X_\ell)_{\ell\geq1}\right]\\
   +&\frac{1}{n}\sum_{\substack{1\leq i,j \leq n \\ i\neq j}}\Big(\E\left[\Tilde{f}_{> M,i}\Tilde{f}_{> M,j}|(X_\ell)_{\ell\geq1}\right]+\E\left[\Tilde{f}_{> M,i}\Tilde{g}_{> M,j}|(X_\ell)_{\ell\geq1}\right]\E\left[\left.D_j\right|(X_\ell)_{\ell\geq1}\right]\\
   &~~~~~~~~~~~~~~~~~~+\E\left[\Tilde{f}_{> M,j}\Tilde{g}_{> M,i}|(X_\ell)_{\ell\geq1}\right]\E\left[\left.D_i\right|(X_\ell)_{\ell\geq1}\right]+\E\left[\Tilde{g}_{> M,i}\Tilde{g}_{> M,j}|(X_\ell)_{\ell\geq1}\right]\E\left[D_iD_j|(X_\ell)_{\ell\geq1}\right]\Big)\\
   =&\frac{1}{n}\sum_{i=1}^n \E\left[\left.\Tilde{f}_{> M,i}^2\right|X_i\right]+\E\left[\left.\left(2\Tilde{f}_{> M,i}\Tilde{g}_{> M,i}+\Tilde{g}_{> M,i}^2\right)\right|X_i\right]\pi_i\\
   +&\frac{1}{n}\sum_{\substack{1\leq i,j \leq n \\ i\neq j}}\Big(\E\left[\Tilde{f}_{> M,i}|X_i\right]\E\left[\Tilde{f}_{> M,j}|X_j\right]+\E\left[\Tilde{f}_{> M,i}|X_i\right]\E\left[\Tilde{g}_{> M,j}|X_j\right]\pi_j\\
   &~~~~~~~~~~~~~~~~~~+\E\left[\Tilde{f}_{> M,j}|X_j\right]\E\left[\Tilde{g}_{> M,i}|X_i\right]\pi_i+\E\left[\Tilde{g}_{> M,i}|X_i\right]\E\left[\Tilde{g}_{> M,j}|X_j\right]\E\left[D_iD_j|(X_\ell)_{\ell\geq1}\right]\Big)\\
   =&\frac{1}{n}\sum_{i=1}^n \E\left[\Tilde{f}_{> M,i}^2+\left(2\Tilde{f}_{> M,i}\Tilde{g}_{> M,i}+\Tilde{g}_{> M,i}^2\right)\pi_i|X_i\right]
\end{align*}
The second equality holds because $(f_1,\ldots,f_n,g_1,\ldots,g_n)\indep (D_1,\ldots,D_n) |X_1,\ldots,X_n.$ by Assumption \ref{hyp:sample}. The third equality holds because $(f_i,g_i,X_i)_{i\geq 1}$ are independent across $i$ by Assumption \ref{hyp:iid+mom} and $\E\left[D_i|(X_{\ell})_{\ell>1}\right]=\pi_i$ by Assumption \ref{hyp:sample}. The fourth equality holds because $\E\left[\Tilde{f}_{> M,\ell}|X_\ell\right]=\E\left[\Tilde{g}_{> M,\ell}|X_\ell\right]=0$.\\
The SLLN ensures that $\frac{1}{n}\sum_{i=1}^n \E\left[\Tilde{f}_{> M,i}^2+\left(2\Tilde{f}_{> M,i}\Tilde{g}_{> M,i}+\Tilde{g}_{> M,i}^2\right)\pi_i|X_i\right]$ converges almost-surely to $\E\left[\Tilde{f}_{> M,1}^2+\left(2\Tilde{f}_{> M,1}\Tilde{g}_{> M,1}+\Tilde{g}_{> M,1}^2\right)\pi_1\right]$.
It follows that by Cauchy-Schwarz inequality:
\begin{align}
&\lim\sup_n \E\left[\left|\frac{1}{\sqrt{n}}\sum_{i=1}^n(\Tilde{f}_{> M,i}+\Tilde{g}_{> M,i}D_i)\right|\Big| (X_\ell)_{\ell\geq1}\right]\notag\\
&\leq \lim\sup_n \E\left[\left.\left|\frac{1}{\sqrt{n}}\sum_{i=1}^n(\Tilde{f}_{> M,i}+\Tilde{g}_{> M,i}D_i)\right|^2 \right|  (X_\ell)_{\ell\geq1}\right]^{1/2}\notag\\
&= \E\left[\Tilde{f}_{> M,1}^2+\left(2\Tilde{f}_{> M,1}\Tilde{g}_{> M,1}+\Tilde{g}_{> M,1}^2\right)\pi_1\right]^{1/2}\label{eq:ineg_limsup}
\end{align}
which, by dominated convergence, is arbitrarily small for a sufficiently large $M$.\\
Let $h$ a bounded Lipschitz function of constant $c_h$, $V(M)=\E\left[\Tilde{f}_{\leq M,1}^2+\left(2\Tilde{f}_{> M,1}\Tilde{g}_{\leq M,1}+\Tilde{g}_{\leq M,1}^2\right)\pi_1\right]$, and $N\sim \mathcal{N}(0,1)$. We have by triangle and Lipschitz inequlities, and the fact that $f_i+g_iD_i=\Tilde{f}_{\leq M,i}+\Tilde{g}_{\leq M,i}D_i+\Tilde{f}_{> M,i}+\Tilde{g}_{> M,i}D_i$:
  \begin{align*}
 	&\left|\E\left[h\left(\frac{1}{\sqrt{n}}\sum_{i=1}^nf_i+g_iD_i\right)\Big|(X_\ell)_{\ell\geq1}\right]-\E\left[h\left(V_0^{1/2}N\right)\right]\right|\\
 	&\leq  \left|\E\left[h\left(\frac{1}{\sqrt{n}}\sum_{i=1}^nf_i+g_iD_i\right)\Big|(X_\ell)_{\ell\geq1}\right]-\E\left[h\left(\frac{1}{\sqrt{n}}\sum_{i=1}^n\Tilde{f}_{\leq M,i}+\Tilde{g}_{\leq M,i}D_i\right)\Big|(X_\ell)_{\ell\geq1}\right]\right|\\
 	&+\left|\E\left[h\left(\frac{1}{\sqrt{n}}\sum_{i=1}^n\Tilde{f}_{\leq M,i}+\Tilde{g}_{\leq M,i}D_i\right)\Big|(X_\ell)_{\ell\geq1}\right]-\E\left[h\left(V(M)^{1/2}N\right)\right]\right|\\
 	&+\left|\E\left[h\left(V(M)^{1/2}N\right)\right]-\E\left[h\left(V_0^{1/2}N\right)\right]\right|\\
 	&\leq c_h\E\left[\left|\frac{1}{\sqrt{n}}\sum_{i=1}^n\Tilde{f}_{> M,i}+\Tilde{g}_{> M,i}D_i\right|\Big| (X_\ell)_{\ell\geq1}\right]\\
 	&+\left|\E\left[h\left(\frac{1}{\sqrt{n}}\sum_{i=1}^n\Tilde{f}_{\leq M,i}+\Tilde{g}_{\leq M,i}D_i\right)\Big|(X_\ell)_{\ell\geq1}\right]-\E\left[h\left(V(M)^{1/2}N\right)\right]\right|\\
 	&+c_h\left|V(M)^{1/2}-V_0^{1/2}\right|\E(|N|).
 \end{align*}
The first step of the proof and \eqref{eq:ineg_limsup} ensure that for any value of $M>0$:
\begin{align*}
    &\limsup_n \left|\E\left[h\left(\frac{1}{\sqrt{n}}\sum_{i=1}^nf_i+g_iD_i\right)\Big|(X_\ell)_{\ell\geq1}\right]-\E\left[h\left(V_0^{1/2}N\right)\right]\right|\\
    &\leq  c_h\left(\E\left[\Tilde{f}_{> M,1}^2+\left(2\Tilde{f}_{> M,1}\Tilde{g}_{> M,1}+\Tilde{g}_{> M,1}^2\right)\pi_1\right]^{1/2}+\left|V(M)^{1/2}-V_0^{1/2}\right|\right).
\end{align*}
By dominated convergence, $\lim_M V(M)=V_0$ and $\lim_M\E\left[\Tilde{f}_{> M,1}^2+\left(2\Tilde{f}_{> M,1}\Tilde{g}_{> M,1}+\Tilde{g}_{> M,1}^2\right)\pi_1\right]=0$. Next, considering $M$ tending to $\infty$, dominated convergence ensures
$$\limsup_n \left|\E\left[h\left(\frac{1}{\sqrt{n}}\sum_{i=1}^nf_i+g_iD_i\right)\Big|(X_\ell)_{\ell\geq1}\right]-\E\left[h\left(V_0^{1/2}N\right)\right]\right|=0.$$
This achieves the proof.
\newpage
\section{Additional Tables and Figures}
\begin{table}[H]
	\caption{Empirical application: Standard deviation of ATE estimators}
	\label{tb:empsd}
	\centering\scriptsize
	\begin{threeparttable}
		\begin{tabularx}{\textwidth}{cY|YYYY}
			\specialrule{.1em}{.05em}{.05em}\specialrule{.1em}{.05em}{.05em}
			&Number of covariates   & Complete Randomization & Stratified Randomization & Matched \quad\quad\quad Pairs & Cube \quad\quad\quad Method \\  
			&		& (1)         & (2)                      & (3)                     & (4)                 \\\specialrule{.1em}{.05em}{.05em}
			&1 & 2.958 & 2.662 & 2.363 & 2.382 \\ 
			&2 & --   & 2.673 & 2.357 & 2.438   \\   
			&3 &  --  & 2.658 & 2.390 & 2.448   \\   
			$n=100$&5 &  --  & 2.816 & 2.427 & 2.428   \\   
			&7 &  --  & 3.153 & 2.459 & 2.434 \\   
			&9 &  --  & 3.745 & 2.530 & 2.389 \\   
			&12 &  --  & 6.770 & 2.563 & 2.391\\\specialrule{.1em}{.05em}{.05em}
			&1 & 1.825 & 1.675 & 1.455 & 1.478 \\ 
			&  2 & -- & 1.664 & 1.456 & 1.491  \\ 
			&  3 &  --  & 1.673 & 1.459 & 1.485 \\ 
			$n=256$ &5&--& 1.673 & 1.506 & 1.483\\ 
			&  7 & -- & 1.771 & 1.532 & 1.492\\ 
			& 9 & -- & 2.004 & 1.545 & 1.468\\ 
			& 12 & -- &2.818 & 1.593 & 1.505\\  
			\specialrule{.1em}{.05em}{.05em}
			&1 & 1.299 & 1.189 & 1.052 & 1.055  \\
			&  2 & -- & 1.174 & 1.052 & 1.048 \\ 
			&  3 & -- & 1.173 & 1.040 & 1.060  \\ 
			$n=500$&5&--& 1.188 & 1.050 & 1.057  \\ 
			& 7 & -- & 1.227 & 1.055 & 1.059 \\ 
			&  9 & -- & 1.317 & 1.085 & 1.063\\ 
			&  12 & -- & 1.691 & 1.107 & 1.055  \\ 
			\specialrule{.1em}{.05em}{.05em}
			&1 &0.918 & 0.836 & 0.738 & 0.751  \\ 
			&  2 & -- & 0.835 & 0.735 & 0.739 \\ 
			&  3 & -- &0.837 & 0.736 & 0.734 \\ 
			$n=1000$&  5 & -- &0.832 & 0.741 & 0.750\\ 
			&  7 & -- & 0.844 & 0.745 & 0.748 \\ 
			& 9 & -- & 0.898 & 0.761 & 0.748  \\ 
			& 12 & -- & 1.059 & 0.778 & 0.744 \\ 
			\specialrule{.1em}{.05em}{.05em}\specialrule{.1em}{.05em}{.05em}
		\end{tabularx}
		\begin{tablenotes}[flushleft,para]
			\scriptsize
			This table shows the standard deviation of PATE estimators for different allocation designs and experimental sample sizes. For each allocation design, the standard deviation estimates are computed over $10,000$ simulations. For columns 1, 3, and 4, the estimator used is the Horvitz-Thompson algorithm. In column, 1 the design used is complete randomization. For column 3, we assign treatment using a matched pairs design, pairing individuals to the closest unit and using the Mahalanobis distance whenever more than one covariate is balanced. Column 4 shows the results for the cube method with two moments for each variable. For column 2, we run stratified randomization. We use median values for continuous variables and we estimate the PATE using OLS with strata fixed-effects in accordance with \citet{bugni_inference_2018}.
		\end{tablenotes}
	\end{threeparttable}
\end{table}

\begin{table}[ht]
	\caption{Empirical application: Average bias of ATE estimators}
	\label{tb:empbias}
	\centering\scriptsize
	\begin{threeparttable}
		\begin{tabularx}{\textwidth}{cY|YYYY}
			\specialrule{.1em}{.05em}{.05em}\specialrule{.1em}{.05em}{.05em}
			&Number of covariates   & Complete Randomization & Stratified Randomization & Matched \quad\quad\quad Pairs & Cube \quad\quad\quad Method
            \\  
			&		& (1)         & (2)                      & (3)                     & (4)                  \\\specialrule{.1em}{.05em}{.05em}
			&1 & 0.046 & -0.065 & -0.007 & 0.032 \\ 
			&2 & --   & -0.025 & -0.028 & 0.023  \\   
			&3 &  --  & -0.047 & -0.058 & -0.031 \\   
			$n=100$&5 &  --  & -0.040 & -0.064 & -0.051 \\   
			&7 &  --  & -0.010 & -0.011 & -0.003\\   
			&9 &  --  & 0.115 & -0.036 & 0.008\\   
			&12 &  --  & 0.359 & -0.008 & -0.043 \\\specialrule{.1em}{.05em}{.05em}
			&1 & -0.016 & -0.028 & -0.006 & -0.038  \\ 
			&  2 & -- & -0.010 & -0.019 & -0.027 \\ 
			&  3 &  --  & -0.014 & -0.019 & -0.019 \\ 
			$n=256$ &5&--& -0.032 & -0.022 & -0.004 \\ 
			&  7 & -- & -0.015 & -0.010 & -0.025 \\ 
			& 9 & -- & 0.048 & -0.033 & -0.030 \\ 
			& 12 & -- & 0.222 & -0.030 & -0.018\\  
			\specialrule{.1em}{.05em}{.05em}
			&1 & -0.010 & -0.005 & 0.009 & 0.007   \\
			&  2 & -- & 0.011 & -0.027 & 0.000\\ 
			&  3 & -- & -0.013 & 0.005 & -0.011 \\ 
			$n=500$&5&--& 0.028 & -0.002 & 0.006  \\ 
			& 7 & -- & 0.003 & -0.007 & 0.002 \\ 
			&  9 & -- & 0.074 & -0.019 & -0.005  \\ 
			&  12 & -- & 0.146 & -0.024 & -0.015  \\ 
			\specialrule{.1em}{.05em}{.05em}
			&1 & -0.015 & -0.008 & -0.013 & -0.002  \\ 
			&  2 & -- & -0.000 & 0.002 & -0.006 \\ 
			&  3 & -- & -0.002 & -0.008 & -0.002 \\ 
			$n=1000$&  5 & -- & 0.003 & -0.006 & 0.002 \\ 
			&  7 & -- & 0.004 & -0.009 & 0.004\\ 
			& 9 & -- & 0.021 & -0.010 & -0.003 \\ 
			& 12 & -- & 0.064 & -0.004 & -0.008\\ 
			\specialrule{.1em}{.05em}{.05em}\specialrule{.1em}{.05em}{.05em}
		\end{tabularx}
		\begin{tablenotes}[flushleft,para]
			\scriptsize
				This table shows the average bias of PATE estimators for different allocation designs and experimental sample sizes. For each allocation design, the average bias estimates are computed over $10,000$ simulations. For columns 1, 3, and 4, the estimator used is the Horvitz-Thompson algorithm. In column 1, the design used is complete randomization. For column 3, we assign treatment using a matched pairs design, pairing individuals to the closest unit and using the Mahalanobis distance whenever more than one covariate is balanced. Column 4 shows the results for the cube method with two moments for each variable. For column 2, we run stratified randomization. We use median values for continuous variables and we estimate the PATE using OLS with strata fixed-effects in accordance with \citet{bugni_inference_2018}.
		\end{tablenotes}
	\end{threeparttable}
\end{table}

\begin{table}[ht]
	\caption{Empirical application: Coverage Rates}
	\label{tb:empcoverage}
	\centering\scriptsize
	\begin{threeparttable}
		\begin{tabularx}{\textwidth}{cY|YYYY}
			\specialrule{.1em}{.05em}{.05em}\specialrule{.1em}{.05em}{.05em}
			&Number of covariates   & Complete Randomization & Stratified Randomization & Matched \quad\quad\quad Pairs & Cube \quad\quad\quad Method \\  
			&		& (1)         & (2)                      & (3)                     & (4)               \\\specialrule{.1em}{.05em}{.05em}
		  &1 & 0.945 & 0.943 & 0.948 & 0.940  \\ 
		  &2 & --   & 0.937 & 0.949 & 0.932 \\   
		  &3 &  --  & 0.937 & 0.950 & 0.931  \\   
		  $n=100$&5 &  --  & 0.910 & 0.953 & 0.927   \\   
		  &7 &  --  & 0.884 & 0.956 & 0.924  \\   
		  &9 &  --  & 0.865 & 0.956 & 0.928  \\   
		  &12 &  --  & 0.776 & 0.955 & 0.927 \\\specialrule{.1em}{.05em}{.05em}
			&1 & 0.951 & 0.946 & 0.951 & 0.944  \\ 
			&  2 & -- & 0.947 & 0.953 & 0.939  \\ 
			&  3 &  --  & 0.943 & 0.954 & 0.943 \\ 
			$n=256$ &5&--& 0.938 & 0.952 & 0.943  \\ 
			&  7 & -- & 0.918 & 0.952 & 0.939 \\ 
			& 9 & -- & 0.902 & 0.955 & 0.941 \\ 
			& 12 & -- & 0.858 & 0.955 & 0.930 \\  
			\specialrule{.1em}{.05em}{.05em}
			&1 & 0.952 & 0.951 & 0.949 & 0.950  \\
			&  2 & -- & 0.950 & 0.948 & 0.949  \\ 
			&  3 & -- & 0.951 & 0.954 & 0.950 \\ 
			$n=500$&5&--& 0.943 & 0.955 & 0.945 \\ 
			& 7 & -- & 0.931 & 0.957 & 0.946 \\ 
			&  9 & -- & 0.919 & 0.957 & 0.941  \\ 
			&  12 & -- & 0.883 & 0.958 & 0.944 \\ 
			\specialrule{.1em}{.05em}{.05em}
			&1 & 0.955 & 0.952 & 0.953 & 0.945  \\ 
			&  2 & -- & 0.953 & 0.953 & 0.950  \\ 
			&  3 & -- & 0.952 & 0.954 & 0.947  \\ 
			$n=1000$&  5 & -- & 0.947 & 0.954 & 0.947 \\ 
			&  7 & -- & 0.939 & 0.957 & 0.948  \\ 
			& 9 & -- & 0.927 & 0.959 & 0.946 \\ 
			& 12 & -- & 0.902 & 0.955 & 0.946  \\ 
			\specialrule{.1em}{.05em}{.05em}\specialrule{.1em}{.05em}{.05em}
		\end{tabularx}
		\begin{tablenotes}[flushleft,para]
			\scriptsize
		This table shows the coverage rate of 95\%-confidence intervals of PATE estimators for different allocation designs and experimental sample sizes. For each allocation design, the coverage rate estimates are computed over $10,000$ simulations. For columns 1, 3, and 4, the estimator used is the Horvitz-Thompson algorithm. In column 1, the design used is complete randomization and the confidence intervals are constructed using White standard errors. For column 3, we assign treatment using a matched pairs design, pairing individuals to the closest unit and using the Mahalanobis distance whenever more than one covariate is balanced. Confidence intervals are constructed as described in \citep{bai_optimality_2022}. Column 4 shows the results for the cube method with two moments for each variable. Confidence intervals follow the asymptotic-based procedure described in Section \ref{sec:inf}. For column 2, we run stratified randomization. We use median values for continuous variables and we estimate the PATE using OLS with strata fixed-effects in accordance with \citet{bugni_inference_2018}.
		\end{tablenotes}
	\end{threeparttable}
\end{table}

\begin{table}[ht]
	\caption{Empirical application: Test Power}
	\label{tb:emppower}
	\centering\scriptsize
	\begin{threeparttable}
		\begin{tabularx}{\textwidth}{cY|YYYY}
			\specialrule{.1em}{.05em}{.05em}\specialrule{.1em}{.05em}{.05em}
			&Number of covariates   & Complete Randomization & Stratified Randomization & Matched \quad\quad\quad Pairs & Cube \quad\quad\quad Method  \\  
			&		& (1)         & (2)                      & (3)                     & (4)             \\\specialrule{.1em}{.05em}{.05em}
			&1 & 0.151 & 0.191 & 0.223 & 0.226  \\ 
			&2 & --   & 0.194 & 0.215 & 0.234   \\   
			&3 &  --  & 0.207 & 0.209 & 0.249  \\   
			$n=100$&5 &  --  & 0.233 & 0.197 & 0.256   \\   
			&7 &  --  &0.235 & 0.181 & 0.257  \\   
			&9 &  --  & 0.220 & 0.180 & 0.255  \\   
			&12 &  --  & 0.251 & 0.171 & 0.262 \\\specialrule{.1em}{.05em}{.05em}
			&1 & 0.325 & 0.385 & 0.465 & 0.477 \\ 
			&  2 & -- &0.389 & 0.464 & 0.475 \\ 
			&  3 &  --  & 0.393 & 0.457 & 0.476\\ 
			$n=256$ &5&--& 0.416 & 0.448 & 0.478 \\ 
			&  7 & -- & 0.429 & 0.423 & 0.486 \\ 
			& 9 & -- &0.382 & 0.403 & 0.496 \\ 
			& 12 & -- & 0.288 & 0.394 & 0.499\\  
			\specialrule{.1em}{.05em}{.05em}
			&1 & 0.552 & 0.638 & 0.740 & 0.739   \\
			&  2 & -- & 0.637 & 0.749 & 0.743 \\ 
			&  3 & -- & 0.642 & 0.735 & 0.749\\ 
			$n=500$&5&--& 0.640 & 0.725 & 0.743  \\ 
			& 7 & -- & 0.663 & 0.714 & 0.750\\ 
			&  9 & -- & 0.605 & 0.692 & 0.752 \\ 
			&  12 & -- & 0.483 & 0.681 & 0.754 \\ 
			\specialrule{.1em}{.05em}{.05em}
			&1 & 0.847 & 0.907 & 0.961 & 0.957 \\ 
			&  2 & -- &0.904 & 0.960 & 0.958 \\ 
			&  3 & -- &  0.904 & 0.959 & 0.959   \\ 
			$n=1000$&  5 & -- & 0.912 & 0.957 & 0.959 \\ 
			&  7 & -- & 0.914 & 0.951 & 0.956\\ 
			& 9 & -- & 0.893 & 0.945 & 0.961 \\ 
			& 12 & -- &0.810 & 0.932 & 0.959 \\ 
			\specialrule{.1em}{.05em}{.05em}\specialrule{.1em}{.05em}{.05em}
		\end{tabularx}
		\begin{tablenotes}[flushleft,para]
			\scriptsize
				This table shows the rejection power of 95\%-confidence intervals of PATE estimators for different allocation designs and experimental sample sizes. For each allocation design, the power estimates are computed over $10,000$ simulations. For columns 1, 3, and 4, the estimator used is the Horvitz-Thompson algorithm. In column 1, the design used is complete randomization and the confidence intervals are constructed using White standard errors. For column 3, we assign treatment using a matched pairs design, pairing individuals to the closest unit and using the Mahalanobis distance whenever more than one covariate is balanced. Confidence intervals are constructed as described in \citep{bai_optimality_2022}. Column 4 shows the results for the cube method with two moments for each variable. Confidence intervals follow the asymptotic-based procedure described in Section \ref{sec:inf}. For column 2, we run stratified randomization. We use median values for continuous variables and we estimate the PATE using OLS with strata fixed-effects in accordance with \citet{bugni_inference_2018}.
		\end{tablenotes}
	\end{threeparttable}
\end{table}

\end{document}